\documentclass[aps,prd,amsmath,amssymb,eqsecnum,nofootinbib]{revtex4}
\usepackage{color}
\usepackage[pdftex]{graphicx}

\newcommand{\leaveout}[1]{}
\newcommand{\fixme}[1]{}
\newcommand{\fixmeAO}[1]{}

\newcommand{\f}{f_{\ell}}

\newcommand{\fb}{\hat f_{\ell}}

\newcommand{\g}{g_{\ell}}
\newcommand{\gt}{g_{\ell}}
\newcommand{\gp}{g_{\ell+}}
\newcommand{\gpt}{g_{\ell+}}
\newcommand{\gpmt}{g_{\ell\pm}}
\newcommand{\gm}{g_{\ell-}}
\newcommand{\gmt}{g_{\ell-}}

\newcommand{\Dgt}{\Delta \gt}

\newcommand{\Aout}{A^{out}_{\ell,\omega}}

\newcommand{\Ain}{A^{in}_{\ell,\omega}}
\newcommand{\Ainp}{A^{in}_{\ell,\omega+}}
\newcommand{\Ainm}{A^{in}_{\ell,\omega-}}

\newcommand{\nb}{\bar\nu}
\newcommand{\ob}{\bar{\omega}}
\newcommand{\rb}{\bar r}

\newcommand{\nMB}{\nu_0}
\newcommand{\DGw}[3]{\Delta G_{\ell}(#1,#2;#3)}
\newcommand{\F}{F_{\ell}}
\newcommand{\Fb}{\hat F_{\ell}}
\newcommand{\iv}{u}
\newcommand{\W}[1]{W\left(#1\right)}
\newcommand{\Wb}[1]{\hat W\left(#1\right)}
\newcommand{\Wp}[1]{W_+\left(#1\right)}

\newcommand{\pF}{\psi^F}
\newcommand{\tb}{t}
\newcommand{\re}[1]{\text{Re}\left(#1\right)}
\newcommand{\im}[1]{\text{Im}\left(#1\right)}
\renewcommand{\aa}{\alpha_1}
\newcommand{\ab}{\alpha_2}

\newcommand{\ba}{\beta_1}
\newcommand{\bb}{\beta_2}

\newcommand{\dsa}{d_1}

\newcommand{\ca}{c} 

\newcommand{\wQNM}{\omega_{ln}}
\newcommand{\wbQNM}{\ob_{ln}}
\newcommand{\nuQNM}{\nu_{ln}}

\newcommand{\A}{A}

\begin{document}
\global\parskip 6pt


\author{Marc Casals}
\email{mcasals@perimeterinstitute.ca; mcasals@uoguelph.ca; marc.casals@ucd.ie}
\affiliation{Perimeter Institute for Theoretical Physics, Waterloo, Ontario, Canada N2L 2Y5}
\affiliation{Department of Physics,
University of Guelph,
Guelph, Ontario, Canada N1G 2W1}
\affiliation{School of Mathematical Sciences and Complex \& Adaptive Systems
Laboratory, University College Dublin, Belfield, Dublin 4, Ireland}

\author{Adrian Ottewill}
\email{adrian.ottewill@ucd.ie}
\affiliation{School of Mathematical Sciences and Complex \& Adaptive Systems
Laboratory, University College Dublin, Belfield, Dublin 4, Ireland}

\title{The Branch Cut and Quasi-normal Modes at large imaginary frequency\\ in Schwarzschild space-time}

\begin{abstract}
The `retarded' Green function for fields propagating on a Schwarzschild black hole spacetime possesses a branch cut on the complex frequency plane.
Classically, the branch cut is important, for example, in order to fully determine the response of the  black hole to a linear field perturbation.
The branch cut is also useful for the calculation of the self-force on a point particle moving in the Schwarzschild background.
In this paper we use techniques of analytic-continuation to the complex plane of the radial coordinate in order
to calculate the branch cut contribution to the Green function in the limit of large imaginary frequency.
It is expected that the contribution of this frequency regime to the perturbation response and to the self-force will be mostly for short time intervals.
We also determine the highly-damped quasinormal mode frequencies for electromagnetic perturbations in Schwarzschild for the first time 
(previously only the leading imaginary part was known), which 
seem to have a `deep connection' with the branch cut.
We find that these frequencies behave like 
$ \wQNM
= -\dfrac{in}{2}-\dfrac{i\left[\ell(\ell+1)\right]^2}{2n}+\dfrac{\pi^{1/2}(1-i)\left[\ell(\ell+1)\right]^3}{2^{3/2}n^{3/2}}
+O\left(n^{-2}\right)$.
The highly-damped quasinormal modes are particularly interesting for theories of quantum gravity
in that they are believed to probe the
small scale structure of the spacetime. 
\end{abstract}

\date{\today}
\maketitle



\section{Introduction}

In the investigation of spin-field perturbations on a curved spacetime, the 
 `retarded' Green function plays a crucial r\^{o}le. 
  For example, the full evolution in time of some initial data may be determined by integrating over space
  the `retarded' Green function convolved with the initial data.
 Also, the self-force acting on a point particle 
 moving on a background spacetime may be calculated via an integration of the  `retarded' Green function
 over the whole past worldline of the particle~\cite{Poisson:2011nh}.
Similarly in quantum field theory the covariant commutation relations are determined by
the `advanced' and `retarded' Green functions. 
While in some contexts such as renormalization a knowledge of the short-distance behaviour of the Green 
function is sufficient,  in other contexts the knowledge of the global behaviour of the Green function is important.

The `retarded' Green function for linear field perturbations of the Schwarzschild black hole spacetime can be expressed as a multipole decomposition
together with a Fourier integral 
over the frequencies just above the real axis.
In the impressive work of~\cite{Leaver:1986}, Leaver deforms the frequency integral into a contour in the complex-frequency plane and he investigates
three distinct contributions to the Green function coming from: (1) a high-frequency arc, (2) the poles 
(quasinormal modes) in the lower plane, and (3) a branch cut (BC) along the negative imaginary axis (NIA).
Under a linear field perturbation, the high-frequency arc leads to a prompt response; 
the poles lead to the well-known quasinormal mode `ringing'
(this `ringing' was observed for the first time in~\cite{Vishveshwara:1970zz});
the BC leads to, at least, a power-law tail decay with time~\cite{Price:1971fb,Price:1972pw}.
While the quasinormal modes (QNMs) have been extensively investigated, most work on the BC has been limited to small frequency on the NIA
(and usually also large radial coordinates), which is precisely
the regime that gives the power-law tail - see, e.g.~\cite{Leaver:1986,Andersson:1997}.
The BC, however, is also expected~\cite{Leaver:1986} to have a significant contribution at `early' times, just after the start of the prompt response from the  high-frequency arc.
In fact, at `very early' times, i.e., before the prompt response, the separate contributions from the high-frequency arc, QNMs and BC might
all separately diverge with their divergences cancelling each  other out so that the complex-frequency contour integration is still valid.
It is expected that these BC contributions at `early' and `very early' times come from the mid- and high-frequency regimes of the BC on the NIA, which
might also bring about other yet-unsuspected contributions.
To the best of our knowledge, these  mid- and high-frequency regimes of the BC have only been investigated, respectively,
in~\cite{Leung:2003ix,Leung:2003eq,MaassenvandenBrink:2000ru} 
and in Maassen van den Brink's~\cite{MaassenvandenBrink:2003as}, and it was done solely for the
case of gravitational perturbations.
In~\cite{MaassenvandenBrink:2003as}, Maassen van den Brink investigated the BC in Schwarzschild for high-frequencies using
a  method based on an analytic continuation to the complex plane of the radial coordinate $r$. This method is mirrored on the method
used in~\cite{Motl&Neitzke,Neitzke:2003mz} for the calculation of highly-damped QNMs, i.e., QNMs for large
overtone index $n$, which therefore lie far down in the lower frequency plane.

On the quantum side, highly-damped QNMs have attracted considerable attention since attempts~\cite{Hod:1998vk,Dreyer2003,Maggiore:2007nq} have been made at linking them to the area quantization
 of a black hole~\cite{Bekenstein:1974jk,Bekenstein:1995ju}.
For example,
in~\cite{Padmanabhan:2003fx},  the imaginary part of the highly-damped QNMs is shown to be related to the exponential redshift of the wave modes close to the horizon.
In~\cite{Keshet:2007be}, highly-damped QNMs in Kerr spacetime have been interpreted as semiclassical bound states along a specific contour
in the complex-$r$ plane; they speculate that QNMs and another set of modes (the so-called totally-transmitted modes) correspond to different sets of microscopic
degrees of freedom which, when they interact, produce Hawking radiation.
Separately, in~\cite{Babb:2011ga} they have shown that highly-damped QNMs probe the short length scale structure of the
Schwarzschild black hole spacetime by calculating them in the context of a `quantum-corrected' black hole.

To set the current work in context we here give a succinct review of the -mostly- analytic results obtained in the literature for highly-damped QNM frequencies in
the  Schwarzschild spacetime
for various spin-$s$ fields.
Motl and Neitzke~\cite{Motl&Neitzke,Neitzke:2003mz} originally calculated analytically the leading-order of the highly-damped QNMs for $s=0$ and $2$
(\cite{Neitzke:2003mz} also predicted the dependence of their next-to-leading order term on the `angular momentum number' $\ell$).
Subsequently, Maassen van den Brink~\cite{MaassenvandenBrink:2003as} for $s=2$ and Musiri and Siopsis~\cite{Musiri:2003bv} for $s=0$ and $2$
confirmed the leading-order behaviour and obtained the next-to-leading order highly-damped
QNMs.
In~\cite{Motl:2002hd,Musiri:2003bv} they also give the leading-order for the imaginary part -though not for the real part- of the QNM frequencies for $s=1/2$, $1$ and $3/2$.
Finally, in~\cite{Musiri:2007zz} they obtain the leading-order and next-to-leading order QNMs for $s=1/2$ and $s=5/2$.
To the best of our knowledge, the leading-order for the real part of the QNM frequencies for $s=1$ and $s=3/2$ remains unknown,
and only the leading order of their imaginary parts is known.
The case of electromagnetic ($s=1$) perturbations is particularly interesting since the real part of the  spin-1 highly-damped QNM frequencies
have been expected to approach the NIA, thus hinting at an unexplored possible connection between the highly-damped QNMs and the BC.
Indeed, 
in~\cite{Musiri:2003bv,Musiri:2007zz} they show that the real part of the electromagnetic QNM frequencies can only go at most like $n^{-1}$ for $n\to\infty$,
and in~\cite{Cardoso:2003vt} they find numerical indications that it goes like $n^{-3/2}$,
with a coefficient which is a 3rd order polynomial in $\ell (\ell+1) $ (with undetermined polynomial coefficients). We will confirm this behaviour and determine the polynomial 
coefficients in Sec.\ref{sec:QNMs}.
This behaviour is in contrast with the $O(1)$ behaviour of the corresponding QNM frequencies for $s=0$ and $2$ and faster-decaying than the $O(n^{-1/2})$ for $s=1/2$ and $5/2$.
See~\cite{Berti:2009kk} for a thorough and recent review of QNMs in different spacetimes; see also~\cite{Konoplya:2011qq}.

As mentioned, the contributions of the mid- and high-frequency regimes of the BC still remain largely unexplored.
In the present paper we 
use the method in~\cite{MaassenvandenBrink:2003as} (which was restricted to $s=2$) to
carry out an asymptotic calculation  in the high-frequency regime of 
the BC contribution to the `retarded' Green function in Schwarzschild for fields
of spin $s=0$ (scalar) and $1$ (electromagnetic).
We also calculate the highly-damped QNM frequencies for electromagnetic perturbations
for the  first time in the literature (other than the leading-order of the imaginary part, which is known):
we calculate these frequencies  up to order $n^{-2}$, that is, leading-order 
for the real
part and up to two orders after leading-order for the imaginary part.
The feature that these QNMs approach the NIA `so fast'
 has made them very unyielding:
like we show in this paper, in order to obtain the leading-order behaviour for large-frequency for spin-$1$
 we are required to go up to two higher orders in perturbation theory than we are required for spins $0$ and $2$.
Electromagnetic QNMs are increasingly important, since the detection of the electromagnetic counterpart of the gravitational wave emission by astrophysical sources might play a valuable r\^ole for localizing the source and obtaining further information 
about it~\cite{Schnittman:2010wy}.
For completeness, we also reproduce in this paper the results in~\cite{MaassenvandenBrink:2003as} for the case of  $s=2$ (gravitational) 
and in~\cite{Musiri:2003bv} for the corresponding spin-$0$ QNM frequencies.

One particular physical application of the BC contribution to the `retarded' Green function
is the calculation of the corresponding contribution to the response of the Schwarzschild black hole to some initial perturbation.
In this paper we investigate the r\^ole played by the high-frequency part of both the BC and the QNM contributions to such a response in the case of 
a spin-$s$(=0,1,2) field sourced by: (1) initial data of compact support and (2) a non-compact Gaussian distribution
in the `tortoise' radial coordinate.

In two other papers~\cite{Casals:Ottewill:2011smallBC,Casals:Ottewill:2011midBC}, 
we will investigate, using completely different methods, the BC in Schwarzschild in the mid-frequency 
and the small-frequency regimes along the NIA.
In most of the figures in the present paper we plot the various quantities required for the calculation of the BC contribution to the Green function.
In these figures we compare the high-frequency asymptotics obtained here with the results
 using the independent methods which we will present in~\cite{Casals:Ottewill:2011midBC}.
The method in~\cite{Casals:Ottewill:2011midBC}
 has good convergence in a `mid'-frequency regime which overlaps with the `high'-frequency regime, as shown here;
 it also overlaps with the `small'-frequency regime, as we will show in~\cite{Casals:Ottewill:2011smallBC}.

This paper is organized as follows:
In Sec.\ref{sec:BC GF} we give the general formulae for the BC contribution to the `retarded' Green function.
In Sec.\ref{sec:large-nu} we perform the asymptotic analysis of the BC Green function for high-frequency along the NIA. 
In Sec.\ref{sec:QNMs} we calculate the highly-damped QNMs.
In Sec.\ref{Perturbation} we investigate the contribution from the high-frequency regime of both the BC and the QNMs to
the black hole response to an initial perturbation.
In Sec.\ref{sec:conclusions} we present some conclusions.


\section{Branch cut contribution to the Green function} \label{sec:BC GF}

The `retarded' Green function for linear field perturbations in the Schwarzschild spacetime can be expressed in terms of a multipole decomposition
together with a Fourier transform as
\begin{align} \label{eq:Green}
&
G_{ret}(x,x')=
\sum_{\ell=0}^{\infty}(2\ell+1)P_{\ell}(\cos\gamma)G^{ret}_{\ell}(r,r';\Delta t),\quad
G^{ret}_{\ell}(r,r';\Delta t)\equiv
\frac{1}{2\pi}
\int_{-\infty+ic}^{\infty+ic} d\omega\ G_{\ell}(r,r';\omega)e^{-i\omega \Delta t},
\quad \Delta t\equiv t-t',
\\& 
G_{\ell}(r,r';\omega)=\frac{\f(r_<,\omega)\g(r_>,\omega)}{\W{\omega}},
\quad r_>\equiv \max(r,r'),\ r_<\equiv \min(r,r')
\nonumber
\\&
\W{\omega}\equiv 
W\left[\g(r,\omega),\f(r,\omega)\right]=\g(r,\omega)\frac{d\f(r,\omega)}{dr_*}-\f(r,\omega)\frac{d\g(r,\omega)}{dr_*}
\nonumber
\end{align}
where $c>0$, 
$t$ and $r$ are -- respectively -- the time and radial Schwarzschild coordinates,
$r_*=r+r_h\ln\left(\bar r-1\right)$ is the so-called  `tortoise' radial coordinate,
$r_h=2M$ is the radius of the event horizon,
$M$ is the mass of the black hole
and $\gamma$ is the angle between the spacetime points $x$ and $x'$.
Note that the physical region $r\in (r_h,\infty)$ corresponds to $r_*\in (-\infty,\infty)$.
A bar over a quantity indicates that the quantity has been made dimensionless via an appropriate factor of $r_h$, e.g.,
  $\rb \equiv r/r_h$, $\ob\equiv \omega r_h$,  $\bar{t}\equiv t/r_h$, etc. 
The function $\W{\omega}$ is the Wronskian of the 
two
 linearly independent solutions 
 $\f$ and $\g$
 of the following homogeneous radial ODE:
\begin{align} \label{eq:radial ODE}
&\left[\frac{d^2}{dr_*^2}+\omega^2-V(r)\right]
\psi_{\ell}(r,\omega)
=0
\\
&V(r)\equiv \left(1-\frac{r_h}{r}\right)\left[\frac{\lambda }{r^2}+\frac{r_h(1-s^2)}{r^3}\right],
\quad \lambda\equiv \ell (\ell+1)
\nonumber
\end{align}
The parameter $s$ denotes the spin of the field: 
$s=2$ corresponds to axial -- also called `odd' -- gravitational perturbations (in which case Eq.(\ref{eq:radial ODE}) becomes the Regge-Wheeler equation~\cite{Regge:1957td}), 
$s=1$ to electromagnetic perturbations~\cite{Wheeler:1955zz} 
and $s=0$ to scalar perturbations~\cite{Price:1971fb,Price:1972pw}.
The ODE Eq.(\ref{eq:radial ODE}) has two regular singular points at $r=0$, $r_h$ and an irregular singular point at $r=\infty$.
These singularities in the ODE generally cause the radial solutions to have branch points at $r=0$ and $r_h$
in the complex-$r$ plane.
For real $\omega$, the solutions $\f$ and $\g$ are uniquely determined by the boundary conditions:
\begin{align} \label{eq:f,near hor}
&
\f(r,\omega) \sim
e^{-i\omega r_*}
, \quad
r_*\to -\infty, 
\\& \gt(r,\omega) \sim  e^{+i\omega r_*}, \quad r_*\to +\infty \nonumber.
\end{align}
Strictly speaking, the limits in Eq.(\ref{eq:f,near hor}) should actually be $\bar{r}_* \to \mp \infty$, but following standard conventions
we denote them by $r_* \to \mp \infty$, with a certain abuse of language.
We also have
\begin{equation} \label{eq:f inf}
\f(r,\omega) \sim \Aout e^{+i\omega r_*}+\Ain e^{-i\omega r_*}, \quad  r_*\to +\infty, 
\end{equation}
where $\Aout \in\mathbb{C}$ and $\Ain\in\mathbb{C}$ are reflection and incidence coefficients, respectively.
It is immediate that the Wronskian is equal to $\W{\omega}=-2i\omega \Ain$.
The boundary condition (\ref{eq:f,near hor}) 
also define $\f$ and $\gt$ unambiguously for $\text{Im}(\omega)\ge 0$ when $r_*\in\mathbb{R}$;
these solutions are then defined for  $\text{Im}(\omega)< 0$ by analytic continuation
(see~\cite{MaassenvandenBrink:2000ru} for details on the region of validity of these boundary conditions including $r_*$ as well as $\omega$).
On the NIA, the boundary condition (\ref{eq:f,near hor}) becomes meaningless since it does not allow one to exclude the exponentially-decaying solution,
$ e^{+i\omega r_*}$ for $\f$ and $ e^{-i\omega r_*}$ for $\g$.
We are therefore motivated to analytically continue to the complex-$r$ plane, 
and impose these boundary conditions in the regions $\text{Re}(-i\omega r_*)\le 0$ for $\f $ and $\text{Re}(i\omega r_*)\le 0$ for $\g$, 
where they 
define the solutions uniquely (see also~\cite{Andersson:2003fh,1993CQGra..10..735A}).

 Leaver~\cite{Leaver:1986a} has shown that $\g(r,\omega)$ has a branch cut (BC) which can be naturally taken to run along the line $\omega r:0\to -\infty\cdot i$.
If $r>0$, then $\g(r,\omega)$ has a branch point at $\omega=0$ and a BC along the NIA, $\omega:0\to -\infty\cdot i$.
\fixme{This means that if we take $r$ into the complex plane then the cut on the frequency plane is not along the NIA anymore?}
In~\cite{Ching:1994bd,Ching:1995tj} they have shown that the existence of a BC in the complex-$\omega$ plane
 is linked to the asymptotic behaviour
of the radial potential: 
the exponentially-decaying Schwarzschild potential $V(r)\sim e^{\bar r_*-1}\left[\lambda+1-s^2\right]/r_h^2$ as $r_*\to -\infty$
leads to poles in $\f$ on the NIA (these poles, however, are cancelled out in $G_{\ell}$ by
the corresponding poles in $\W{\omega}$), whereas its slower-than-exponential decay
(with the exception of the centrifugal barrier)
$V(r)-\lambda/r_*^2\sim 2\lambda r_h\ln(\bar r_*)/r_*^3$ as $r_*\to \infty$ leads to the BC in $\g$.
The function $\f$ does not have a BC in the complex-$\omega$ plane~\cite{Leaver:1986a,Leaver:1986}, and so the BC of $G_{\ell}(r,r';\omega)$ along the NIA is 
due to the corresponding BC in $\g$.
Note that the Wronskian 
$
\W{\omega}$, and therefore also $\Ain$, inherits the BC on the NIA from $\g$
(see, e.g., Eq.128~\cite{Leaver:1986a} and Eq.34~\cite{Leaver:1986}).

It is convenient to define 
$\nu\equiv i\omega\in \mathbb{C}$ (and $\nb \equiv \nu r_h$), which is positive along the NIA. 
We also define
$\Delta A(-i\nu)\equiv  A_+(-i\nu)-A_-(-i\nu)$
for any function $A=A(\omega)$ possessing a BC along the NIA,
where
 $ A_{\pm}(-i\nu)\equiv \lim_{\epsilon\to 0^+}A(\pm\epsilon-i\nu)$, with $\nu>0$.
That is, $\Delta A$ is the discontinuity of $A(\omega)$ across the NIA.

As mentioned above, $\f$ has poles on the NIA; these lie at 
$\nb=k/2$, $\forall k\in\mathbb{N}$
(with the exception for $s=2$ of the algebraically-special frequency $\nb=\nb_{AS}$,
defined below, which is not a pole of $\f$~\cite{MaassenvandenBrink:2000ru,Leung:2003ix}), and so it is convenient to define
a new radial function:  $\fb (r,-i\nu) \equiv -\sin\left(2\pi\nb\right)\f(r,-i\nu)$.
Accordingly, we define the Wronskian $\Wb{\omega}\equiv W\left[\g(r,\omega),\fb(r,\omega)\right]$.

From the radial ODE Eq.(\ref{eq:radial ODE}) and the boundary conditions Eq.(\ref{eq:f,near hor}), there follow the symmetries
\begin{align} \label{eq:symms f,g}
\gt(r,\omega)&=\gt^*(r,-\omega^*)\quad  \f(r,\omega)=\f^*(r,-\omega^*)
\quad \text{if} \quad r_*\in \mathbb{R},
\\
\W{\omega}&=W^*\left(-\omega^*\right),\quad \forall\omega\in\mathbb{C}
\nonumber
\end{align}
These symmetries lead to 
\begin{align}\label{eq:symms g,W on NIA}
\gm (r,-i\nu)&=\gp^*(r,-i\nu)
\quad \text{if} \quad r_*\in \mathbb{R},
\\
W_-(-i\nu)&=W^*_+(-i\nu),
\quad \forall \nu>0,
\nonumber
\end{align}
so that the discontinuity of $\g$ across the NIA is only in its imaginary part.
Note also that although $\Ain$ has a cut, $|\Ain|$ does not have a cut.

The BC contribution to the `retarded' Green function is given by
\begin{equation} \label{eq: G^BC integral}
G^{BC}(x,x')=
\sum_{\ell=0}^{\infty}(2\ell+1)P_{\ell}(\cos\gamma)G^{BC}_{\ell}(r,r';\Delta t),\quad
G_{\ell}^{BC}(r,r';\Delta t)\equiv
\frac{1}{2\pi i}
\int_{0}^{\infty } d\nu\ 
\DGw{r}{r'}{-i\nu}
e^{-\nu \Delta t},
\end{equation}

The functions $\gpt$ and  $\gmt$ satisfy the same homogeneous, linear 2nd order differential equation (namely, Eq.(\ref{eq:radial ODE})).
Therefore, $\Dgt$ will also satisfy this same differential equation
and it may be expressed as a linear combination of the solutions $\gt (r,-i\nu)$ and $\gt (r,+i\nu)$.
In addition, both $\gpt$ and  $\gmt$ satisfy the same boundary condition (\ref{eq:f,near hor}) as $r\to \infty$: $\gpmt(r,-i\nu)\sim e^{\nu r_*}$, and so
$\Dgt$ will not satisfy this boundary condition. 
To understand this, note that the behaviour for $r\to \infty$ of $\gpmt(r,-i\nu)\sim e^{\nu r_*}$ is dominant over that of $\gt(r,+i\nu)\sim e^{-\nu r_*}$, 
in  $\Dgt$ the boundary conditions determine that the dominant terms cancel  so that $\Dgt$ must be proportional to $\gt(r,+i\nu)$.
Finally, $\Dgt$ must be purely-imaginary from Eq.(\ref{eq:symms g,W on NIA}) and, since we know that $\gt(r,+i\nu)$ is real-valued, it must be~\cite{Leaver:1986,Leung:2003ix} that
\begin{equation} \label{eq:dg=qg}
\Dgt(r,-i\nu)=iq(\nu)\gt(r,+i\nu),\qquad \forall\nu>0
\end{equation}
for some real-valued function $q(\nu)$ which entirely characterises the  BC `strength'.

Using Eq.(\ref{eq:dg=qg}), we can express the BC modes as~\cite{Leaver:1986,Leung:2003ix} 
\begin{align} \label{eq:DeltaG in terms of Deltag}
\DGw{r}{r'}{-i\nu}=
-2i\nu q(\nu)
\frac{
\f(r,-i\nu)\f(r',-i\nu)
}{
\Wp{-i\nu}
W_-(-i\nu)
}=
-2i\nu q(\nu)
\frac{
\f(r,-i\nu)\f(r',-i\nu)
}{
\left|W_{\pm}(-i\nu)\right|^2},
\quad \forall r_*,r'_*\in \mathbb{R}
\end{align}

The second step is due to Eq.(\ref{eq:symms g,W on NIA}), and we include the double subindex
in $W_{\pm}$
 because $\left|\Ain\right|$ has no cut.


\section{Large-$\nu$ asymptotics on the branch cut} \label{sec:large-nu}


\subsection{Method  \fixme{To be CHECKED}}

We essentially follow the method of Maassen van den Brink~\cite{MaassenvandenBrink:2003as}, who calculates the high-frequency asymptotics 
of $\f$, $\gp$ and $q(\nu)$
on the NIA for the case $s=2$, to obtain the corresponding asymptotics for the cases $s=0$ and $1$.
For completeness, we will also include the results of~\cite{MaassenvandenBrink:2003as} for $s=2$.
We now give a general description of the method. 

\begin{figure}[h!]
\begin{center}
\includegraphics[width=7cm]{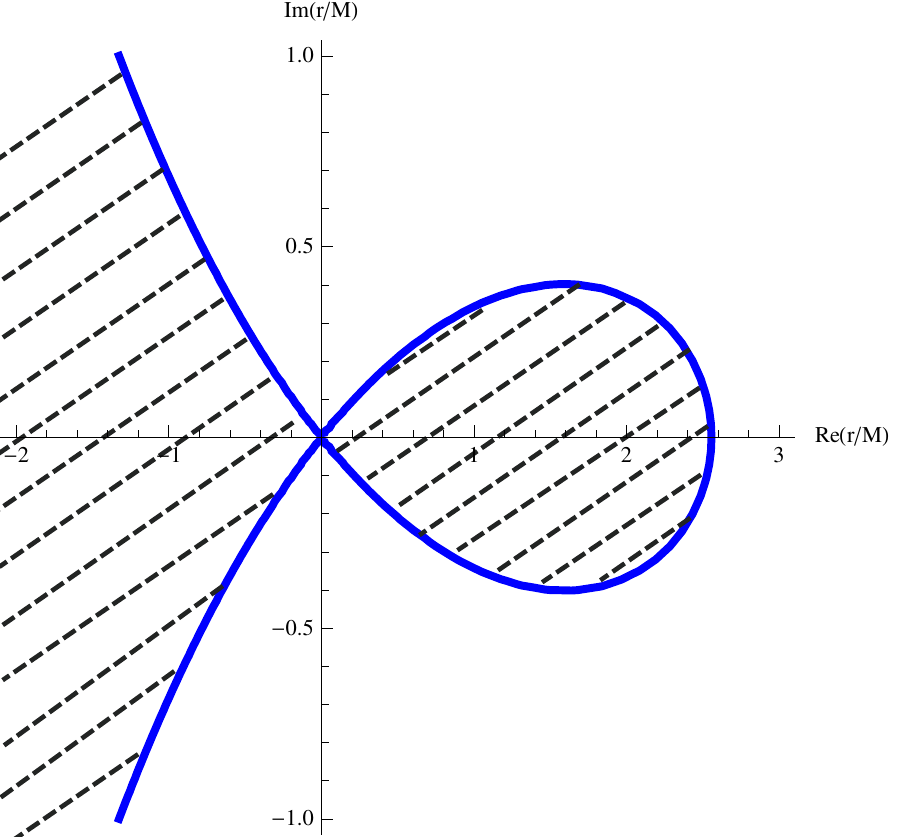} 
\includegraphics[width=7cm]{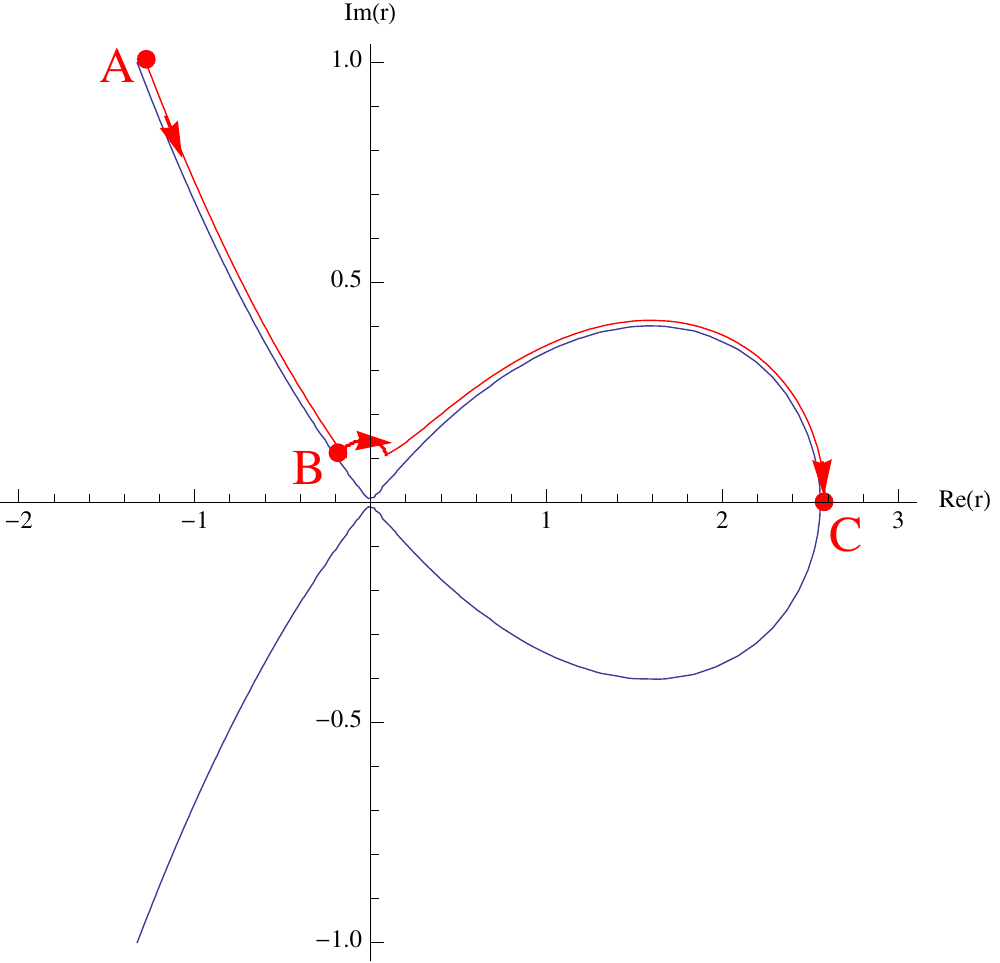} 
\includegraphics[width=7cm]{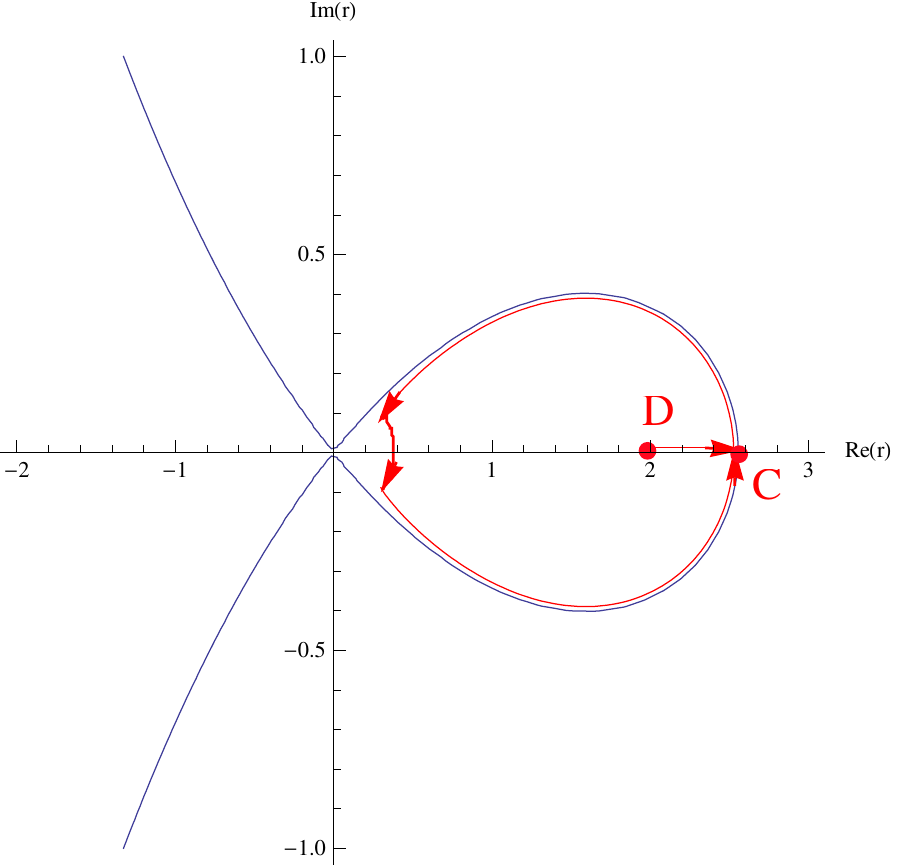} 
\end{center}
\caption{(a) In blue: schematic illustration of anti-Stokes lines (i.e., where $\text{Re}(r_*)=0$) on
the complex-$r$ plane
in the case that $\omega$ is on the NIA 
in Schwarzschild.
[See, e.g., Fig.1~\cite{Andersson:2003fh} for the topology of their anti-Stokes lines for large-frequency very near $r=0$, 
where their method differs only slightly from 
the standard WKB analysis.]
In dashed black lines: the region where $\text{Re}(r_*)<0$.
(b) In red: the contour we follow in order to calculate the large-$\nb$ asymptotics for $\gp$.
(c) In red: the contour followed in order to calculate $\f$.
Point A corresponds to a point for $|\rb|\gg 1$ along the anti-Stokes line that stems from near $r=0$ at $\arg(r)=3\pi/4$;
point B corresponds to a point on that same anti-Stokes line but lies within  $\nb^{-1/2}\ll \rb\ll\nb^{-1/3}$;
point C corresponds to $r_*=0$; 
point D corresponds to a point with $r\gtrsim r_h$ 
but $r$ sufficiently far from $r_h$ so that the WKB expansions $g_a(r,\pm i\nu)$ are valid.
}
\label{fig:antiStokes}
\end{figure}

As described in the previous section it is convenient to impose the boundary conditions on $\g$ and $\f$ in the complex $r$ or $r_*$ plane.
The relationship between these two planes is complicated by the existence of a cut  in the interrelationship $ r_*=r+r_h \ln (r/r_h -1)$.
Of special importance to our analysis are the anti-Stokes lines which for $\omega$ on the NIA are defined as the curves on the complex-$r$ plane where $\text{Re}(r_*)=0$.
In this case there are four anti-Stokes lines emanating from near $r=0$ and they have slopes 
equal to $\pm 1$ (i.e., $\arg r=\pm \pi/4,\pm3\pi/4$). 
The full structure is illustrated in Fig.\ref{fig:antiStokes}(a).
We  have two linearly-independent WKB asymptotic expansions for large-$\nb$, $g_a(r,\mp i\nu)\sim e^{\pm \nu r_*}$; these expansions are valid for $|\rb\sqrt{\nb}|\gg 1$ and away from the singularities  $r=0$ and $r_h$ of the radial ODE. The importance of the anti-Stokes lines is that along them neither expansion,  $g_a(r,+i\nu)$ nor $g_a(r,- i\nu)$, dominates over the other.

The contour chosen for $\g(r, - i\nu)$, illustrated in Fig.\ref{fig:antiStokes}(b),
 is as follows.
By analytic continuation, we can impose the boundary condition for $\gt(r,- i\nu)$  on the anti-Stokes line going to $|r|\to\infty$ on the upper
complex-$r$ plane instead of
imposing it for $r\to \infty$~\cite{Andersson:2003fh,1993CQGra..10..735A}.
One can safely match  $\gt(r,- i\nu)$  to $g_a(r,- i\nu)$
for $|r|\to\infty$ on that anti-Stokes line 
(i.e., at point A of Fig.\ref{fig:antiStokes}(b))
since, there, $g_a(r,- i\nu)$ does not dominate over $g_a(r,+i\nu)$.
One can can then continue this expression from 
$|r|\to \infty$ along the anti-Stokes line down
to a region near $r=0$ with $\arg r=3\pi/4$.
(i.e., down to point B of Fig.\ref{fig:antiStokes}(b)).
However, one cannot continue this expression from 
point B
to $r>r_h$ since
one 
would have to cross into a region where $g_a(r,- i\nu)$ dominates over $g_a(r,+i\nu)$.
Therefore, instead we introduce  two linearly-independent solutions 
$\psi_i$, $i=1,2$, whose behaviour we can determine analytically and which we can match to 
the solutions we are interested in.  
Of course, we cannot determine these solutions exactly but we can obtain them in the limit of large $\nb$
with fixed $\rb\sqrt{\nb}$
and this is sufficient for our purpose.
In this approximation, the  solutions $\psi_i$ are expressed in terms of 
special functions which
one knows how to analytically continue 
from one anti-Stokes line to another.
One then matches the solutions 
$\psi_i$
to $g_a(r,-i\nu)\sim  \gpt (r,-i\nu)$ along  $\arg r=3\pi/4$ 
in a region of overlap, given by $\nb^{-1/2}\ll \rb\ll\nb^{-1/3}$ (a region which includes point B of Fig.\ref{fig:antiStokes}(b)).
Since one knows how to analytically continue 
$\psi_i$
around $r=0$, with the matching done,
 one can analytically-continue $\gpt (r,-i\nu)$ from the anti-Stokes line on $\arg r=3\pi/4$ to another anti-Stokes line
 on $\arg r=\pi/4$ and have it expressed as a linear combination of $g_a(r,\pm i\nu)$. 
 This linear combination is safely valid along the anti-Stokes line clockwise all the way to $r_*=0$ (i.e., point C of Fig.\ref{fig:antiStokes}(b)). 
 The coefficient of $g_a(r,- i\nu)$ in this linear combination is 1, and the coefficient of $g_a(r,+ i\nu)$ yields the asymptotics for 
 the BC `strength' $q(\nu)$.
 From $r_*=0$, the linear combination is asymptotically valid on the physical line, either along $r_*\in (-\infty,0)$ or along $r_*\in (0,\infty)$,
 even though the coefficient of the subdominant solution in the corresponding region becomes meaningless.
 
One can also find, using Frobenius method, two linearly-independent series expansions, 
$\pF_1(r)$ and $\pF_2(r)$,
about $r=0$.
These expansions are valid for $|\rb^2\nb|\ll 1$ and
so there is no overlap with the region of validity of $g_a(r,\pm i\nu)$; the expansions $\pF_i$ do however provide a check for the solutions $\psi_i$.

The contour chosen for $\f(r,-i\nu)$,
illustrated in Fig.\ref{fig:antiStokes}(c),
is as follows. First, from its boundary condition as $r_*\to -\infty$ we know that at 
$r_*=0$ the solution
 $\f(r,-i\nu)$ has to be a linear combination of $g_a(r,\pm  i\nu)$ with the coefficient
of $g_a(r,+i\nu)$ (which is the dominant WKB expansion on $r_*\in (-\infty,0)$)  being equal to 1.
This linear combination of $g_a(r,\pm  i\nu)$ is valid 
anticlockwise along the anti-Stokes line from $r_*=0$ up to $\arg r=\pi/4$.
In order to obtain the coefficient of $g_a(r,-i\nu)$, one re-expresses this linear combination as a linear combination
of 
$\psi_1$ and $\psi_2$, on $\arg r=\pi/4$, using the matching described above. One knows how to analytically continue this
expression onto the anti-Stokes line on $\arg r=-\pi/4$, and there it
can be re-expressed as a new linear combination of $g_a(r,\pm  i\nu)$.
One can then continue this linear combination anticlockwise along the anti-Stokes line back to $r_*=0$,
thus yielding the asymptotic monodromy of $\f$ around $r=r_h$.
The obtained asymptotic monodromy is then compared with the exact monodromy 
$\f\left((r-r_h)e^{2\pi i},-i\nu\right)=e^{-2\pi i\nb}\f(r-r_h,-i\nu)$, which follows from the boundary 
condition on $\f$ at the horizon.
The comparison then yields the coefficient of $g_a(r,- i\nu)$ that we wanted.

Let us note that in~\cite{Andersson:2003fh} they use a similar, but different, method to the one we use here.
In~\cite{Andersson:2003fh} they make use of the Stokes phenomenon, instead of the solutions 
$\psi_i$,
in order to continue from one anti-Stokes line to another. 
For that reason, they require the knowledge of the topology of the anti-Stokes lines very near $r=0$, down to $r=O\left(\nb^{-1/2}\right)$, where
their WKB expansion breaks down - see Fig.1~\cite{Andersson:2003fh}.
In our analysis, however, the WKB expansions never reach a region so near $r=0$.
See also~\cite{Motl&Neitzke} for a description of similar contours followed.

\subsection{Radial solution $\gt(r,-i\nu)$}

Firstly, we find WKB asymptotic expansions for the solution of the radial ODE for $|\ob|\gg 1$: 
 \begin{align} \label{eq:g_a large-nu}
 &
g_a(r,\omega)\equiv e^{i\omega r_*}\left\{1+\frac{g_1(r)}{\ob}+\frac{g_2(r)}{\ob^2}+\dots\right\}
\\&
g_1(r)=\frac{1}{2i}\int_{\infty}^rdv \frac{vV(v)}{v-1}=\frac{i\left[2\lambda \rb+1-s^2\right]}{4\rb^2}
\nonumber
\\&
g_2(r)=\frac{V(r)}{4}-\frac{1}{8}\left[\int_{\infty}^rdv \frac{vV(v)}{v-1}\right]^2=
\frac{8(\rb-1)\left[\lambda \rb+1-s^2\right]-\left[2\lambda \rb+1-s^2\right]^2}{32\rb^4}
\nonumber
 \end{align}
These expansions $g_a(r,\omega)$ are valid away from the singularities 
$r=0$ (specifically, $|\rb\sqrt{\nb}|\gg 1$ is required) and $r_h$ of the radial ODE and, 
they are such that neither expansion,  $g_a(r,+i\nu)$ nor $g_a(r,- i\nu)$, dominates over the other along an anti-Stokes line.

Secondly, we use the Frobenius method to find two linearly-independent solutions about $r=0$ of the radial ODE
(re-expressed with $r$, not $r_*$, as the independent variable).
The characteristic exponents are: $1\pm s$, and so they differ by an integer number when $s=0,1,2$. 
We first find a power series solution $\pF_1(r)$ about $\rb=0$:
\begin{align} \label{eq:psi1 r=0}
s=0:\qquad
\pF_1(r)&=\rb-\lambda \rb^2+O(\rb^3)\nonumber\\
s=1:\qquad
\pF_1(r)&=\rb^2-\frac{
(\lambda-2)
}{3}\rb^3+\frac{
(\lambda-2)(\lambda-6)
}{24}\rb^4+O(\rb^5)\\
s=2:\qquad
\pF_1(r)&=\rb^3
-\frac{(\lambda-6)
}{5}\rb^4+O(\rb^5) \nonumber
\end{align}
and the second, linearly independent solution $\pF_2(r)$ is given by
\begin{align} \label{eq:psi2 r=0}
s=0:\quad
\pF_2(r)&=\left[1+2\lambda \right]\rb^2+\frac{
(2+2\lambda-3\lambda^2)
}{4}\rb^3+O(\rb^4)+
\pF_1(r)
\ln \rb \nonumber\\
s=1:\quad
\pF_2(r)&=1+\lambda \rb+O(\rb^2)
-\frac{\lambda ^2}{2}\pF_1(r)
\ln \rb \\\
s=2:\quad
\pF_2(r)&=\frac{1}{\rb}+\frac{
2\nMB
}{3}+\frac{\nb_{AS}}{2}\rb+O(\rb^2)+\frac{(\nb^2-\nb_{AS}^2)}{4}\pF_1(r)\ln \rb \nonumber
\end{align}
\fixme{\cite{MaassenvandenBrink:2003as} says that these solutions are valid for $|r^2\omega|\ll 1$, why
is that? is it true $\forall s$? - I don't see why its not just r}
where $\nb=\nb_{AS}\equiv 
\lambda(\lambda-2)/6
$ is
the so-called algebraically special frequency (note that we use this term to refer both to $\nb_{AS}$ and to $\omega_{AS}=-i\nu_{AS}$). 
Note the exact monodromies:
\begin{align} \label{eq:psi2 r=0 monodromy}
s=0:\quad
\pF_2(r e^{2\pi i})&=\pF_2(r)+2\pi i \pF_1(r)
\nonumber\\
s=1:\quad
\pF_2(r e^{2\pi i})&=\pF_2(r)-\lambda^2 \pi i \pF_1(r)
 \\\
s=2:\quad
\pF_2(r e^{2\pi i})&=\pF_2(r)
+\frac{(\nb^2-\nb_{AS}^2)\pi i}{2}
 \pF_1(r)
 \nonumber
\end{align}
and $\pF_1(r e^{2\pi i})=\pF_1(r)$, $\forall s=0,1,2$.

We now rewrite the ODE in terms of the independent variable $t\equiv \rb\sqrt{\nb}$ and then group terms in different powers of $\nb$:
\begin{align}
&
\hat L_0
\psi=
\frac{1}{\sqrt{\nb}}
\hat L_1
\psi+
\frac{1}{\nb}
\hat L_2
\psi
\\&
\hat L_0 \equiv  t^2\frac{d^2}{dt^2}-t\frac{d}{dt}-(s^2-1)-t^4,
\quad
\hat L_1\equiv 2t^3\frac{d^2}{dt^2}-t^2\frac{d}{dt}-\left[\lambda +s^2-1\right]t,
\quad 
\hat L_2\equiv -t^4\frac{d^2}{dt^2}+\lambda t^2
\nonumber
\end{align}

Expanding for large-$\nb$, there are two linearly-independent solutions $\psi_i=\psi^{(0)}_i+\psi^{(1)}_i$, $i=1,2$,
where $\psi^{(0)}_i$ denotes the two leading order solutions and $\psi^{(1)}_i$ the next-to-leading order ones:
\begin{align} \label{eq:psi_1,2^1}
\hat L_0\psi^{(0)}_i(t)&=0,
\quad
\psi^{(1)}_i(t)=
\int_0^t d\iv\ G_{\psi}(t,u) R_i(\iv),
\quad
R_i(t)\equiv \frac{1}{\sqrt{\nb}}\hat L_1\psi^{(0)}_i(t),\quad
\forall i=1,2
\\ 
G_{\psi}(t,u) &\equiv \frac{\left[-\psi^{(0)}_1(t)\psi^{(0)}_2(\iv)+\psi^{(0)}_2(t)\psi^{(0)}_1(\iv)\right]}{\iv^2 W\left[\psi^{(0)}_1(\iv),\psi^{(0)}_2(\iv)\right]},\quad
W\left[\psi^{(0)}_1(\iv),\psi^{(0)}_2(\iv)\right]\equiv \psi^{(0)}_1(\iv)\frac{d\psi^{(0)}_2(\iv)}{d\iv}-\psi^{(0)}_2(\iv)\frac{d\psi^{(0)}_1(\iv)}{d\iv}
\nonumber
\end{align}

This order will be sufficient for $s=0$ and $2$. However, for $s=1$ we will require the next order,
$\psi_i=\psi^{(0)}_i+\psi^{(1)}_i+\psi^{(2)}_i$, with:
\begin{align} \label{eq:psi_1,2^2}
\psi^{(2)}_i(t)&=\psi^{(2a)}_i(t)+\psi^{(2b)}_i(t),
\quad i=1,2
\\
\psi^{(2a)}_i(t)&=
\int_0^t
d\iv\
G_{\psi}(t,u)
R_i^{(1a)}(\iv),\quad
R_i^{(1a)}(t)\equiv \frac{1}{\nb}\hat L_2\psi^{(0)}_i(t)
\nonumber
\\ 
\psi^{(2b)}_i(t)&=
\int_0^t
d\iv\
G_{\psi}(t,u)
R_i^{(1b)}(\iv),\quad 
R_i^{(1b)}(t)\equiv \frac{1}{\sqrt{\nb}}\hat L_1\psi^{(1)}_i(t)
\nonumber
\end{align}

 The leading order solutions can be found to be given by
\begin{align} \label{eq:psi^0_1,2}
s&=0:\quad
\psi^{(0)}_1(t)=\frac{t}{\sqrt{\nb}}J_{0}\left(\frac{t^2}{2i}\right),\quad
\psi^{(0)}_2(t)=\frac{\pi t}{4\sqrt{\nb}}Y_{0}\left(\frac{t^2}{2i}\right)
\\
s&=1:\quad
\psi^{(0)}_1(t)=\frac{e^{-\pi i/4}\sqrt{\pi}}{\nb}tJ_{1/2}\left(\frac{t^2}{2i}\right)
=\frac{2\sinh\left(t^2/2\right)}{\nb}
,\quad
\psi^{(0)}_2(t)=\frac{e^{3\pi i/4}\sqrt{\pi}}{2}tY_{1/2}\left(\frac{t^2}{2i}\right)
=\cosh\left(t^2/2\right)
\nonumber
\\
s&=2:\quad
\psi^{(0)}_1(t)=\frac{4it}{\nb^{3/2}}J_1\left(\frac{t^2}{2i}\right),\quad
\psi^{(0)}_2(t)=\frac{i\pi\nb^{1/2}t}{4}Y_1\left(\frac{t^2}{2i}\right)
\nonumber
\end{align}
We have chosen the normalization constants so that, as indicated below, these solutions agree with the corresponding solutions $\pF_i$.
As for the quantities in the next-to-leading order solutions (and also the following order for $s=1$), we find the following.
For $s=0$,
\begin{align}
&W\left[\psi^{(0)}_1(\iv),\psi^{(0)}_2(\iv)\right]=\frac{\iv}{\nb}
\\
&
R_1(t)=\frac{t^2\left[it^2J_1\left(\frac{t^2}{2i}\right)-(
\lambda
-2t^4)J_0\left(\frac{t^2}{2i}\right)\right]}{\nb}
\nonumber
\\
&
R_2(t)=\frac{\pi t^2\left[it^2Y_1\left(\frac{t^2}{2i}\right)-(
\lambda
-2t^4)Y_0\left(\frac{t^2}{2i}\right)\right]}{4\nb}
\nonumber
\end{align}

For $s=1$,
\begin{align} \label{eq:explicit psi's R's W's s=1}
&W\left[\psi^{(0)}_1(\iv),\psi^{(0)}_2(\iv)\right]=-\frac{2 \iv}{\nb}
\\
&
R_1(t)=\frac{2t\left[t^2\cosh\left(t^2/2\right)-(
\lambda
-2t^4)\sinh\left(t^2/2\right)\right]}{\nb^{3/2}}
\nonumber
\\
&
\psi_1^{(1)}(t)=-\frac{4t^3\cosh(t^2/2)+3\sqrt\pi \lambda  \left[e^{-t^2/2}\text{erfi} (t)-\text{erf}(t)e^{t^2/2}\right]}{6\nb^{3/2}}
\nonumber
\\
&
R_2(t)=\frac{t\left[t^2\sinh\left(t^2/2\right)-(\lambda -2t^4)\cosh\left(t^2/2\right)\right]}{\nb^{1/2}}
\nonumber
\\
&
\psi^{(1)}_2(t)=
-
\psi^{(0)}_1(t)\left\{\int_0^t\frac{d\iv}{\iv^2}
\left[
\frac{\psi^{(0)}_2(\iv)R_2(\iv)}{W\left[\psi^{(0)}_1(\iv),\psi^{(0)}_2(\iv)\right]}
-\frac{\sqrt{\nb}\lambda }{2}\right]
-\frac{\sqrt{\nb}\lambda }{2t}\right\}
+
\psi^{(0)}_2(t)\int_0^t\frac{d\iv}{\iv^2}
\frac{\psi^{(0)}_1(\iv)R_2(\iv)}{W\left[\psi^{(0)}_1(\iv),\psi^{(0)}_2(\iv)\right]}=
\nonumber
\\
&
\frac{12\lambda \sinh(t^2/2)+e^{-t^2/2}
\left\{2\left(e^{t^2/2}-1\right)\left[t^4-3\lambda \right]+3\lambda \sqrt{\pi}t 
\left[e^{t^2}\text{erf}(t)+\text{erfi}(t)\right]\right\}}{12t\nb^{1/2}},
\nonumber
\end{align}
For $s=1$, we also required the following order. The first solution for $s=1$ to following
order is given by
\begin{align} 
&
R_1^{(1a)}(t)=\frac{2t^2\left[-t^2\cosh(t^2/2)+(\lambda -t^4)\sinh(t^2/2)\right]}{\nb^2}
\\
&
R_1^{(1b)}(t)=
\frac{e^{-\frac{t^2}{2}} t \sqrt{\pi } \ell  \left\{(\ell+1) e^{t^2} \text{erf}(t)
   \left(\lambda -2 t^4-t^2\right)- \text{erfi}(t) \left[\ell^3+2
   \ell^2+\ell \left(-2 t^4+t^2+1\right)-2 t^4+t^2\right]
\right\}  }{2 \nb ^2}-
\nonumber
\\
&
   \frac{e^{-\frac{t^2}{2}} t^2\left\{
   \lambda  \left[t^2+e^{t^2}
   \left(t^2+3\right)-3\right]-t^2 \left[2 t^4-13
   t^2+e^{t^2} \left(2 t^4+13 t^2+9\right)+9\right]\right\}}{3\nb ^2}
   \nonumber
 \\ &
 \psi_1^{(2a)}(t)=\frac{-\cosh\left(t^2/2\right)\left\{-2\gamma_E\lambda +t^4+2\lambda \left[\text{chi}(t^2)-2\ln t\right]\right\}+2\lambda \sinh\left(t^2/2\right)\text{shi}(t^2)}{4\nb^2}
\nonumber
\end{align}
while the second solution is given by
\begin{align}
&
R_2^{(1a)}(t)=\frac{t^2\left[(\lambda -t^4)\cosh(t^2/2)-t^2\sinh(t^2/2)\right]}{\nb}
 \\
&
R_2^{(1b)}(t)=
 \frac{-e^{-\frac{t^2}{2}} t \sqrt{\pi }\ell\left\{(\ell+1) e^{t^2} \text{erf}(t)
   \left(\ell^2+\ell-2 t^4-t^2\right)+\text{erfi}(t) \left[\ell^3+2
   \ell^2+\ell \left(-2 t^4+t^2+1\right)-2 t^4+t^2\right]\right\} }  
   {4 \nb }
   -
\nonumber   \\&
   \frac{e^{-\frac{t^2}{2}}  t^2 \left\{\lambda  \left[-t^2+e^{t^2}   \left(t^2+3\right)+3\right]-t^2 \left[-2 t^4+13
   t^2+e^{t^2} \left(2 t^4+13 t^2+9\right)-9\right]\right\}}{6 \nb }
   \nonumber
   \\
&
\psi^{(2a)}_2(t)=
-
\psi^{(0)}_1(t)\left\{\int_0^t\frac{d\iv}{\iv}
\left[
\frac{\psi^{(0)}_2(\iv)R_2^{(1a)}(\iv)}{\iv W\left[\psi^{(0)}_1(\iv),\psi^{(0)}_2(\iv)\right]}
+\frac{\lambda }{2}\right]
-\frac{\lambda }{2}\ln t\right\}
+
\psi^{(0)}_2(t)\int_0^t\frac{d\iv}{\iv^2}
\frac{\psi^{(0)}_1(\iv)R_2^{(1a)}(\iv)}{W\left[\psi^{(0)}_1(\iv),\psi^{(0)}_2(\iv)\right]}
\nonumber
   \\
&
\psi^{(2b)}_2(t)=
-
\psi^{(0)}_1(t)\left\{\int_0^t\frac{d\iv}{\iv}
\left[
\frac{\psi^{(0)}_2(\iv)R_2^{(1b)}(\iv)}{\iv W\left[\psi^{(0)}_1(\iv),\psi^{(0)}_2(\iv)\right]}
-\frac{\lambda (\lambda +1)}{2}\right]
+\frac{(\lambda +1)\lambda }{2}\ln t\right\}
+
\nonumber
   \\
&
\psi^{(0)}_2(t)\int_0^t\frac{d\iv}{\iv^2}
\frac{\psi^{(0)}_1(\iv)R_2^{(1b)}(\iv)}{W\left[\psi^{(0)}_1(\iv),\psi^{(0)}_2(\iv)\right]}
\nonumber
\end{align}
\fixme{Check signs}
where $\text{erf}(z)$ and $\text{erfi}(z)=\text{erf}(iz)/i$ are, respectively, the error function and the imaginary error function,
and $\text{chi}(z)$ and $\text{shi}(z)$ are, respectively, the hyperbolic cosine integral and the hyperbolic sine integral functions.
Note that while $\text{shi}(z)$  is an entire function of $z\in\mathbb{C}$, $\text{chi}(z)$ has a branch cut along the negative real axis.
Finally, the next-to-leading order solutions for $s=2$ are given by
\begin{align}
&W\left[\psi^{(0)}_1(\iv),\psi^{(0)}_2(\iv)\right]=-\frac{4\iv}{\nb}
\\
&
\psi^{(1)}_1(t)=\frac{\pi}{4}t\int_0^t\frac{d \iv}{\iv ^2}\left[Y_1\left(\frac{t^2}{2i}\right)J_1\left(\frac{\iv ^2}{2i}\right)-J_1\left(\frac{t^2}{2i}\right)Y_1\left(\frac{\iv ^2}{2i}\right)\right]R_1(\iv),
\nonumber
\\&
R_1(\iv)=\frac{4i}{\nb^2}\left\{2\left[\iv ^6-
\nMB
\iv ^2\right]J_1\left(\frac{\iv ^2}{2i}\right)-i\iv ^4J_0\left(\frac{\iv ^2}{2i}\right)\right\}
\nonumber
\\
&
\psi^{(1)}_2(t)=\frac{\pi}{4}tY_1\left(\frac{t^2}{2i}\right)\int_0^t\frac{d \iv}{\iv ^2}J_1\left(\frac{\iv ^2}{2i}\right)R_2(s)+
\frac{\pi}{4}tJ_1\left(\frac{t^2}{2i}\right)\left\{\int_0^td \iv\left[\frac{8i\nMB}{\pi \iv ^4}-Y_1\left(\frac{\iv ^2}{2i}\right)\frac{R_2(\iv)}{\iv ^2}\right]+\frac{8i\nMB}{3\pi t^3}\right\},
\nonumber
\\&
R_2(\iv)=\frac{i\pi}{4}\left\{2\left[\iv ^6-\nMB s^2\right]Y_1\left(\frac{\iv ^2}{2i}\right)-i \iv ^4Y_0\left(\frac{\iv ^2}{2i}\right)\right\}
\nonumber
\end{align}
Note that the integrand for $\psi^{(1)}_2$ given by Eq.(\ref{eq:psi_1,2^1}) leads to an integral divergent at the lower limit, both for $s=1$ and $s=2$.
We therefore regularize it by integrating by parts and dropping the contribution at the lower limit which corresponds to adding an (infinite) multiple
of the leading-order first solution, $\psi^{(0)}_1$. A similar divergence occurs for $\psi^{(2a)}_2$ and $\psi^{(2b)}_2$ for $s=1$, and we regularize them similarly.
One can check that by taking an expansion as $t\to 0$
for $\psi^{(0)}_{i}(t)+\psi^{(1)}_{i}(t)$, $i=1,2$, both
the leading order and the next-to-leading-order terms as $r\to 0$ in 
Eqs.(\ref{eq:psi1 r=0}) 
and (\ref{eq:psi2 r=0})
are recovered for $s=0$, $1$ and $2$.
In the $s=1$ case we have also checked that the leading order term for $t\to 0$ of $\psi^{(2)}_{1}(t)$ recovers
the $O(\rb^4)$ term of $\pF_1(r)$ in Eq.(\ref{eq:psi1 r=0});
similarly, the leading order term for $t\to 0$ of $\psi^{(2)}_{2}(t)$ recovers
the $O(\rb^4\ln\rb)$ term of $\pF_2(r)$ in Eq.(\ref{eq:psi2 r=0}).


Using the formulae from~\cite{bk:onlineAS}, 
\begin{align}\label{eq:Bessel func cont}
&J_{\mu}\left(ze^{m\pi i}\right)=e^{m \mu \pi i}J_{\mu}\left(z\right) 
\\
&Y_{\mu}\left(ze^{m\pi i}\right)=e^{-m \mu \pi i}Y_{\mu}\left(z\right)+2i\sin(m \mu\pi)\cot(\mu\pi)J_{\mu}\left(z\right), \nonumber \\
&Y_{k}\left(ze^{m\pi i}\right)=(-1)^{m k}\left[Y_{k}\left(z\right)+2imJ_{k}\left(z\right)\right],
\nonumber
\end{align}
valid $ \forall z, \mu\in\mathbb{C}$ and $\forall m,k\in\mathbb{Z}$, we obtain the following analytic continuations for the various terms in $\psi_i$.
For $s=0$:
\begin{align} \label{eq:psi_1,2^0,1(it) s=0}
\psi^{(0)}_1(it)&=i\psi^{(0)}_1(t),\quad &\psi^{(0)}_2(it)&=i\psi^{(0)}_2(t)-\frac{\pi}{2}\psi^{(0)}_1(t),
\\
\psi^{(1)}_1(it)&=-\psi^{(1)}_1(t),\quad &\psi^{(1)}_2(it)&=-\psi^{(1)}_2(t)-\frac{\pi i}{2}\psi^{(1)}_1(t)
\nonumber
\end{align}
For $s=1$:
\begin{align}  \label{eq:psi_1,2^0,1(it) s=1}
\psi^{(0)}_1(it)&=-\psi^{(0)}_1(t),\quad & \psi^{(0)}_2(it)&=\psi^{(0)}_2(t),
\\
\psi^{(1)}_1(it)&=-i\psi^{(1)}_1(t),\quad &\psi^{(1)}_2(it)&=i\psi^{(1)}_2(t)
\nonumber
\\
\psi^{(2a)}_1(it)&=\psi^{(2a)}_1(t),\quad &
\psi^{(2a)}_2(it)&=-\psi^{(2a)}_2(t)-\frac{\pi i \lambda }{4}\psi^{(0)}_1(t),
\nonumber
\\
\psi^{(2b)}_1(it)&=\psi^{(2b)}_1(t),\quad
 &
 \psi^{(2b)}_2(it)&=-\psi^{(2b)}_2(t)+\frac{\pi i \lambda (\lambda +1)}{4}\psi^{(0)}_1(t)
\nonumber
\\
& &
\psi^{(2)}_2(it)&=-\psi^{(2)}_2(t)+\A\psi^{(0)}_1(t),
\qquad \A \equiv \frac{\pi i \lambda^2}{4}
\nonumber
\end{align}
Note the terms with $\psi^{(0)}_1(t)$ in $\psi^{(2a)}_2(it)$, $\psi^{(2b)}_2(it)$ and $\psi^{(2)}_2(it)$ arising from the regularization terms.
\\
For $s=2$:
\begin{align} \label{eq:psi_1,2^0,1(it)}
\psi^{(0)}_1(it)&=-i\psi^{(0)}_1(t),\quad &\psi^{(0)}_2(it)&=-i\psi^{(0)}_2(t)+\frac{\pi \nb^2}{8}\psi^{(0)}_1(t),
\\
\psi^{(1)}_1(it)&=\psi^{(1)}_1(t),\quad &\psi^{(1)}_2(it)&=\psi^{(1)}_2(t)+i\frac{\pi \nb^2}{8}\psi^{(1)}_1(t)
\nonumber
\end{align}

It is also easy to check that, along $\arg t=\pi/4$, we have
\begin{align} \label{eq:phase psi_i argt=pi/4}
&s=0:\quad
\psi^{(0)}_{1,2}e^{-\pi i/4}\in \mathbb{R},
\quad
\psi^{(1)}_{1,2}e^{-\pi i/2}\in \mathbb{R}
\nonumber
\\
&s=1:\quad
\psi^{(0)}_{1}e^{\pi i/2},\psi^{(0)}_{2}\in \mathbb{R},
\quad
\psi^{(1)}_{1}e^{\pi i/4},\psi^{(1)}_{2}e^{-\pi i/4}\in \mathbb{R},
\quad\\
&\qquad \psi^{(2)}_{1}
\in \mathbb{R},
\qquad \left(\psi_2^{(2)}
+\frac{\A }{2}\psi_1^{(0)}\right) e^{\pi i/2} \in \mathbb{R}
\nonumber \\ &
s=2:\quad
\psi^{(0)}_{1,2}e^{-3\pi i/4}\in \mathbb{R},
\quad
\psi^{(1)}_{1,2}\in \mathbb{R}
\nonumber
\end{align}
Next, we carry out an asymptotic series expansion for large-$|t|$ 
along $\arg t=\pi/4$ of $\psi^{(0)}_{i}$ and $\psi^{(1)}_{i}$ (and $\psi^{(2)}_{i}$ for $s=1$).
In the corresponding 
expressions which we write next, we only
show the $t^0$ terms in each of these large-$|t|$ expansions.
Using Eqs.(\ref{eq:phase psi_i argt=pi/4}), we can then write
 for $s=0$:
\begin{align} \label{eq:psi_1,2 for 1<<t<<nu^1/6 s=0}
\psi_1(t)&\sim \frac{1}{\sqrt{\pi\nb}}\left\{e^{t^2/2}\left[1+\frac{\aa}{\sqrt{\nb}}\right]+
e^{-t^2/2}\left[i-\frac{\aa^*}{\sqrt{\nb}}\right]
\right\}
\\
\psi_2(t)&\sim \frac{\sqrt\pi}{4\sqrt{\nb}}\left\{e^{t^2/2}\left[-i+\frac{\ba}{\sqrt{\nb}}\right]+
e^{-t^2/2}\left[-1-\frac{\ba^*}{\sqrt{\nb}}\right]
\right\},
\qquad \arg t=\pi/4
\nonumber
\end{align}
 for $s=1$:
\begin{align} \label{eq:psi_1,2 for 1<<t<<nu^1/6 s=1}
\psi_1(t)&\sim \frac{1}{\nb}\left\{e^{t^2/2}\left[1+\frac{\aa}{{\nb}^{1/2}}+\frac{\ab}{\nb}\right]+
e^{-t^2/2}\left[-1-\frac{i\aa^*}{\sqrt{\nb}}+\frac{\ab^*}{\nb}\right]
\right\}
\\
\psi_2(t)&\sim \frac{1}{2}\left\{e^{t^2/2}\left[1+\frac{\ba}{\nb^{1/2}}+\frac{\bb-\A }{\nb}\right]+
e^{-t^2/2}\left[1+\frac{i\ba^*}{\sqrt{\nb}}+\frac{-\bb^*+\A }{\nb}\right]
\right\} ,
\qquad \arg t=\pi/4
\nonumber
\end{align}
and  for $s=2$:
\begin{align} \label{eq:psi_1,2 for 1<<t<<nu^1/6}
\psi_1(t)&\sim \frac{4}{\sqrt{\pi\nb^3}}\left\{e^{t^2/2}\left[1+\frac{\aa}{\sqrt{\nb}}\right]+
e^{-t^2/2}\left[-i+\frac{\aa^*}{\sqrt{\nb}}\right]+O\left(t^{-2}\right)+O\left(\frac{t^3}{\sqrt{\nb}}\right)+O\left(\nb^{-1}\right)\right\}
\\
\psi_2(t)&\sim \frac{\sqrt{\pi\nb}}{4}\left\{e^{t^2/2}\left[-i+\frac{\ba}{\sqrt{\nb}}\right]+
e^{-t^2/2}\left[1+\frac{\ba^*}{\sqrt{\nb}}\right]+O\left(t^{-2}\right)+O\left(\frac{t^3}{\sqrt{\nb}}\right)+O\left(\nb^{-1}\right)\right\} ,
\qquad \arg t=\pi/4
\nonumber
\end{align}
which is valid for $1\ll t\ll\nb^{1/6}$ \fixme{Is that also the case for $s=0,1$?}
We can obtain the corresponding asymptotics for $\psi_{1,2}(t)$ at $\arg t=3\pi/4$ by using Eqs.(\ref{eq:psi_1,2 for 1<<t<<nu^1/6 s=0})--(\ref{eq:psi_1,2 for 1<<t<<nu^1/6}) along $\arg t=\pi/4$ together with the analytic continuation Eqs.(\ref{eq:psi_1,2^0,1(it) s=0})--(\ref{eq:psi_1,2^0,1(it)}). 
The dominant term of such asymptotics at $\arg t=3\pi/4$ (i.e., the term with $e^{t^2/2}$) must agree with the corresponding term in $\psi_{1,2}(it)$, obtained from 
Eqs.(\ref{eq:psi_1,2 for 1<<t<<nu^1/6 s=0})--(\ref{eq:psi_1,2 for 1<<t<<nu^1/6}) for $\arg t=\pi/4$.
Similarly, we can take a linear combination of the asymptotics for $\psi_{1,2}(t)$ at $\arg t=3\pi/4$ such that the dominant term is zero, and then require that the
remaining, subdominant term (i.e., the term with $e^{-t^2/2}$) agrees with the corresponding term in the linear combination of $\psi_{1}(it)$ and  $\psi_{2}(it)$, 
which are obtained from Eqs.(\ref{eq:psi_1,2 for 1<<t<<nu^1/6 s=0})--(\ref{eq:psi_1,2 for 1<<t<<nu^1/6}) for $\arg t=\pi/4$.
The result of these requirements is:
\begin{align} \label{eq:beta=beta(alpha)}
s=0\ \&\ 2&:\quad \aa\in\mathbb{R},\quad \ba=-(2+i)\aa
\\
s=1&:\quad \aa, \ab \in\mathbb{R},\quad \ba=-\aa,\quad 
\text{Im}(\bb)=\aa^2.
\nonumber
\end{align}

The actual values of $\aa$ can be obtained by directly expanding 
$\psi_1^{(1)}(t)$ in Eq.(\ref{eq:psi_1,2^1}) for large $|t|$ with $\arg t=\pi/4$.
We obtain
\begin{align} 
&s=0:\quad \aa=-\frac{\Gamma\left(\frac{1}{4}\right)^4}{48\pi^{3/2}}\left[1+3\lambda \right] 
\nonumber \\
&s=1:\quad \aa=-\frac{\lambda \sqrt\pi}{2}
 \label{eq:alpha values}
\\
&s=2:\quad \aa=\frac{\Gamma\left(\frac{1}{4}\right)^4}{48\pi^{3/2}}\left[1-\lambda \right]
\nonumber
\end{align}
From the boundary condition Eq.(\ref{eq:f,near hor}) and the leading-order behaviour of Eq.(\ref{eq:g_a large-nu}) it follows  that 
\begin{equation}\label{eq:g+ ga}
\gpt (r,-i\nu)\sim g_a(r,-i\nu),\quad \nb\gg 1,
\quad \text{for}\ |\arg \tb-\pi|<3\pi/4,
\end{equation}
\fixme{why is there an equal sign instead of a $\sim$ above Eq.24MvdB?} 
where we are only neglecting
exponentially-small corrections.
Performing now a 
power series in $\tb/\nb^b$ with $b\geq 1/6$ for 
$g_a(r,\pm i\nu)e^{\pm i\pi\nb}e^{\mp t^2/2}$
(note that we must replace $e^{\pm i\pi\nb}\to e^{\mp i\pi\nb}$ when $\arg(\rb-1)\in(0,\pi] \to \arg(\rb-1)\in(-\pi,0)$, since
this exponential comes from the $\ln(\rb-1)$ in $r_*=r_*(r)$),
we obtain
\begin{align}\label{eq:g t/nu^1/6<<1}
\forall s=0,1,2:
\quad
&g_a(r,\pm i\nu) \sim e^{\mp i\pi\nb}e^{\pm t^2/2}
\left[1+
0\frac{t^0}{\sqrt{\nb}}
\mp
\dsa\frac{t^0}{\nb}
+
\dots
\right],
\quad 
\text{if}\ 
\arg(\rb-1)\in(0,\pi) \ \text{and}\ \tb\ll \nb^{1/6}
\\&g_a(r,\pm i\nu) \sim e^{\pm i\pi\nb}e^{\pm  t^2/2}
\left[1+
0\frac{t^0}{\sqrt{\nb}}\mp \dsa\frac{\tb ^0}{\nb}
+\dots\right],\quad
\text{if}\   \arg(\rb-1)\in(-\pi,0) \ \text{and}\ \tb\ll \nb^{1/6}
\nonumber
\end{align}
We will only require the coefficient $\dsa$ for the case $s=1$, and in that case it is
$\dsa=-\lambda /12$.
\leaveout{
I think the leading order of Eq.(\ref{eq:g t/nu^1/6<<1}) is obtained as follows.
At $r\to -\infty$ (and just above the axis) we have $\gpt (r, \omega)\sim g_a(r, \omega)\sim e^{i\omega r_*}=e^{i\omega \left[r+r_h\ln(1-\rb)+i\pi\right]}$; this
is valid along (just above) the negative axis down to near $r=0$.
We then expand the exponent about $r=0$ and obtain $\gpt (r, \omega)\sim e^{i\pi\nb}e^{-t^2/2}$.
Similarly, for  $\gmt (r, \omega)$ we would start at $r\to -\infty$ but just below the axis and we would obtain $\gmt (r, \omega)\sim e^{-i\pi\nb}e^{-t^2/2}$ instead.
Similarly, we have $g_a(\omega=\pm\epsilon+i\nu)\sim e^{\mp i\pi\nb}e^{t^2/2}$ just above/below the negative $r$-axis.
\fixme{But for $\omega\in PIA$ we don't need to include $\pm\epsilon$, so what dictates the sign in the factor $e^{\mp i\pi\nb}$?}
}
We can then express the right-hand side of Eq.(\ref{eq:g t/nu^1/6<<1}) as a linear combination of $\psi_1(t)$ and $\psi_2(t)$ for $\arg t=3\pi/4$, by using the asymptotics
of Eqs.(\ref{eq:psi_1,2 for 1<<t<<nu^1/6 s=0})--(\ref{eq:psi_1,2 for 1<<t<<nu^1/6}) at $\arg t=\pi/4$ together with the analytic continuation 
Eqs.(\ref{eq:psi_1,2^0,1(it) s=0})--(\ref{eq:psi_1,2^0,1(it)}). 
The result is
\begin{align} 
&s=0:\quad \gpt (r,-i\nu) e^{-i\pi\nb}\sim -\frac{\sqrt{\pi\nb}}{2}\left[3i+
\frac{(2-3i)\aa}{\sqrt{\nb}}
+O\left(\nb^{-1}\right)\right]\psi_1(t)-
\frac{2\sqrt{\nb}}{\sqrt\pi}
\left[-1+
\frac{\aa}{\sqrt{\nb}}
+O\left(\nb^{-1}\right)\right]\psi_2(t)
\nonumber
\\
&s=1:\quad \gpt (r,-i\nu) e^{-i\pi\nb}\sim 
\nonumber
\\ &
\frac{\nb}{2}\left[-1-
\frac{\aa}{\sqrt{\nb}}
+
  \frac{\lambda +12i\aa^2-6\left(\bb+\bb^*\right)+36\A }{12\nb}
 +
O\left(\nb^{-3/2}\right)\right]\psi_1(t)+
\nonumber
\\ &
\left[1-
\frac{\aa}{\sqrt{\nb}}
+
    \frac{-\lambda -12i\aa^2+12\ab+6\left(\bb-\bb^*\right)}{12\nb}+
O\left(\nb^{-3/2}\right)\right]\psi_2(t)\nonumber
\\
&s=2:\quad \gpt (r,-i\nu) e^{-i\pi\nb}\sim \frac{\sqrt{\pi}}{8}\left[3i\nb^{3/2}+(2-3i)\aa\nb+O\left(
\nb^{1/2}
\right)\right]\psi_1(t)+
\frac{2}{\sqrt{\pi}}\left[-\frac{1}{\sqrt{\nb}}+\frac{\aa}{\nb}+O\left(\nb^{-3/2}\right)\right]\psi_2(t)
\nonumber
\end{align}
which are valid $\forall \arg t\in[0,2\pi)$, by analytic continuation.
Using the asymptotics of  Eqs.(\ref{eq:psi_1,2 for 1<<t<<nu^1/6 s=0})--(\ref{eq:psi_1,2 for 1<<t<<nu^1/6}) for $\psi_{1,2}$ 
and 
rewriting the factors $e^{\pm t^2/2}$ in terms of the $g_a(r,\pm i\nu)$ using Eq.(\ref{eq:g t/nu^1/6<<1}),
we find 
\begin{align} 
\label{eq:g=g(g_a)}
s=0,2:\quad &\gpt (r,-i\nu) \sim g_a(r,-i\nu) +(-1)^{1+s/2}2e^{2\pi i\nb}\left[i+\frac{\aa}{\sqrt{\nb}}\right]g_a(r,+i\nu)
\nonumber
\\
s=1:\quad &\gpt (r,-i\nu) \sim \left[1+
O\left(\nb^{-2}\right)\right]g_a(r,-i\nu) 
+\\
&\qquad
e^{2\pi i\nb}\left[-\frac{2\aa}{\sqrt{\nb}}+\frac{
2i \aa^2
}{\nb}+O\left(\nb^{-3/2}\right)\right]g_a(r,+i\nu)
\nonumber
\end{align}
The WKB expansion $g_a(r,+i\nu)$ is dominant over $g_a(r,-i\nu)$
in the region bounded by anti-Stokes lines which contains $r=r_h$, and so its coefficient can be trusted there.

In Fig.\ref{fig:tilde g,r=2.8} we plot the high-frequency asymptotics for $\gpt$ of Eq.(\ref{eq:g=g(g_a)}) together with
Eq.(\ref{eq:g_a large-nu})  (we only include a token plot, for $s=0$, since $\gpt$ does not appear
explicitly in Eq.(\ref{eq:DeltaG in terms of Deltag})). The figure shows that these asymptotics agree with the completely independent calculation
of~\cite{Casals:Ottewill:2011midBC}.


\begin{figure}[h!]
\begin{center}
 \includegraphics[width=8cm]{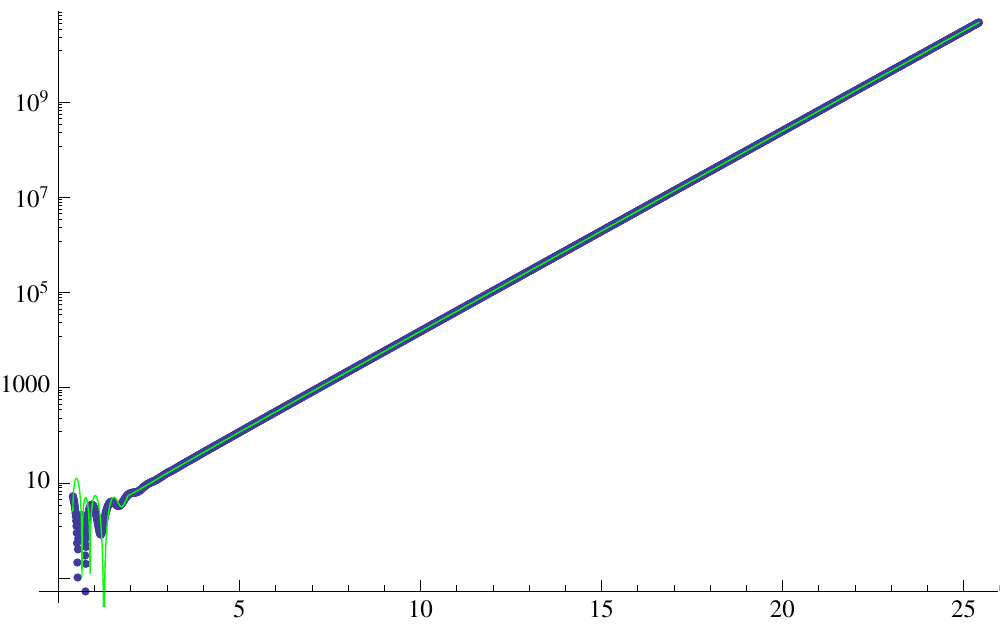}
  \includegraphics[width=8cm]{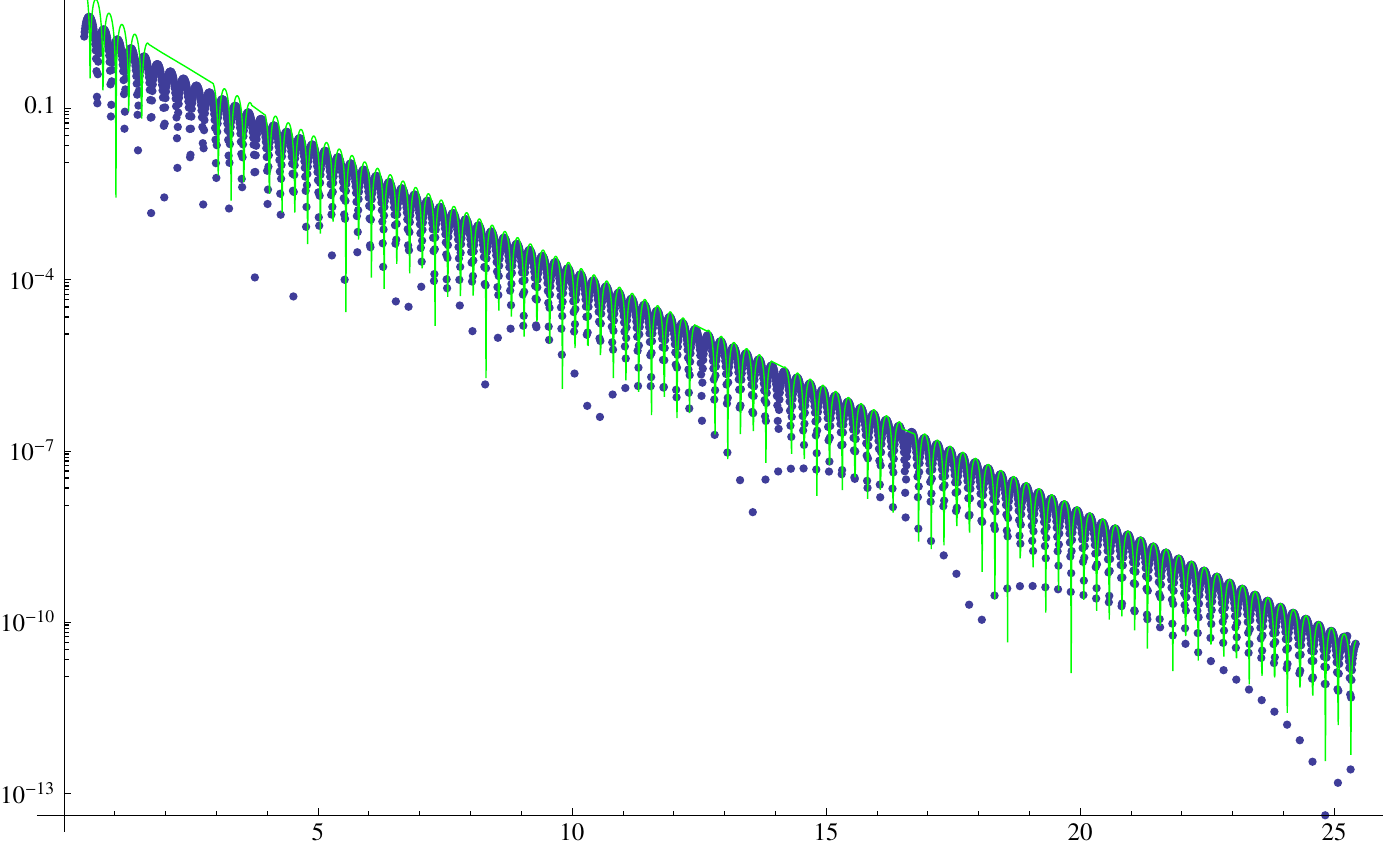}
   \includegraphics[width=8cm]{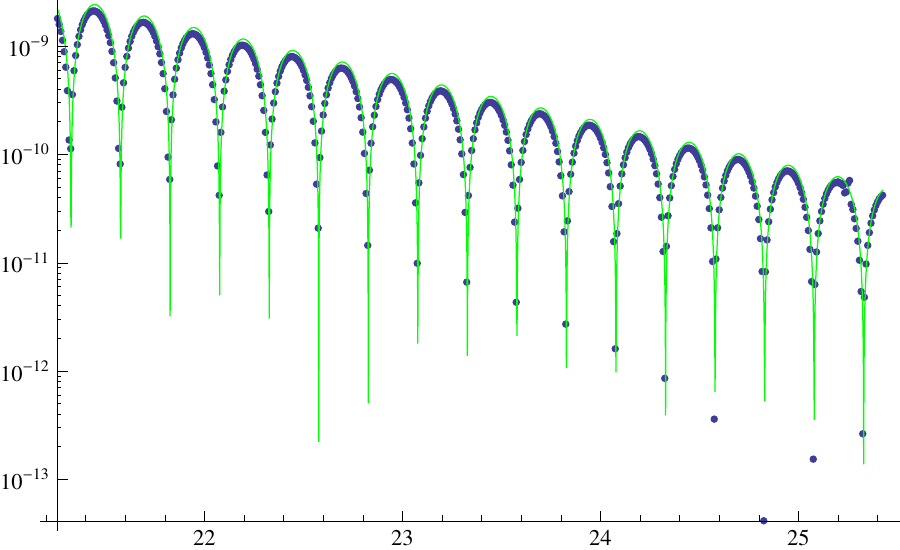}
 \end{center}
\caption{
Log-plots of $\gpt$ as a function of $\nu M$ for  $s=0$, $\ell=1$, $r=2.8M$. 
In blue: solution found using the method of~\cite{Casals:Ottewill:2011midBC}.
In green: large-$\nb$ asymptotics of Eq.(\ref{eq:g=g(g_a)}).
(a) $|\text{Re}(\gpt)|$, (b) $|\text{Im}(\gpt)|$,
(c) `zoomed-in' version of (b).
}
\label{fig:tilde g,r=2.8}
\end{figure}


\subsection{BC `strength' $q(\nu)$}

It follows from Eqs.(\ref{eq:dg=qg}) and (\ref{eq:g=g(g_a)}) that
$
\Delta \gt (r,-i\nu) =2i\text{Im}\ \gt(r,-i\nu)=
iq(\nu)\gt(r,+i\nu)
$
with the following large-$\nb$ asymptotics for the BC `strength':
\begin{align}  \label{eq:q for large-nu}
&s=0,2:
& 
q(\nu)\sim&
(-1)^{1+s/2}4
\left[\cos(2\pi\nb)+\frac{\aa}{\sqrt{\nb}}\sin(2\pi\nb)\right]
+O\left(\nb^{-1}\right)
 \fixme{Check}
\\
&s=1:
 &
q(\nu)\sim&
4\aa \left[-
\frac{\sin(2\pi \nb)}{\sqrt{\nb}}
+\frac{\aa\cos(2\pi \nb)}{\nb}
+O\left(\nb^{-3/2}\right)\right]
\nonumber
\end{align}
Alternatively, considering $q(\nu)$ in modulus-argument form we can rewrite Eq.(\ref{eq:q for large-nu}) as 
\begin{align}  \label{eq:q for large-nu phase}
&s=0,2:&
q(\nu)&\sim
(-1)^{1+s/2}4\left[\cos\left(2\pi \nb-\frac{\aa}{\sqrt{\nb}}\right)
+O\left(\nb^{-1}\right)\right]
\\ 
&s=1:&
q(\nu)&\sim
4\aa \left[-
\frac{1}{\sqrt{\nb}}\sin\left(2\pi \nb-\frac{\aa}{\sqrt{\nb}}\right)
+O\left(\nb^{-3/2}\right)\right]
\nonumber
\end{align}
since $g_a(r,\pm i\nu)\in\mathbb{R}$ and $g_a(r,+i\nu)\sim \gt(r,+i\nu)$, neglecting exponentially-small terms.

Figs.\ref{fig:q,s=0 l=1}--\ref{fig:q,l=2=s} show that the large-$\nb$ asymptotics of Eq.(\ref{eq:q for large-nu}) match with a calculation
of  $q(\nu)$ using the independent method of~\cite{Casals:Ottewill:2011midBC}
(where we will show more clearly that the curve corresponding to the latter method
agrees with Figs.2 in~\cite{Leung:2003ix} and~\cite{Leung:2003eq}).
The asymptotics of Eq.(\ref{eq:q for large-nu phase}) seem to do slightly better than those of  Eq.(\ref{eq:q for large-nu}) for $s=0$,
slightly worse for $s=2$ while, for $s=1$, they seem to do better in the phase but worse in the amplitude.
\fixme{Comment on how Eq.(\ref{eq:q for large-nu phase})  does compared to Eq.(\ref{eq:q for large-nu}) ?}

\clearpage
\begin{figure}[htbp]
\begin{center}
\includegraphics[width=8cm]{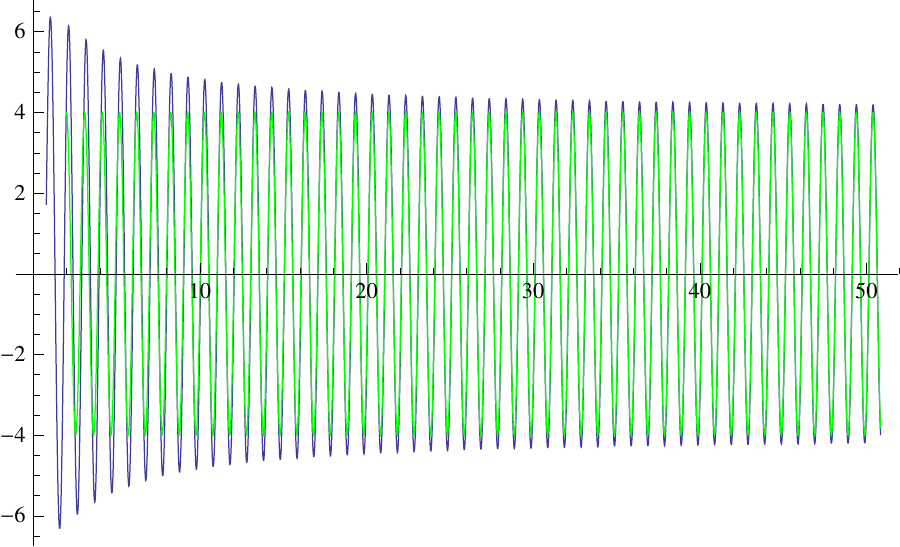}
\includegraphics[width=8cm]{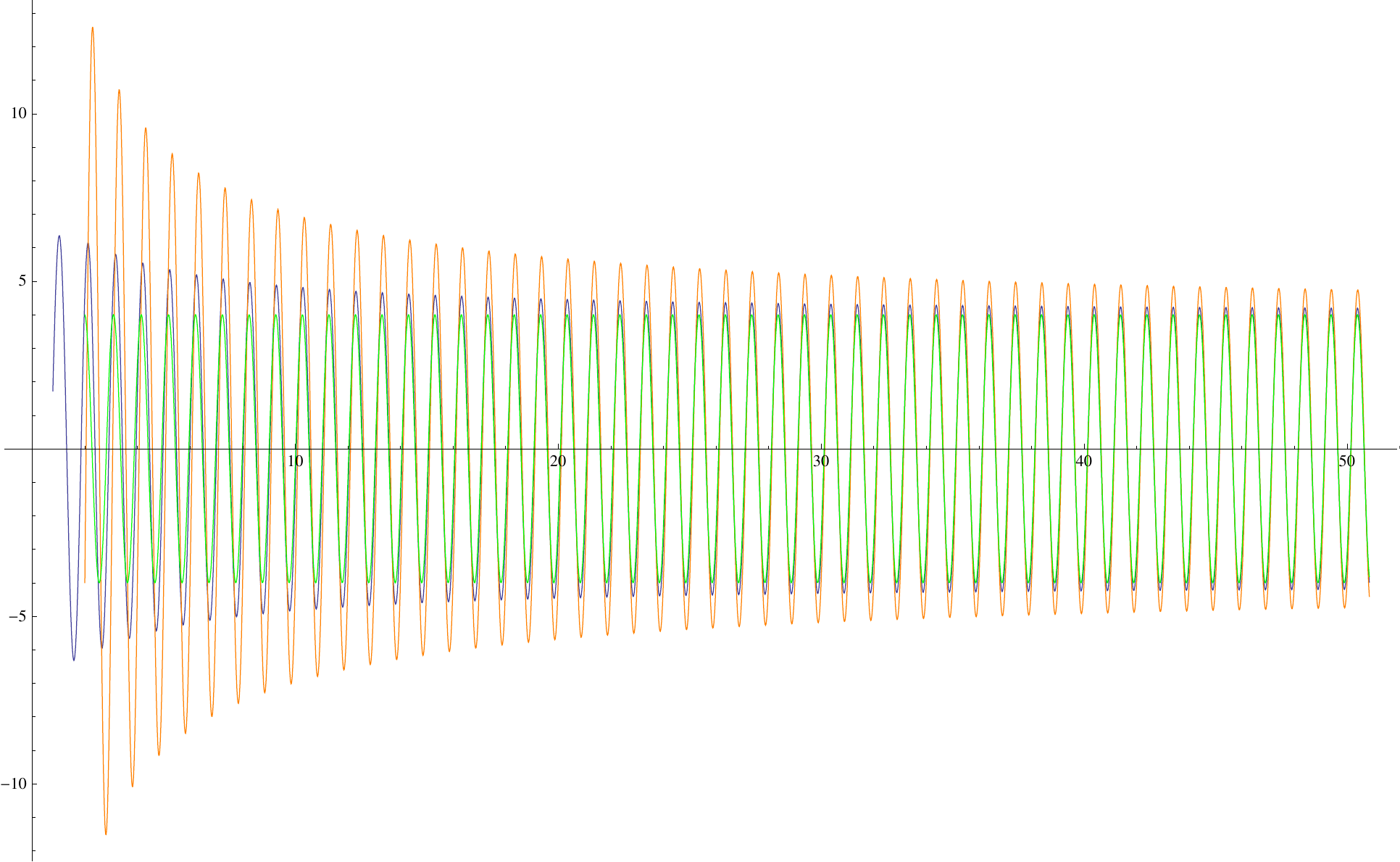}
\end{center}
\caption{BC `strength' $q(\nu)$ of Eq.(\ref{eq:dg=qg}) as a function of $\nb$ for $s=0$, $\ell=1$.
(a) Green curve: large-$\nb$ asymptotics of Eq.(\ref{eq:q for large-nu phase}); blue curve: $q(\nu)$ calculated using the method of~\cite{Casals:Ottewill:2011midBC} via Eq.(\ref{eq:dg=qg}) with the 
choice of value $r=5M$.
(b) Same as (a), where now we also include in orange the asymptotics of Eq.(\ref{eq:q for large-nu}).
}
\label{fig:q,s=0 l=1}
\end{figure}

\begin{figure}[htbp]
\begin{center}
\includegraphics[width=8cm]{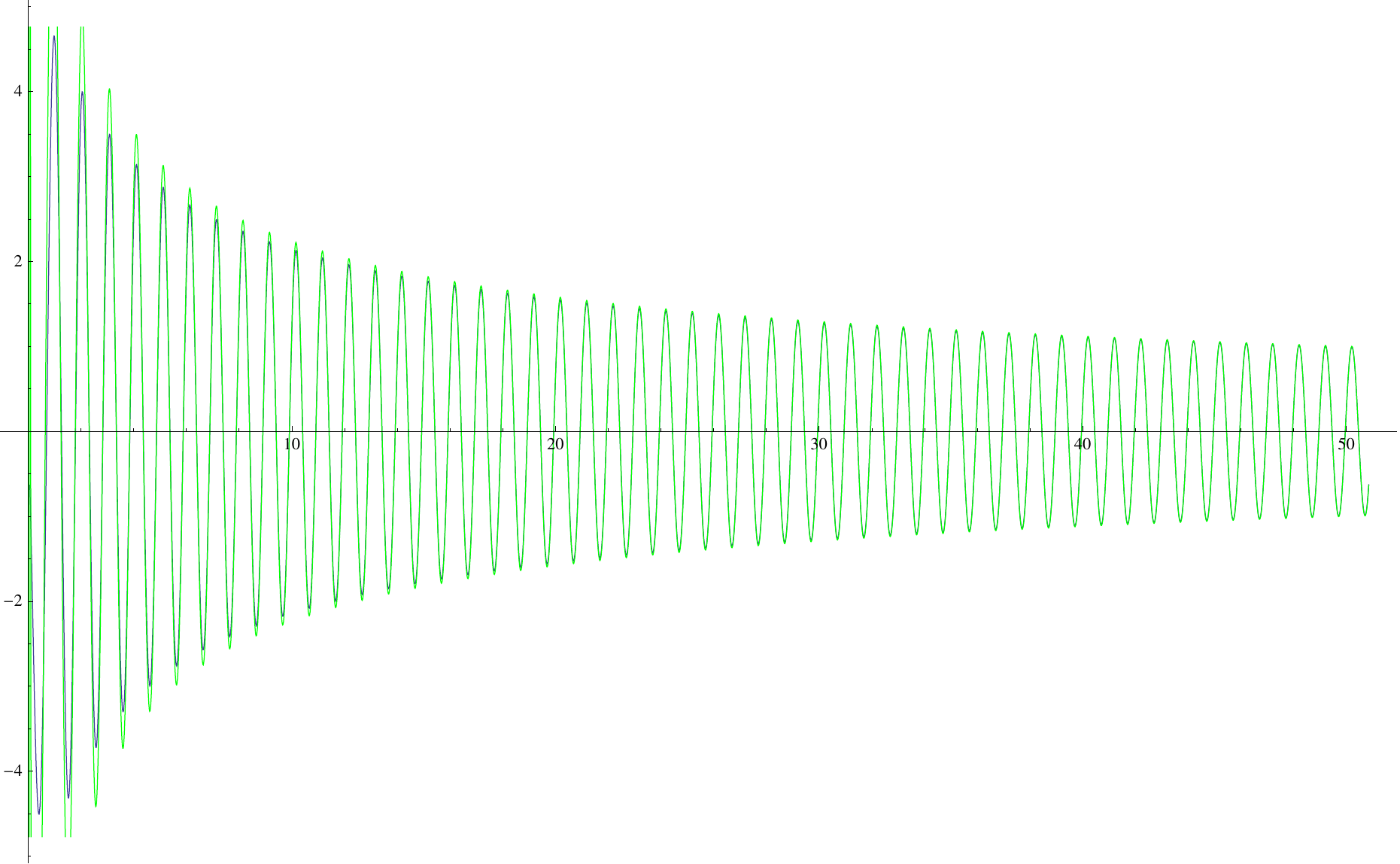}
\includegraphics[width=8cm]{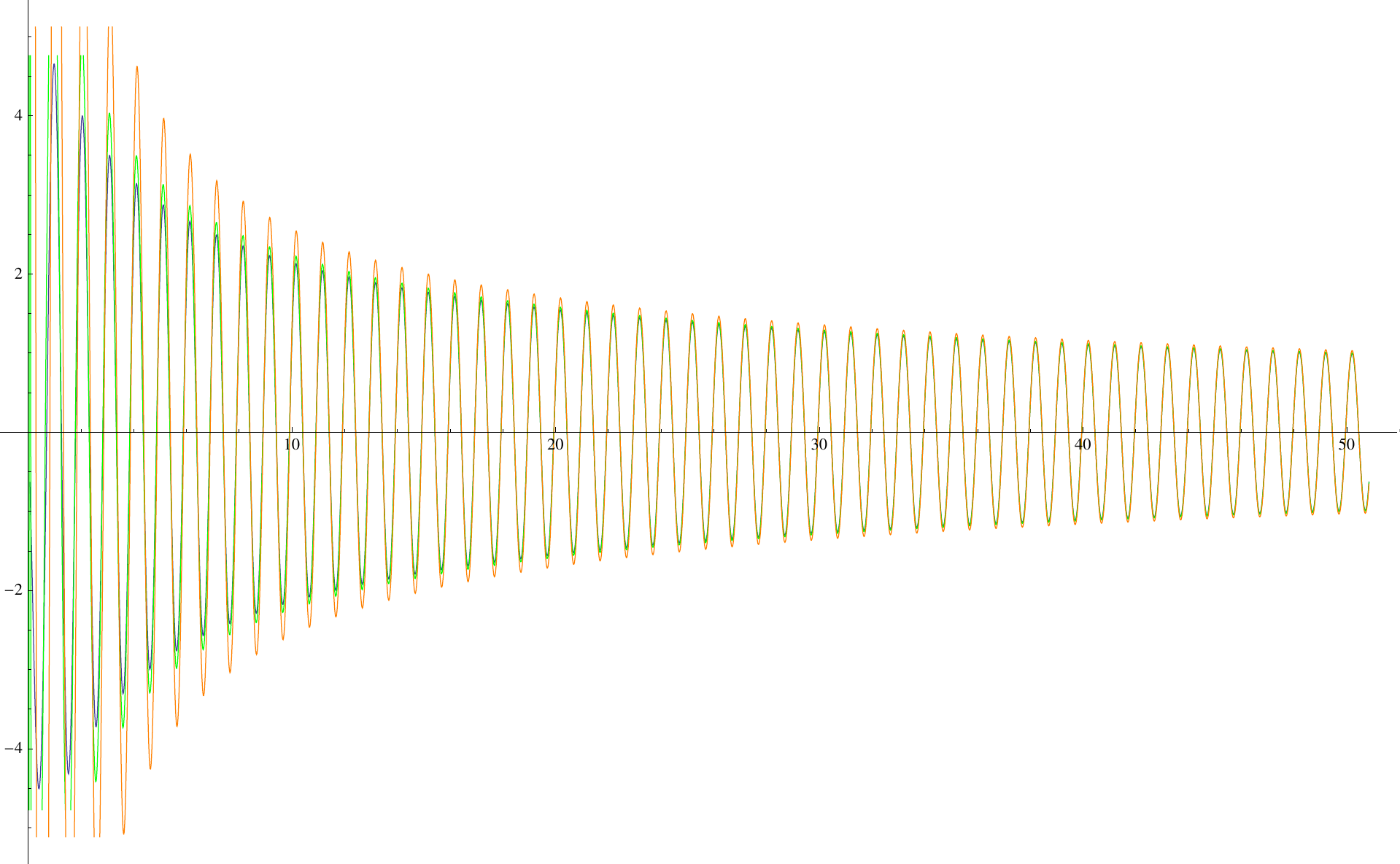}
\end{center}
\caption{ $q(\nu)$ as a function of $\nb$ for $s=1$, $\ell=1$.
See caption in Fig.\ref{fig:q,s=0 l=1} for a description of the curves.}
\label{fig:q,l=1=s}
\end{figure}

\begin{figure}[htbp]
\begin{center}
\includegraphics[width=8cm]{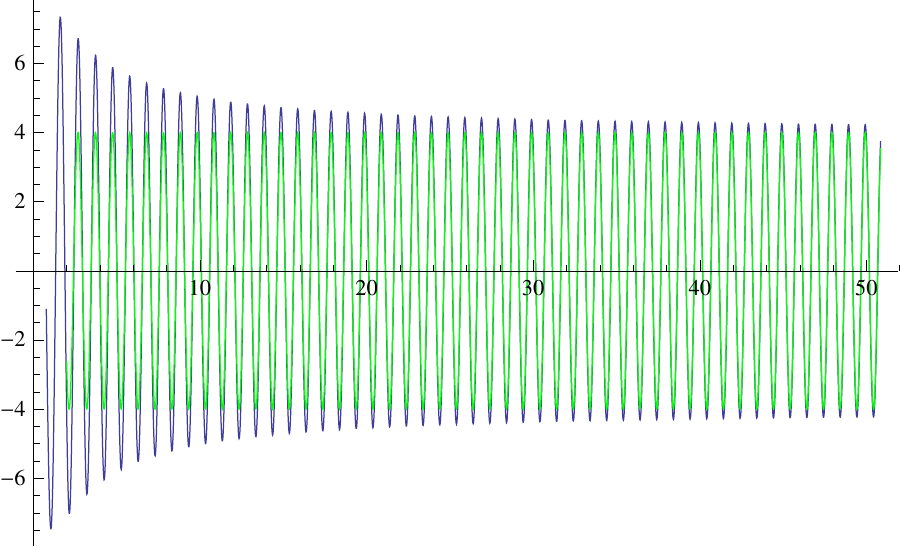}
\includegraphics[width=8cm]{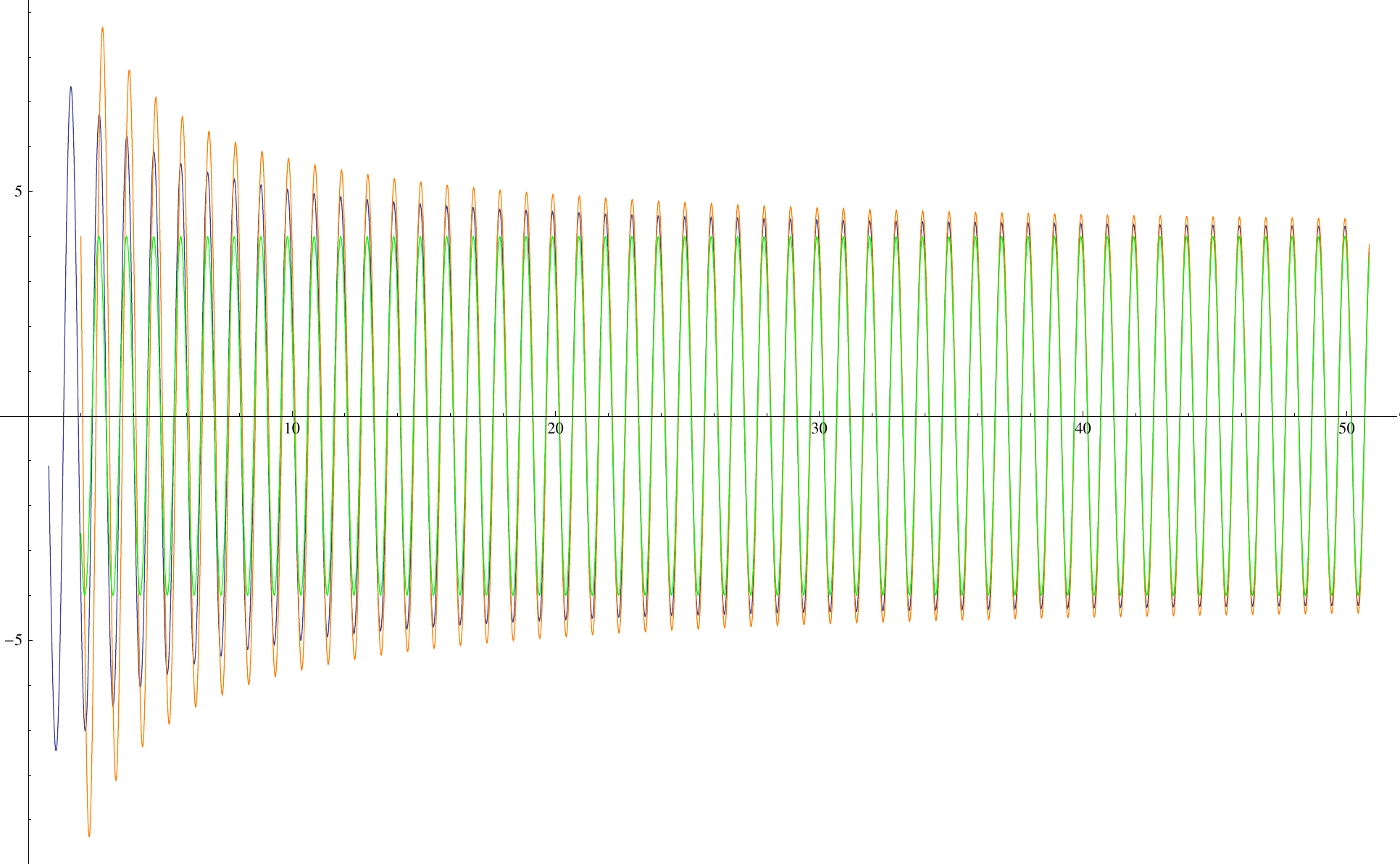}
\end{center}
\caption{ $q(\nu)$ as a function of $\nb$ for $s=2$, $\ell=2$.
See caption in Fig.\ref{fig:q,s=0 l=1} for a description of the curves.
Cf. Fig.2~\cite{Leung:2003ix} (also Fig.2~\cite{Leung:2003eq}).}
\label{fig:q,l=2=s}
\end{figure}


\subsection{Radial solution $\f(r,-i\nu)$}

From the boundary condition Eq.(\ref{eq:f,near hor}) and the relationship $r_*=r_*(r)$, the following exact monodromy around the horizon follows:

\begin{equation} \label{eq:f monodromy}
\f\left((r-r_h)e^{2\pi i},-i\nu\right)=e^{-2\pi i\nb}\f(r-r_h,-i\nu)
\end{equation}
For $r$ near $r_h$ (but far away enough from it
so that the WKB expansions $g_a$ are valid), it is $\f(r,-i\nu)\sim e^{-\nu r_*}\sim g_a(r,i\nu)$ to leading order as  $\nb\gg 1$ and,
since $g_a(r,i\nu)$ is there dominant over $g_a(r,-i\nu)$, we must have
 \begin{align}  \label{eq:f large-nu MvdB}
 &
\f(r,-i\nu)\sim g_a(r,i\nu)+\ca(\nu)g_a(r,-i\nu)
 \end{align}
 \fixme{Why does MvdB'04 write an equal sign?}
for some function $\ca(\nu)$, where in the coefficient of $g_a(r,i\nu)$ we are potentially neglecting exponentially-small terms.
 \fixme{There seems to be a sudden change of behaviour at $r_*=0$ which is where an anti-Stokes line crosses the real axis 
 (and which is an arbitrary value). Is Eq.(\ref{eq:f large-nu MvdB}) not valid at $r_*=0$? or near $r_*=0$? or for $r_*<0$? or for $r_*>0$?}
Eq.(\ref{eq:f large-nu MvdB}) can be continued from
the region of $r$ `near' $r_h$
 to the point $r_*=0$ and then anticlockwise along the anti-Stokes line up to  $\arg(r)=\pi/4$.
From Eq.(\ref{eq:g t/nu^1/6<<1}), we then have that 
 \begin{align} 
 &
\f(r,-i\nu)\sim e^{-i\pi \nb}e^{t^2/2} 
\left[1
-\frac{\dsa}{\nb}+\cdots
\right]
+\ca(\nu)e^{i\pi \nb}e^{-t^2/2}
\left[1+
\frac{\dsa}{\nb}+\cdots
\right]
 \end{align}
for $1/\sqrt{\nb}\ll |\rb|\ll 1$ with $\arg(r)=\pi/4$.
From Eqs.(\ref{eq:psi_1,2 for 1<<t<<nu^1/6 s=0})--(\ref{eq:psi_1,2 for 1<<t<<nu^1/6}) we find that 
 \begin{align} \label{eq:f in terms of psi1,2 large-nu}
s=0:\quad 
& \f(r,-i\nu)\sim 
\frac{\sqrt{\pi\nb}}{2}\left\{e^{-i\pi\nb}\left[1+\frac{(i-2)\aa}{\sqrt{\nb}}\right]-\ca(\nu)e^{i\pi\nb}\left[i+\frac{(i+2)\aa}{\sqrt{\nb}}\right]\right\}\psi_1+
 \nonumber
 \\&
  \frac{2\sqrt{\nb}}{\sqrt{\pi}}\left\{e^{-i\pi\nb}\left[i-\frac{\aa}{\sqrt{\nb}}\right]-\ca(\nu)e^{i\pi\nb}\left[1+\frac{\aa}{\sqrt{\nb}}\right]\right\}\psi_2,
  \nonumber
\\
s=1:\quad 
& \f(r,-i\nu)\sim 
\nonumber  \\ & 
\frac{\nb}{2}\left\{e^{-i\pi\nb}\left[1-\frac{i\aa}{\sqrt{\nb}}
+\frac{i\aa^2-
\re{\bb}
+\A }{\nb}
\right]
\left(1-\frac{\dsa}{\nb}
\right)
-\ca(\nu)e^{i\pi\nb}\left[1-\frac{\aa}{\sqrt{\nb}}
+\frac{i\aa^2+
\re{\bb}
-\A }{\nb}
\right]
\left(1+\frac{\dsa}{\nb}
\right)
\right\}
\psi_1+
 \\&
\left\{e^{-i\pi\nb}\left[1+\frac{i\aa}{\sqrt{\nb}}
+\frac{i\aa^2-\ab
-i\im{\bb}
}{\nb}
\right]
\left(1-\frac{\dsa}{\nb}
\right)
+\ca(\nu)e^{i\pi\nb}\left[1+\frac{\aa}{\sqrt{\nb}}
+\frac{i\aa^2+\ab
-i\im{\bb}
}{\nb}
\right]
\left(1+\frac{\dsa}{\nb}
\right)
\right\}\psi_2,
\nonumber
  \\
s=2:\quad 
& \f(r,-i\nu)\sim 
\frac{\sqrt{\pi\nb^3}}{8}\left\{e^{-i\pi\nb}\left[1+\frac{(i-2)\aa}{\sqrt{\nb}}\right]+\ca(\nu)e^{i\pi\nb}\left[i+\frac{(i+2)\aa}{\sqrt{\nb}}\right]\right\}\psi_1+
 \nonumber
 \\&
  \frac{2}{\sqrt{\pi\nb}}\left\{e^{-i\pi\nb}\left[i-\frac{\aa}{\sqrt{\nb}}\right]+\ca(\nu)e^{i\pi\nb}\left[1+\frac{\aa}{\sqrt{\nb}}\right]\right\}\psi_2,
  \nonumber
  \end{align}
This expression has been obtained with all functions $\f$, $\psi_1$, $\psi_2$ evaluated for $\arg(r)=\pi/4$, but it is valid $\forall \arg(r)$.
In particular, we can express  $\psi_1$ and $\psi_2$ on $\arg(r)=-\pi/4$ in terms of $e^{\pm t^2/2}$ by using 
Eqs.(\ref{eq:psi_1,2 for 1<<t<<nu^1/6 s=0})--(\ref{eq:psi_1,2 for 1<<t<<nu^1/6}) on $\arg(r)=\pi/4$ together with
the analytic continuation Eqs.(\ref{eq:psi_1,2^0,1(it) s=0})--(\ref{eq:psi_1,2^0,1(it)}).
We can match this expression to a linear combination of $g_a(\pm i\nu)$ via Eq.(\ref{eq:g_a large-nu}).
This linear combination can then be continued anticlockwise along the anti-Stokes line all the way back to $r_*=0$, thus yielding
the asymptotic monodromy:
 \begin{align} \label{eq:f monodromy g_a}
s=0: \quad
&
 \f\left((r-r_h)e^{2\pi i},-i\nu\right)=
e^{-2\pi i \nb}
 g_a(r-r_h,i\nu)+\left[-2i\left(1-\frac{\aa}{\sqrt{\nb}}\right)+\ca(\nu)e^{2\pi i \nb}\right]g_a(r-r_h,-i\nu),
 \nonumber
 \\
 s=1: \quad
&
 \f\left((r-r_h)e^{2\pi i},-i\nu\right)=
  \nonumber\\ &
\left\{
 e^{-2\pi i \nb}
  \left[1
 +O\left(\nb^
 {-3/2}
 \right)\right] 
 +\ca(\nu) O\left(\nb^
 {-3/2}
 \right)
 \right\}
  g_a(r-r_h,i\nu)+
  \\ \nonumber
  &
  \left\{
  \frac{2i\aa}{\sqrt{\nb}}
   -\frac{
   2i\aa^2
   }{\nb} 
   +O\left(\nb^
   {-3/2}
   \right)
  +
  \ca(\nu)e^{2\pi i \nb}
 \left[1
 +O\left(\nb^
 {-3/2}
  \right)\right]
 \right\}g_a(r-r_h,-i\nu),
 \nonumber
\\
s=2: \quad
 &
\f\left((r-r_h)e^{2\pi i},-i\nu\right)=e^{-2\pi i \nb}g_a(r-r_h,i\nu)+\left[2i\left(1-\frac{\aa}{\sqrt{\nb}}\right)+\ca(\nu)e^{2\pi i \nb}\right]g_a(r-r_h,-i\nu),
\nonumber
 \end{align}
Finally, comparing with the exact monodromy Eq.(\ref{eq:f monodromy}) and using Eq.(\ref{eq:f large-nu MvdB}), we obtain
 \begin{align} \label{eq:c_a MvdB} 
s&=0,2:&\quad
\ca(\nu)&\sim 
(-1)^{s/2}
\frac{\left(1-\aa/\sqrt{\nb}\right)}{\sin{\left (2\pi\nb\right)}},
\\
s&=1:&\quad 
\ca(\nu)&\sim 
\left[\frac{2i\aa}{\sqrt{\nb}}
-\frac{
2i\aa^2
}{\nb}
+O\left(\nb^
{-3/2}
\right)\right]\left[-2i\sin(2\pi\nb)-
e^{2\pi i\nb}
O\left(\nb^{-3/2}
\right)
\right]^{-1}
\nonumber
 \end{align}
 as well as, for $s=1$ and from the coefficient of $g_a(r-r_h,i\nu)$,
\begin{equation}\label{eq:coeff ga(iNu)}
e^{-2\pi i\nb}=e^{-2\pi i\nb}\left[1+O(\nb^{-3/2})\right]+\ca(\nu)O(\nb^{-3/2})
\end{equation} 
Note that the terms $O(\nb^{-3/2})$ in  Eq.(\ref{eq:coeff ga(iNu)}) may actually decrease faster than $\nb^{-3/2}$ and they
do not necessarily have to decrease both at the same rate. In particular, this implies that, e.g., $\ca(\nu)$ may increase/decrease
with $\nb$ as long as the $O(\nb^{-3/2})$ multiplying it makes up for it by decreasing/increasing faster than the other $O(\nb^{-3/2})$.
 
In Fig.\ref{fig:f wrt nu,s=0,l=1 & l=1,s=1 & l=2,s=2,r*=0.2,0.4} we plot $\fb=-\sin\left(2\pi\nb\right)\f(r,-i\nu)$ as a function of $M\nu$: it shows that
the  large-$\nb$ asymptotics of  Eq.(\ref{eq:f large-nu MvdB})
(together with (\ref{eq:c_a MvdB}) and  (\ref{eq:g_a large-nu})) overlap with a calculation of
 $\fb$ using the so-called Jaff\'e series~\cite{Leaver:1986a,Casals:Ottewill:2011midBC}.

 \begin{figure}[h!]
      \includegraphics[width=8cm]{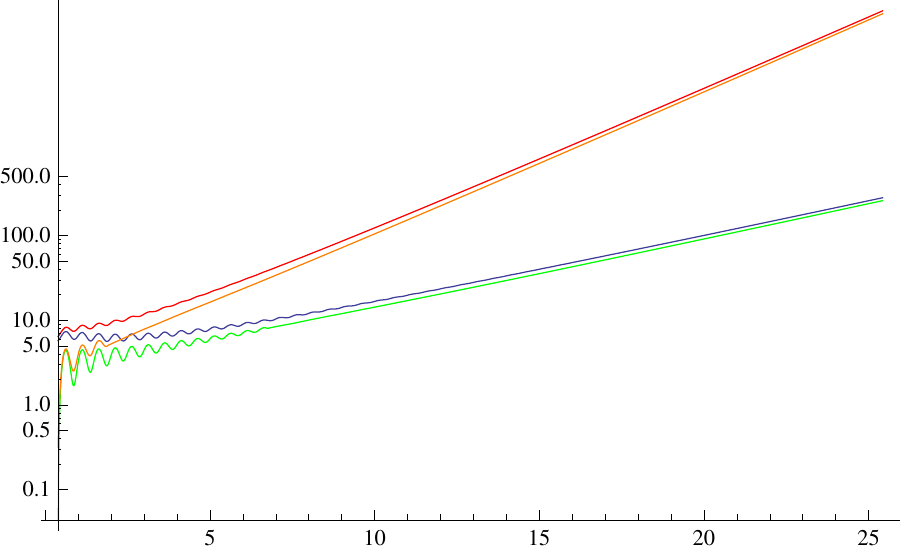}
       \includegraphics[width=8cm]{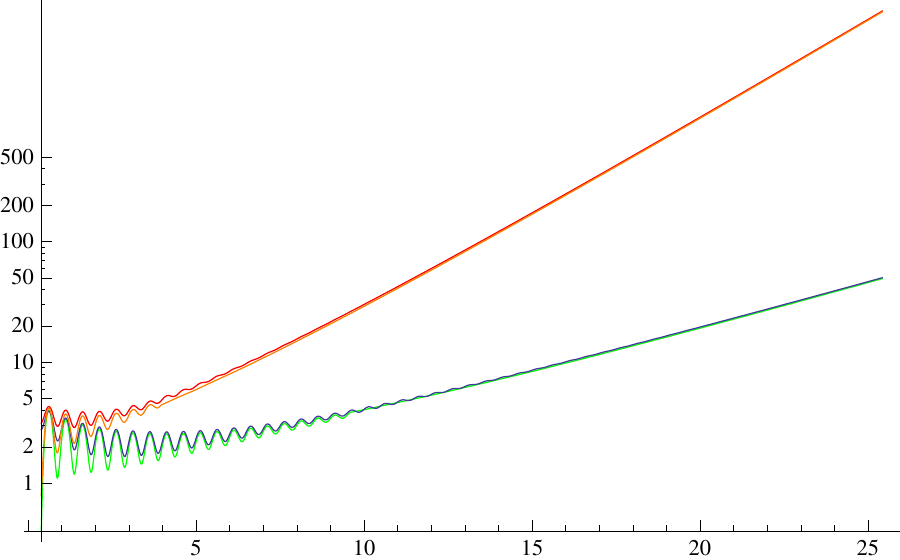}
           \includegraphics[width=8cm]{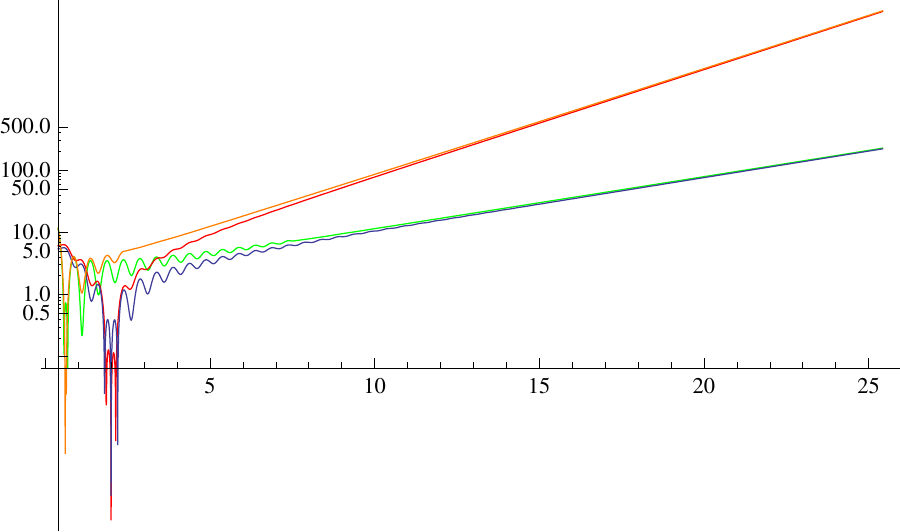}
\caption{
Log-plot of $|\fb|$ as a function of $\nu M$.
The green and orange curves are the large-$\nb$ asymptotics of Eq.(\ref{eq:f large-nu MvdB})
(together with (\ref{eq:c_a MvdB}) and  (\ref{eq:g_a large-nu})) for $r_*=0. 2M$ and $r_*=0.4M$ respectively.
The blue and red curves are respectively calculated using the Jaff\'e series~\cite{Leaver:1986a,Casals:Ottewill:2011midBC}
 for $r_*=0. 2M$ and $r_*=0.4M$.
 (a) For $s=0$, $\ell=1$. (b)  For $s=1$, $\ell=1$.
 (c)  For $s=2$, $\ell=2$;
 note that both curves for the Jaff\'e series give a zero value for $\fb$ at $\nb=\nb_{AS}$
because of the definition of $\fb$ and the fact that, for $s=2$, $\nb_{AS}$ is not a pole.}
\label{fig:f wrt nu,s=0,l=1 & l=1,s=1 & l=2,s=2,r*=0.2,0.4}
\end{figure}


\subsection{Wronskian}

Using Eq.(\ref{eq:g=g(g_a)}) for $\gp$ and Eqs.(\ref{eq:f large-nu MvdB}), (\ref{eq:c_a MvdB}) and (\ref{eq:g_a large-nu})
for $\f$, we find the following large-$\nb$ asymptotics for the Wronskian on the 4th quadrant of the complex-$\omega$
plane infinitesimally close to the NIA:
\begin{align} \label{eq:Wronsk large-nu MvdB}
s=0,2:\quad
 &
-2\nu \Ainp= \Wp{-i\nu}
 \sim \frac{W[g_a(-i\nu),g_a(i\nu)]}{\sin(2\pi\nb)}
 \left[\sin\left(2\pi\nb\right)+2ie^{2\pi i \nb}\left(1-\frac{(1+i) 
 \aa
 }{\sqrt{\nb}}\right)\right],
\\
s=1:\quad
 &
 -2\nu \Ainp=\Wp{-i\nu}
 \sim 
 W[g_a(-i\nu),g_a(i\nu)]\left\{1+O\left(\nb^{-3/2}\right)+\left[\frac{2\aa}{\sqrt{\nb}}-\frac{2
 i\aa^2
 }{\nb}+O\left(\nb^{-3/2}\right)\right]\ca(\nb)e^{2\pi i\nb}\right\}
 \nonumber
 \end{align}
  To leading order for $s=0$ and $s=2$ it yields
$
 W_+(-i\nu)
\sim 2\nu \left[1-2i\cot (2\pi\nb)\right]$.
This agrees with Eq.2.17 in~\cite{Neitzke:2003mz}:
\begin{equation} \label{eq:Ain large-nu}
\Ainm\sim 
\lim_{\epsilon\to 0^+}
\left.\left( \frac{
e^{-4\pi\ob}
-1}{
e^{-4\pi\ob}
+[1+2\cos(\pi s)]}\right)^{-1}\right|_{\ob=-\epsilon-\nb i},\qquad \nb\gg 1,
\end{equation}
on the 3rd quadrant infinitesimally close to the NIA, after using the symmetry Eq.(\ref{eq:symms g,W on NIA}).

Similarly to the `trick' used to go from Eq.(\ref{eq:q for large-nu}) to Eq.(\ref{eq:q for large-nu phase}), we can rewrite 
Eq.(\ref{eq:Wronsk large-nu MvdB}) as
\begin{align} \label{eq:Wronsk large-nu MvdB phase}
s=0,2:\quad
 &
 \Wp{-i\nu}
 \sim \frac{W[g_a(-i\nu),g_a(i\nu)]}{\sin(2\pi\nb)}
 \left[\sin\left(2\pi\nb\right)+2ie^{2\pi i \nb-\frac{(1+i) 
 \aa
 }{\sqrt{\nb}}}\right],
\\
s=1:\quad
 &
 \Wp{-i\nu}
 \sim 
 W[g_a(-i\nu),g_a(i\nu)]\left\{1+O\left(\nb^{-3/2}\right)-\frac{2\aa^2}{\nb \sin\left(2\pi\nb\right)}
 e^{2\pi i\nb-\frac{(1+i)\aa}{\sqrt{\nb}}}
 \right\}
 \nonumber
 \end{align}
where we have used Eq.(\ref{eq:c_a MvdB}) in the $s=1$ case.

In Figs.\ref{fig:wronskian,s=0,l=1,r=2.8}--\ref{fig:wronskian,s=2,l=2,r=2.8} we plot  $\hat W_+(-i\nu) =W\left[\gp(r,-i\nu),\fb(r,-i\nu)\right]$ as a function of $M\nu$ and we show that the
large-$\nb$ asymptotics of Eq.(\ref{eq:Wronsk large-nu MvdB}) agree with the calculation using the method in~\cite{Casals:Ottewill:2011midBC}.
\fixme{Not so good for $s=0$?}

\begin{figure}[h!]
\begin{center}
 \includegraphics[width=15cm]{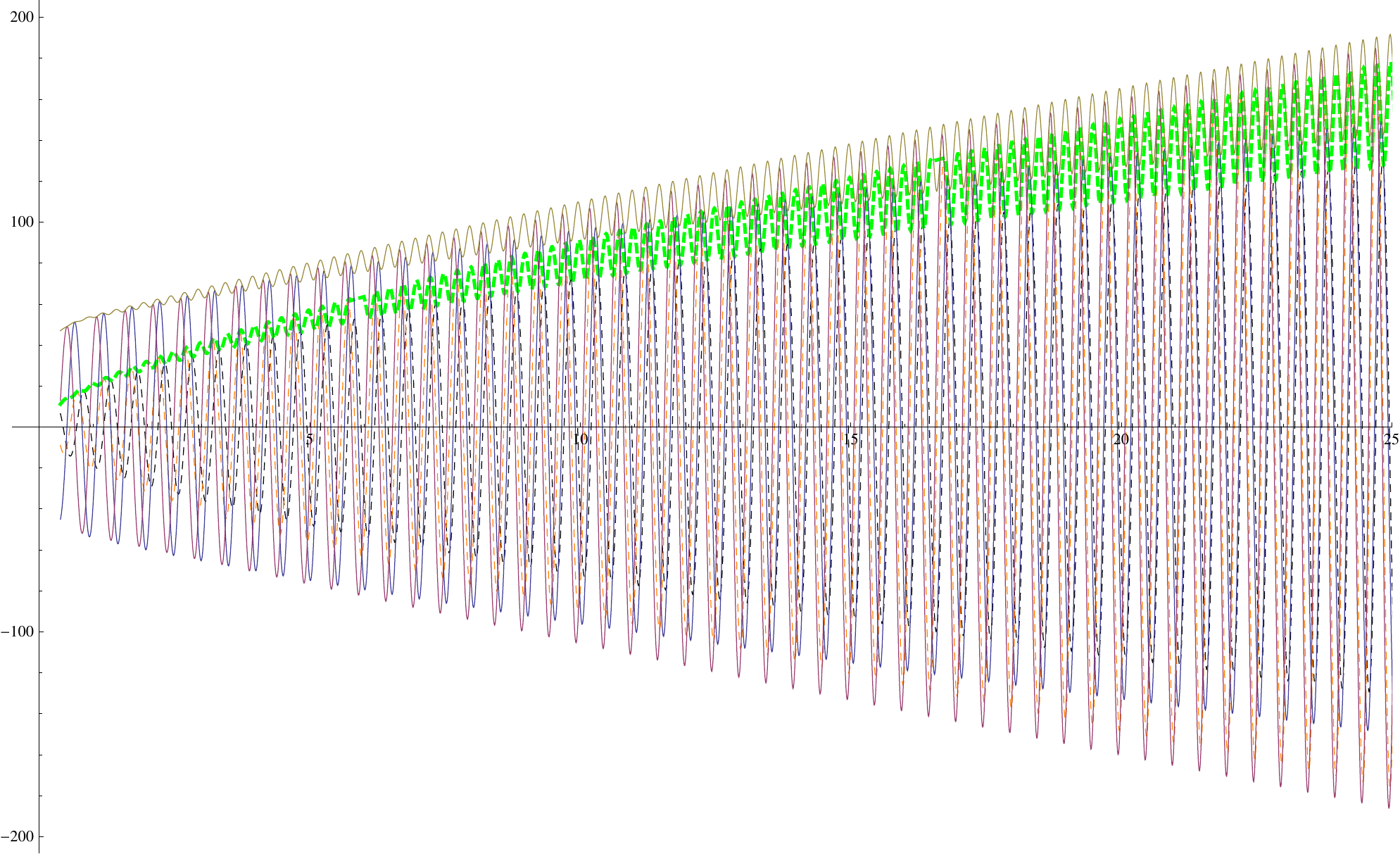}
 \includegraphics[width=15cm]{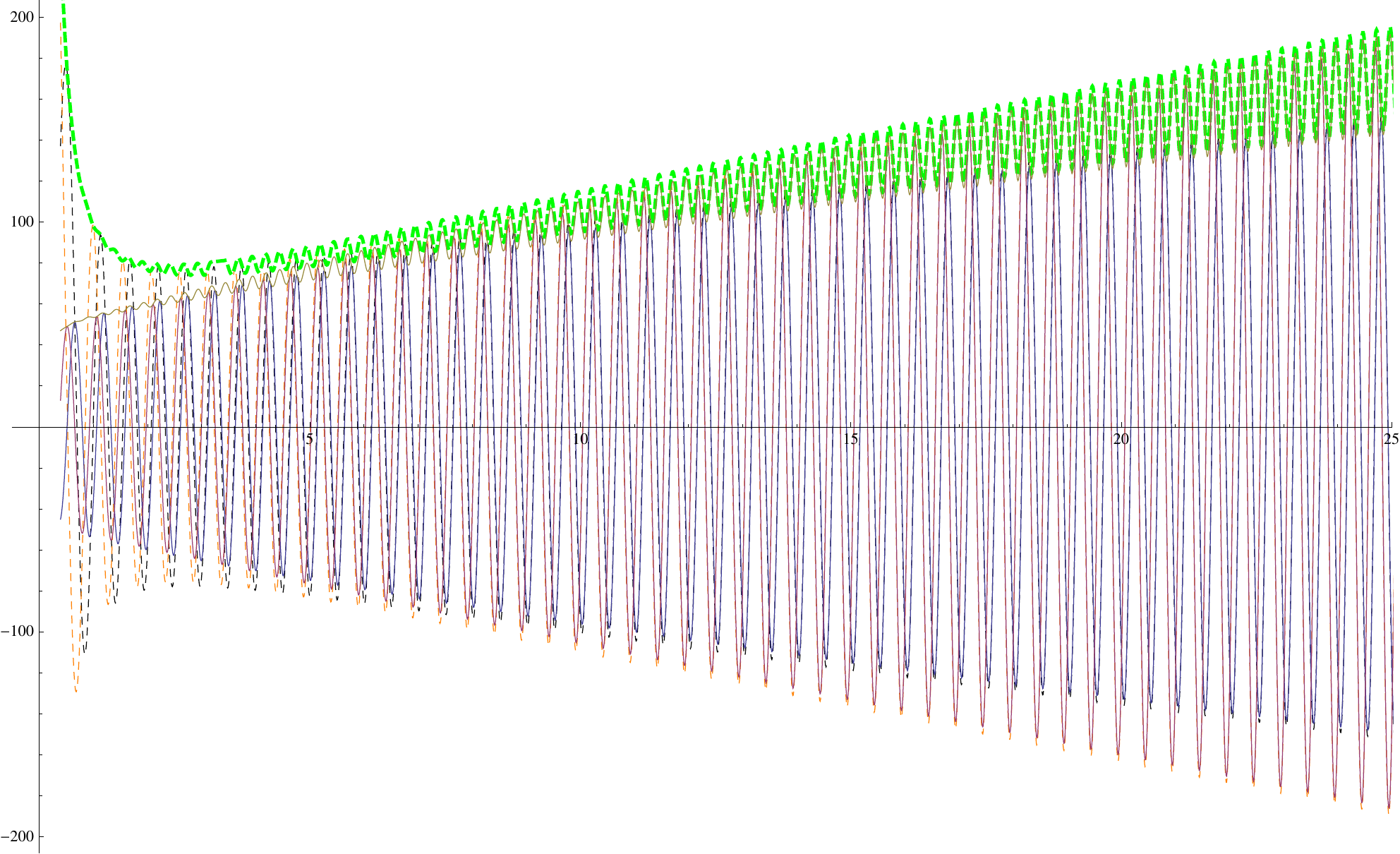}
            \end{center}
\caption{
Real part, imaginary part and absolute value of $\hat W_+$ as functions of $\nu M$.
The asymptotic expressions in Eq.(\ref{eq:Wronsk large-nu MvdB}) are plotted as dashed
black, orange and green curves, corresponding to the 
real part, imaginary part and absolute value of $\hat W_+$, respectively.
The calculation using the method in~\cite{Casals:Ottewill:2011midBC} 
(for this, the  value $r=2.8M$ has been used to calculate the radial solutions $\gp$ and $\fb$) is plotted as continuous 
blue, red and brown curves, corresponding to the 
real part, imaginary part and absolute value of $\hat W_+$, respectively.
This plot is for the values $s=0$, $\ell=1$. 
(a) Asymptotic expressions using Eq.(\ref{eq:Wronsk large-nu MvdB}).
(b) Asymptotic expressions using Eq.(\ref{eq:Wronsk large-nu MvdB phase}).
The odd non-oscillatory intervals in the dashed green curves are just a visual artifact of the computational software program that we used for plotting the curves.
}
\label{fig:wronskian,s=0,l=1,r=2.8}
\end{figure}

\begin{figure}[h!]
\begin{center}
 \includegraphics[width=15cm]{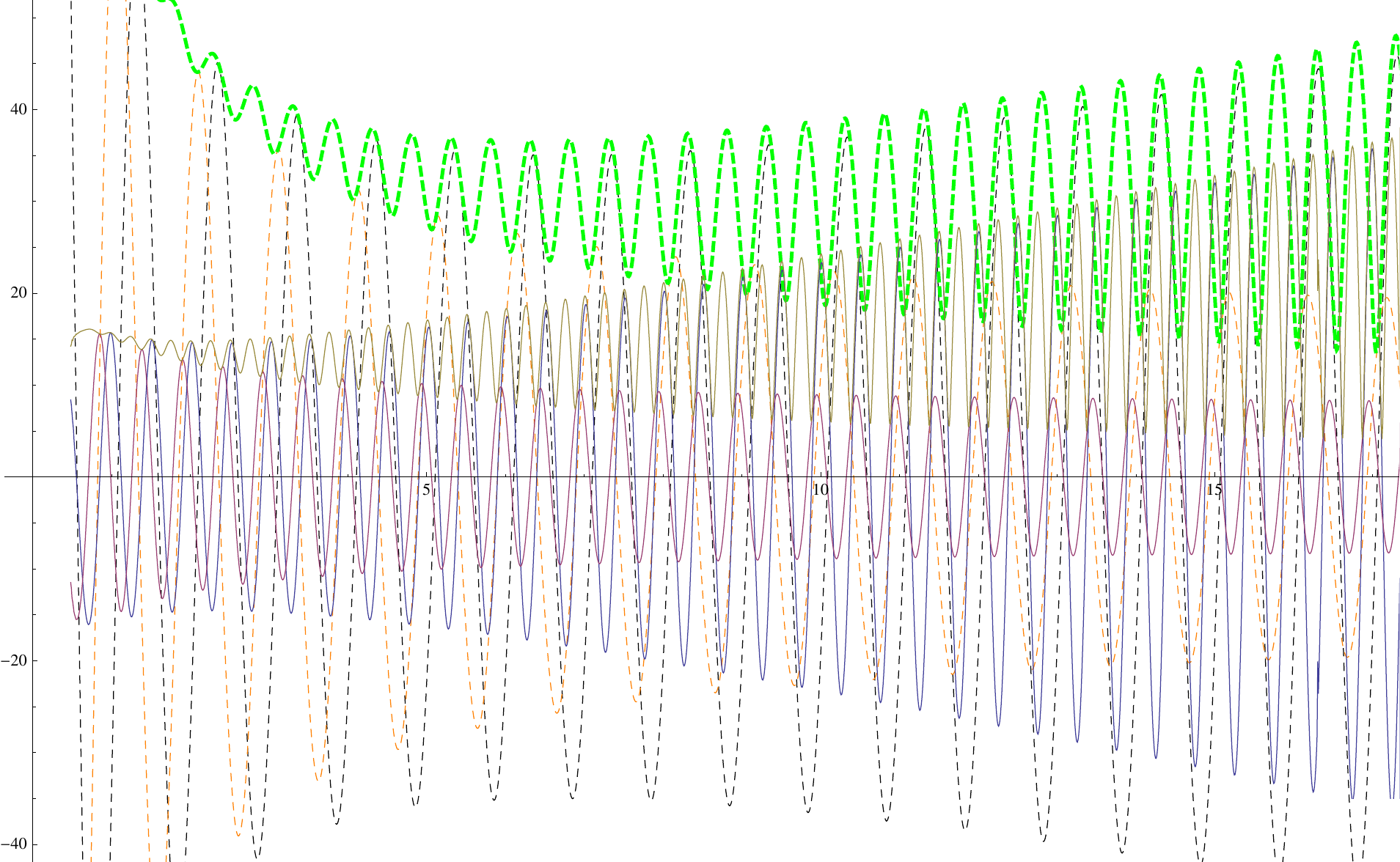} 
            \end{center}
\caption{
Real part, imaginary part and absolute value of $\hat W_+$ as functions of $\nu M$.
This plot is for the values $s=1$, $\ell=1$. 
See caption in Fig.\ref{fig:wronskian,s=0,l=1,r=2.8}(a) for a description of the curves.
Note that, in this case, the method in~\cite{Casals:Ottewill:2011midBC} (corresponding to the blue curve) cannot reach higher values of $\nu M$.
}
\label{fig:wronskian,s=1,l=1,r=2.8}
\end{figure} 

\begin{figure}[h!]
\begin{center}t
 \includegraphics[width=15cm]{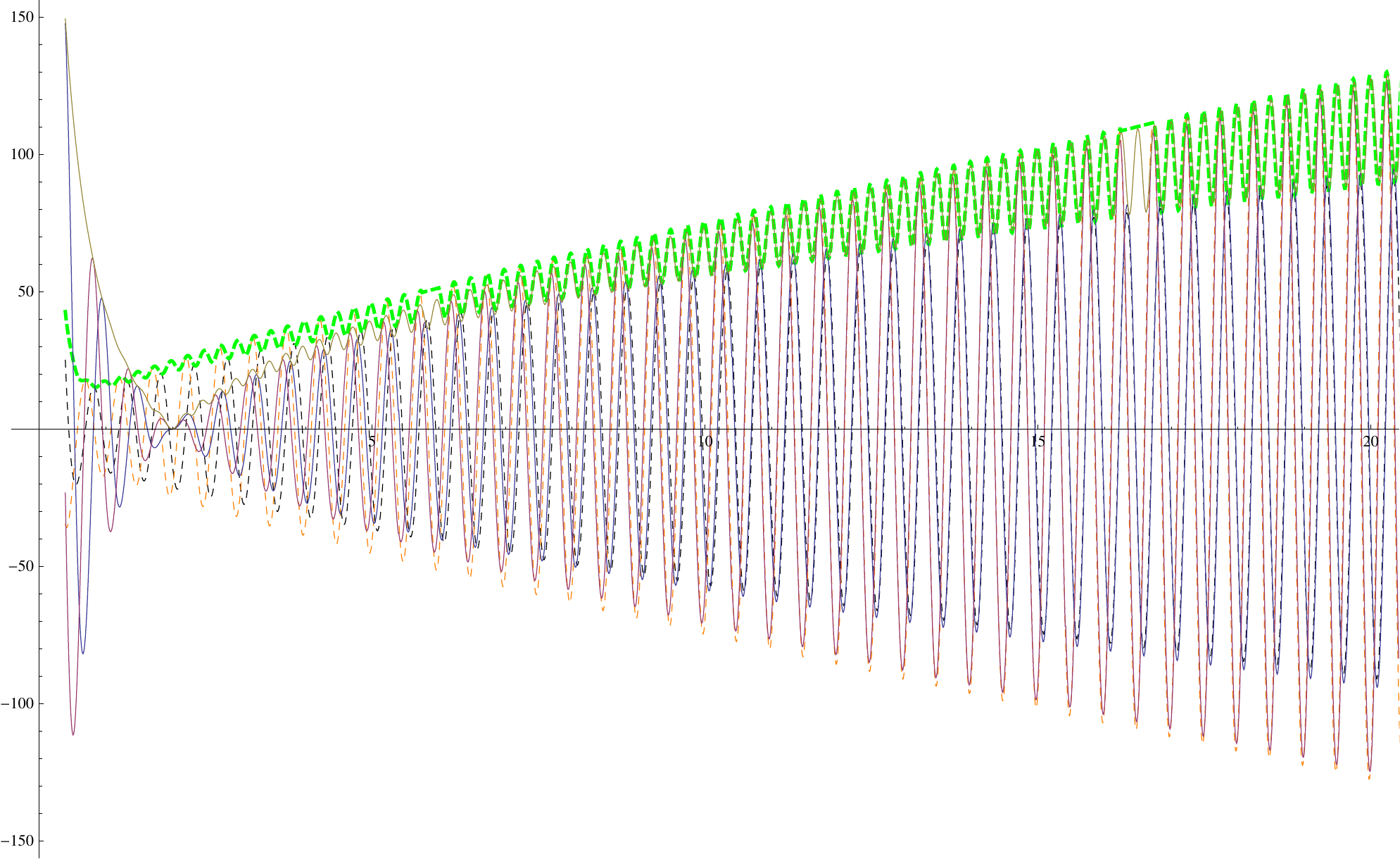}
            \end{center}
\caption{
Real part, imaginary part and absolute value of $\hat W_+$ as functions of $\nu M$.
This plot is for the values $s=2$, $\ell=2$. 
See caption in Fig.\ref{fig:wronskian,s=0,l=1,r=2.8}(a) for a description of the curves.
Note that the blue and black curves are mostly overlapping, and so are the orange and red curves.
The curves obtained with the method in~\cite{Casals:Ottewill:2011midBC} have a zero at $\nu=\nu_{AS}$ due to a corresponding zero in $\fb$.
}
\label{fig:wronskian,s=2,l=2,r=2.8}
\end{figure}


\subsection{
Green function modes}

We can finally give an asymptotic expression for large-$\nb$ for the BC discontinuity of the `retarded' Green function
modes.
From Eq.(\ref{eq:DeltaG in terms of Deltag}) and the leading orders of Eqs.(\ref{eq:Wronsk large-nu MvdB}), (\ref{eq:f large-nu MvdB}) and (\ref{eq:q for large-nu}), we find
\begin{align}\label{eq:DG large-nu MvdB}
\DGw{r}{r'}{-i\nu}&\sim
\frac{(-1)^{s/2}2i}{\nu}
\frac{\cos(2\pi\nb)}{\left[1+3\cos^2(2\pi\nb)\right]}
\left[(-1)^{s/2}e^{\nu r_*}+\sin(2\pi\nb)e^{-\nu r_*}\right]\left[(-1)^{s/2}e^{\nu r'_*}+\sin(2\pi\nb)e^{-\nu r'_*}\right],
&
s&=0,2
\\
\DGw{r}{r'}{-i\nu}&\sim
\frac{-\sqrt{\pi}i\ell (\ell+1)\sin(2\pi\nb)r_h}{\nb^{3/2}}
\left[\dfrac{\sqrt{\pi}\ell(\ell+1)}{2\nb^{1/2}\sin(2\pi\nb)}e^{\nu r_*}+e^{-\nu r_*}\right]
\left[\dfrac{\sqrt{\pi}\ell(\ell+1)}{2\nb^{1/2}\sin(2\pi\nb)}e^{\nu r'_*}+e^{-\nu r'_*}\right],
&
s&=1
\nonumber
\end{align}
\fixme{Like Eq.(\ref{eq:f large-nu MvdB}), is this not valid near $r_*=0$ or for $r_*<0$ or for $r_*>0$? but the plots are for $r_* \gtrsim 0$ so it should be valid there?}
This shows the convergence of the $\nu$-integral in the upper limit of integration in Eq.(\ref{eq: G^BC integral}) 
when $\Delta t>|r_*|+|r'_*|$.
The divergence in the BC when $\Delta t<|r_*|+|r'_*|$ was to be expected, since the QNM series also  seems to diverge at these `very early' times~\cite{Leaver:1986,Andersson:1997}, and one would
expect the divergences in the different contributions to the Green function to cancel each other out (see, e.g.,~\cite{Ching:1996} for the case of a radial potential  which does not lead to a BC, and where the divergences
of the QNM and high-frequency arc contributions at these `very early' times cancel out).
\fixme{We should be able to see this quickly by changing slightly results in Sec.\ref{Perturbation}? but the divergence of QNMs at very early times does not seem to have been proven by anybody - why couldn't others do it like we do in Sec.\ref{Perturbation}?}
In Figs.\ref{fig:DeltaG s=0,l=1}--\ref{fig:DeltaG s=0,l=1 phase}
we plot the large-$\nb$ asymptotics of $\DGw{r}{r'}{\omega}$
together with these modes calculated with the method in~\cite{Casals:Ottewill:2011midBC}.

\begin{figure}[h!]
\begin{center}
\includegraphics[width=8cm]{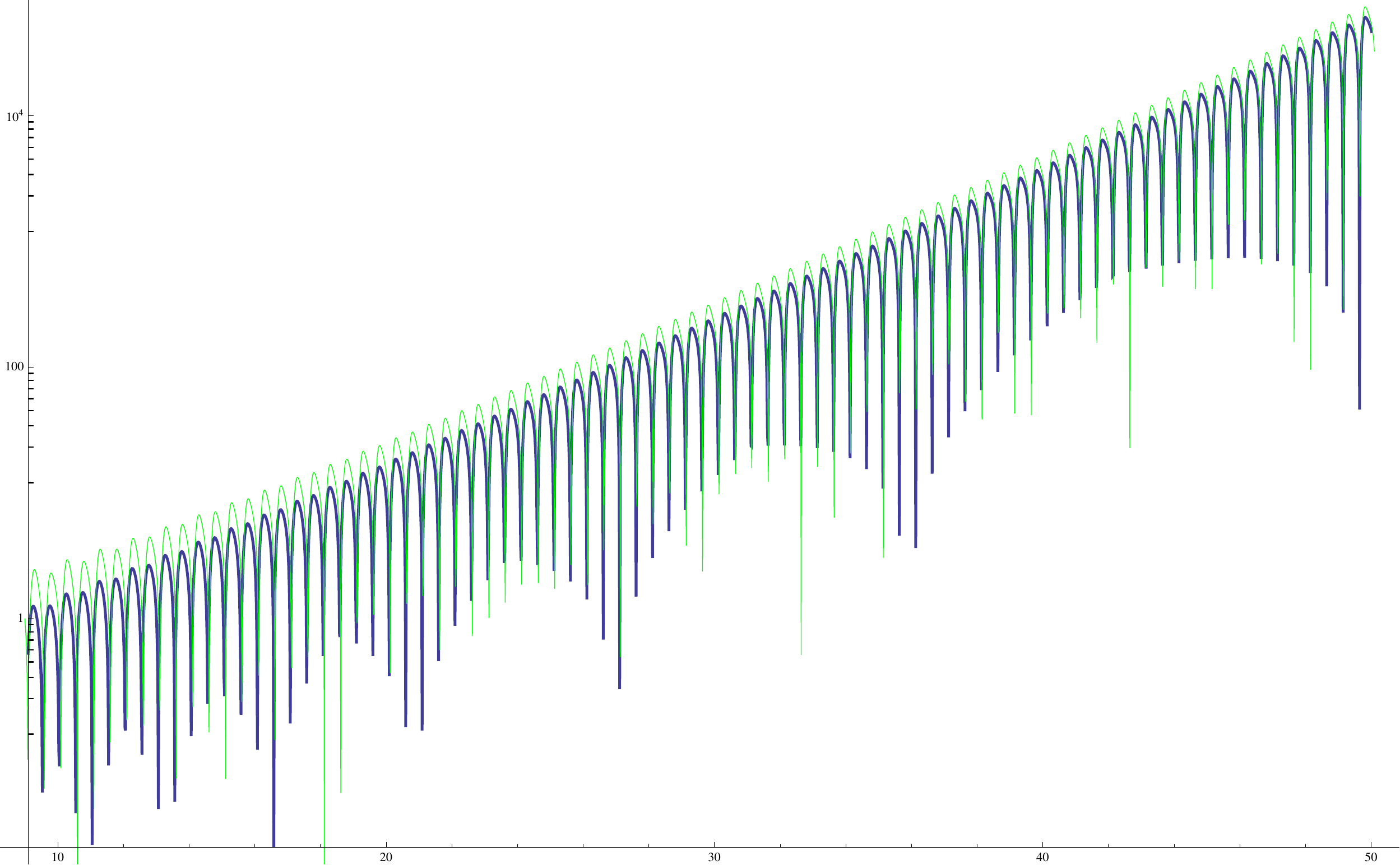}
          \includegraphics[width=8cm]{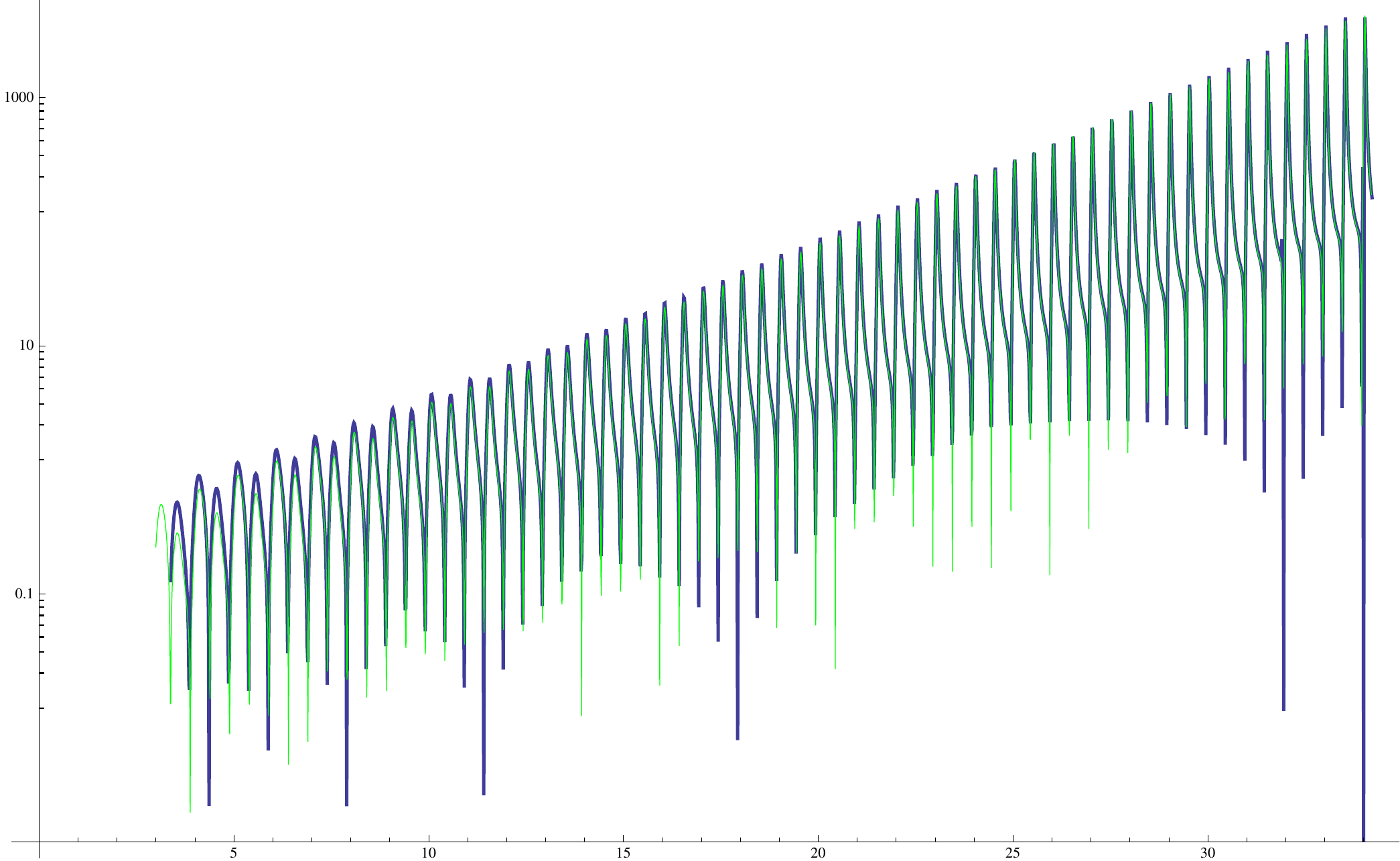}
          \includegraphics[width=8cm]{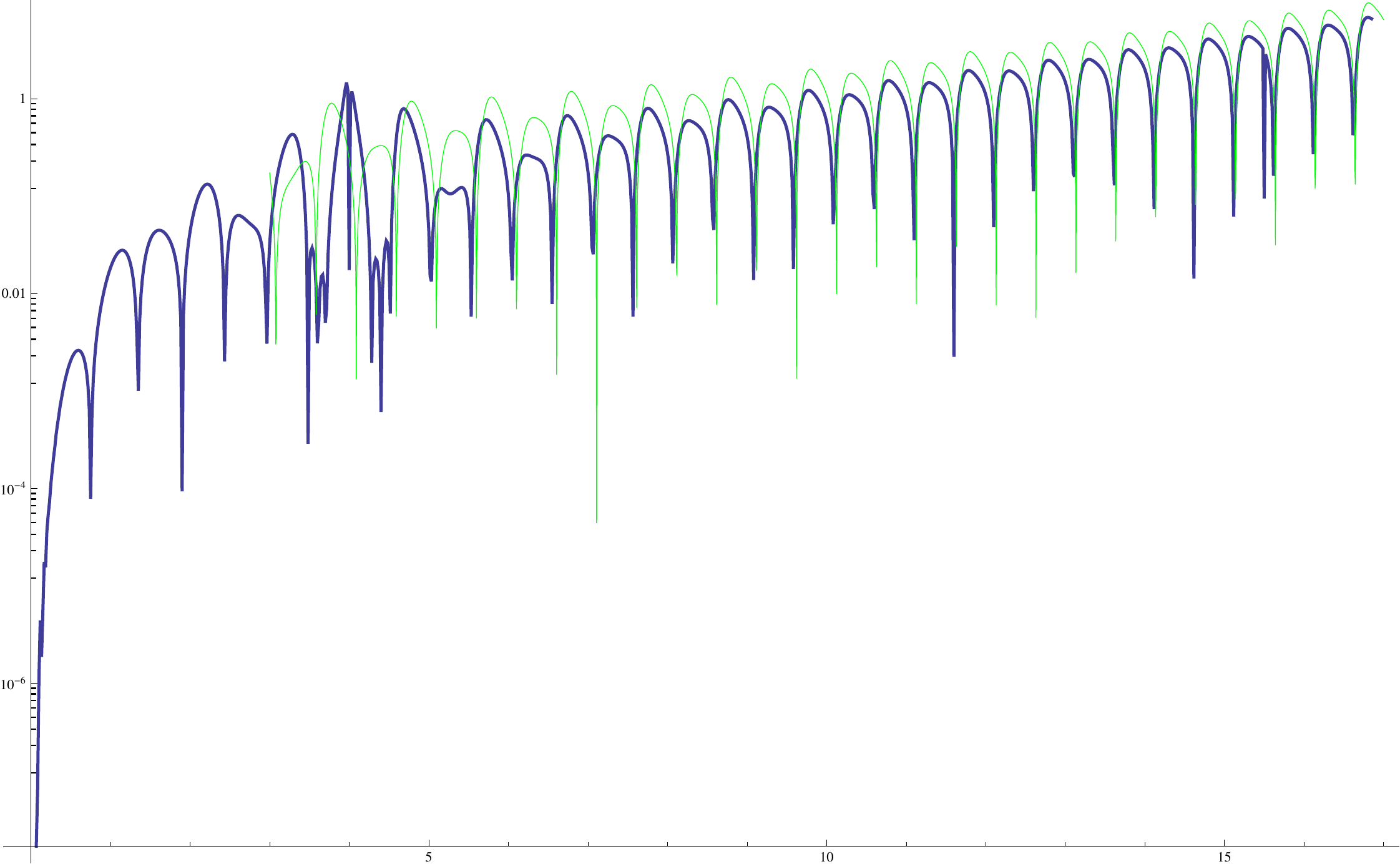}
\end{center}
\caption{
BC Green function modes
$\DGw{r(r_*=0.4)}{r'(r_*=0.2)}{-i\nu}$ of Eq.(\ref{eq:DeltaG in terms of Deltag}) as a function of $\nb$.
The green curves correspond to the large-$\nb$ asymptotics of $\Delta G_{\ell}$ using Eqs.(\ref{eq:q for large-nu}), (\ref{eq:Wronsk large-nu MvdB})
and (\ref{eq:f large-nu MvdB}) (so not just the leading-order Eq.(\ref{eq:DG large-nu MvdB})).
The blue curves have been obtained with the method in~\cite{Casals:Ottewill:2011midBC}
(for which we have used $\fb$ instead of $\f$ everywhere in Eq.(\ref{eq:DeltaG in terms of Deltag}) and
we have chosen to calculate  the radial functions at $r=5M$ and at $r=2.8M$ for the calculation of, respectively, $q(\nu)$
and the Wronskian).
(a) For $s=0$ and $\ell=1$.
(b) For $s=1$ and $\ell=1$.
(c) For $s=2$ and $\ell=2$ (the `particular' behaviour around $\nb=4$ is due to this value being that of the algebraically-special frequency).
\fixme{The large-$\nb$ asymptotics  are much better if multiplied `by hand' by $r_h^2$ for $s=1$?}
}
\label{fig:DeltaG s=0,l=1}
\end{figure} 

\begin{figure}[h!]
\begin{center}
\includegraphics[width=8cm]{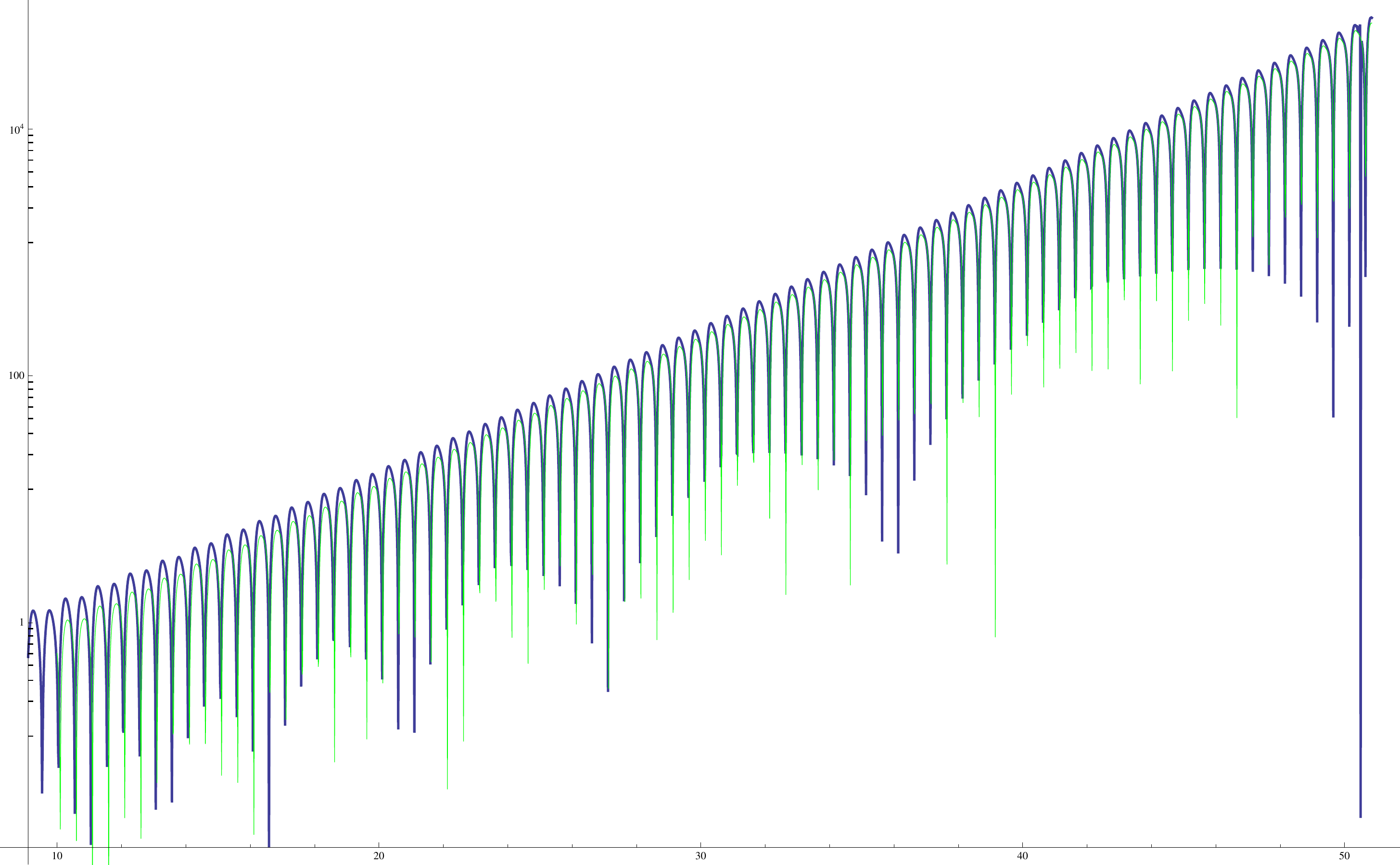}
\includegraphics[width=8cm]{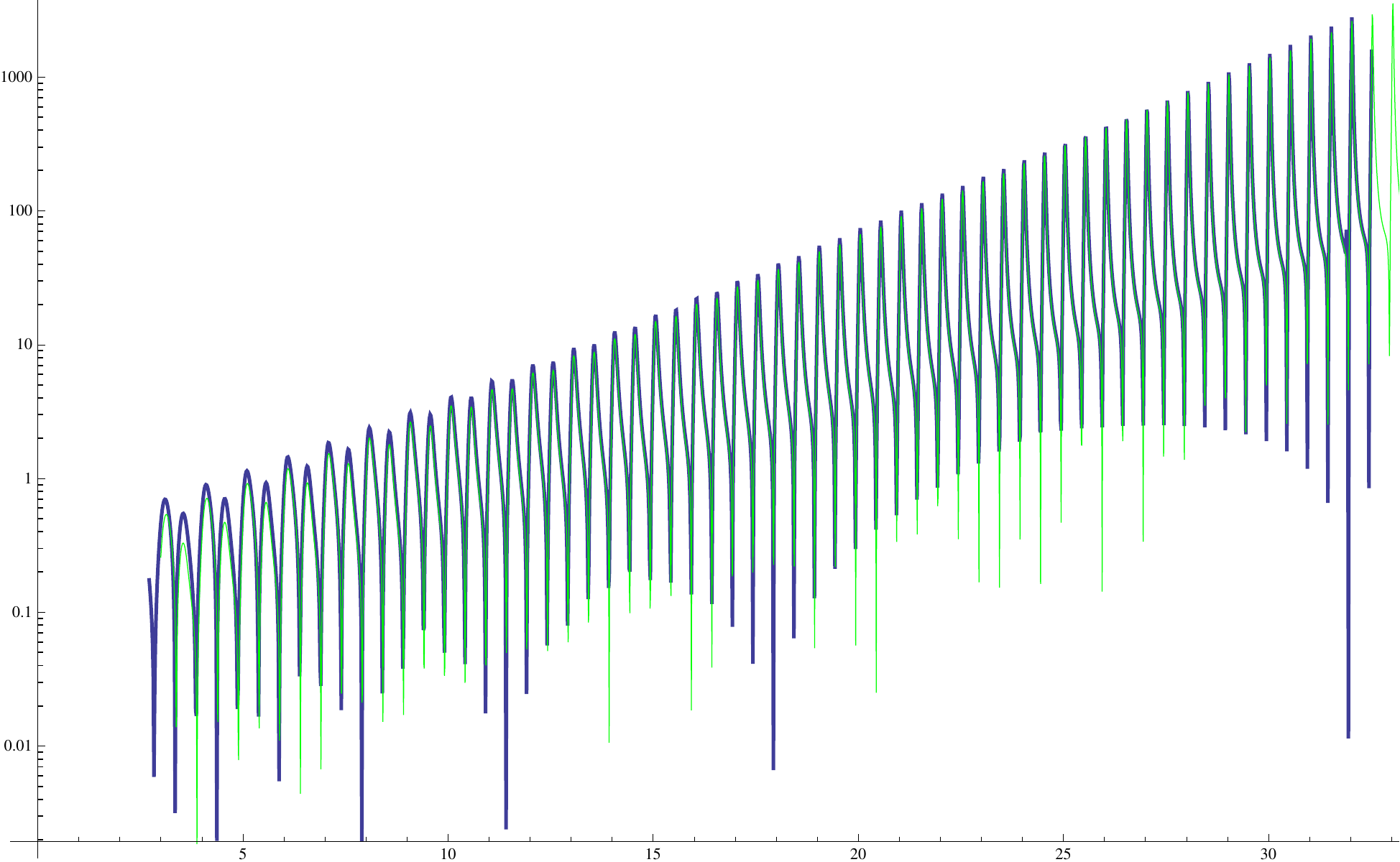}
          \includegraphics[width=8cm]{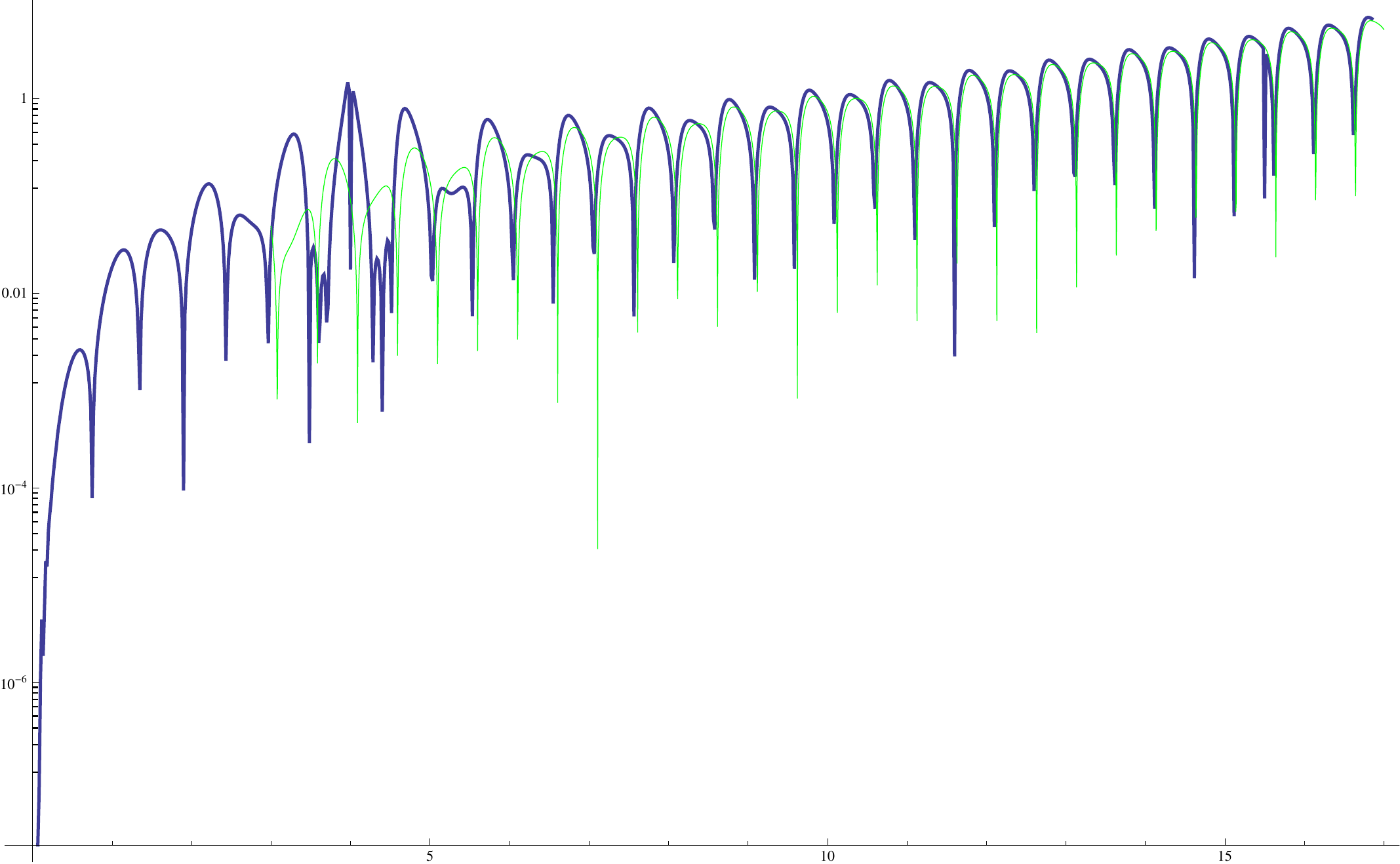}
\end{center}
\caption{
Similar to Figs.\ref{fig:DeltaG s=0,l=1} but using Eq.(\ref{eq:Wronsk large-nu MvdB phase}) instead of Eq.(\ref{eq:Wronsk large-nu MvdB}).
It shows a slight improvement with respect to Figs.\ref{fig:DeltaG s=0,l=1}.}
\label{fig:DeltaG s=0,l=1 phase}
\end{figure}


\section{Highly-damped QNMs} \label{sec:QNMs}

QNMs are poles of the `retarded' Green function in the lower frequency plane.
By requiring that $W_+=-2\nu \Ainp$ in Eq.(\ref{eq:Wronsk large-nu MvdB}) be zero, we obtain the highly-damped QNM frequencies
\begin{align} \label{eq:QNM s=0,2}
s=0:\quad &
\wbQNM\sim \frac{\ln 3}{4\pi}-\left(\frac{n}{2}+\frac{1}{4}\right)i+\frac{\sqrt{2}\Gamma^4(1/4)}{144\pi^{5/2}}(1+i)\frac{3\lambda +1}{\sqrt{n}}+O\left(\frac{1}{n}\right),
\qquad n\to \infty
\\ 
s=2:\quad &
\wbQNM
\sim \frac{\ln 3}{4\pi}-\left(\frac{n}{2}+\frac{1}{4}\right)i+\frac{\sqrt{2}\Gamma^4(1/4)}{144\pi^{5/2}}(1+i)\frac{\lambda -1}{\sqrt{n}}+O\left(\frac{1}{n}\right),
\nonumber
 \end{align}
 where $n$ is the so-called overtone index.
These expressions agree with~\cite{Musiri:2003bv} (for $s=2$ we are just including the result in~\cite{MaassenvandenBrink:2003as} for completeness).

\begin{figure}[h!]
\begin{center}
\includegraphics[width=8cm]{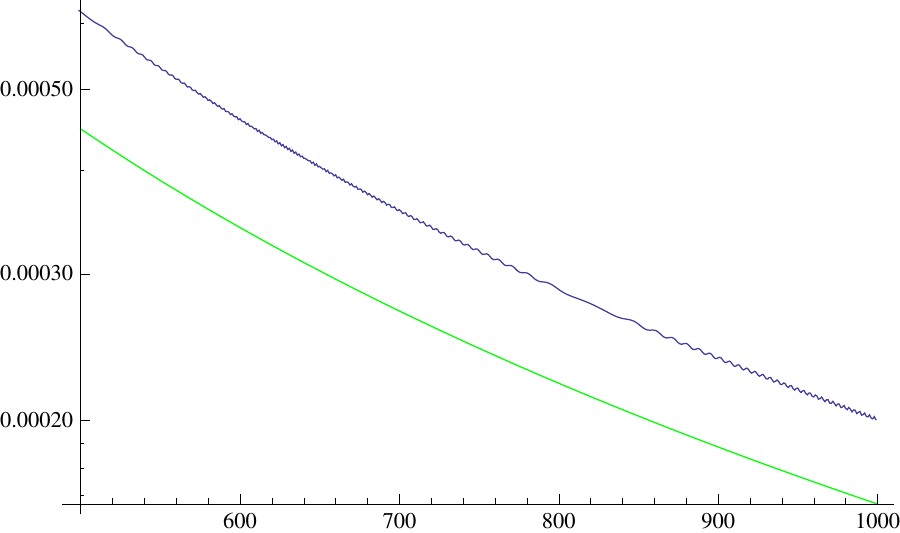}
\includegraphics[width=8cm]{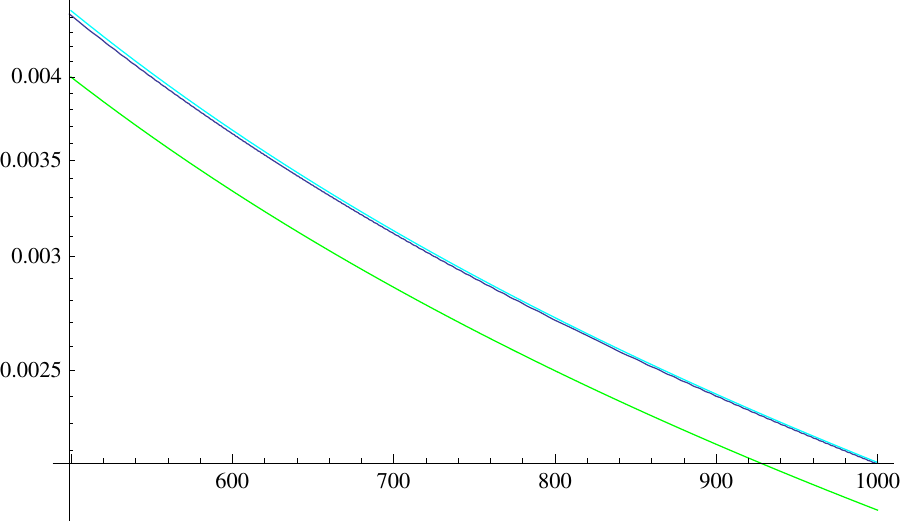}
\includegraphics[width=8cm]{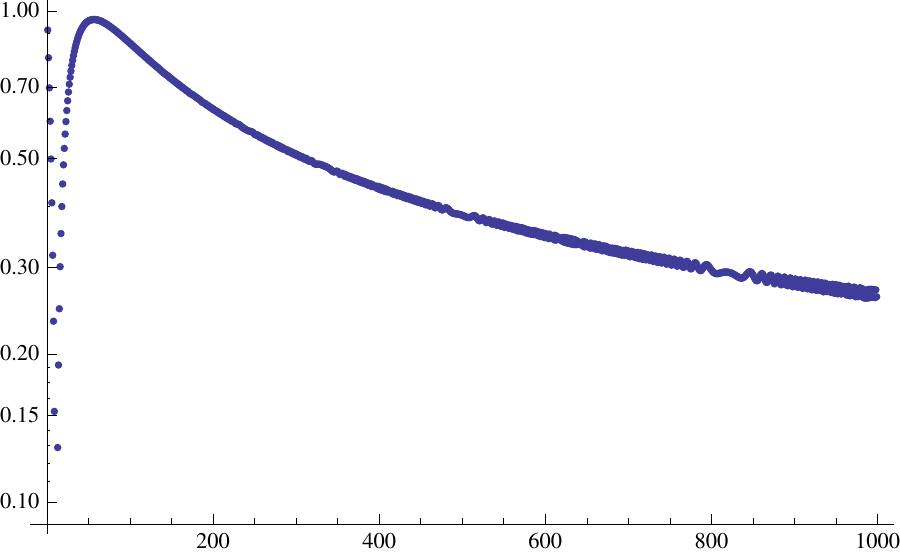}
\includegraphics[width=8cm]{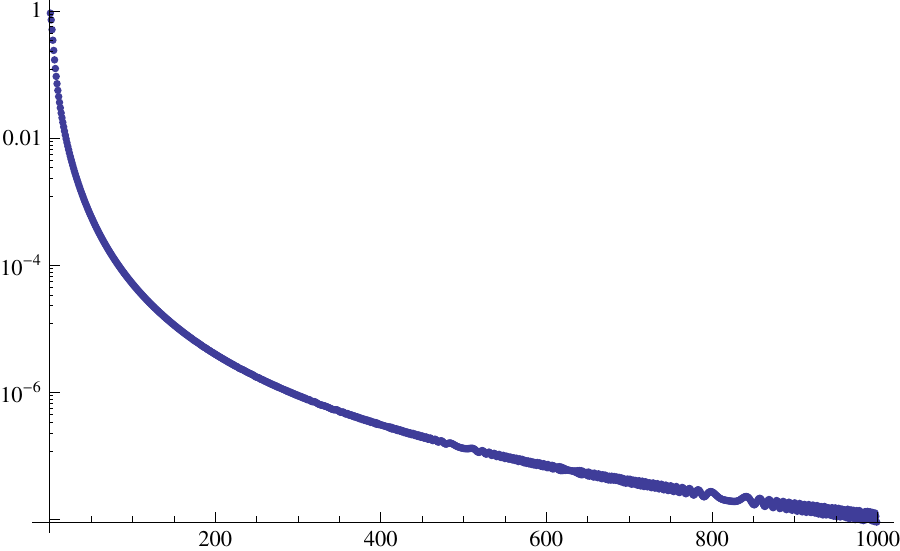}
\includegraphics[width=8cm]{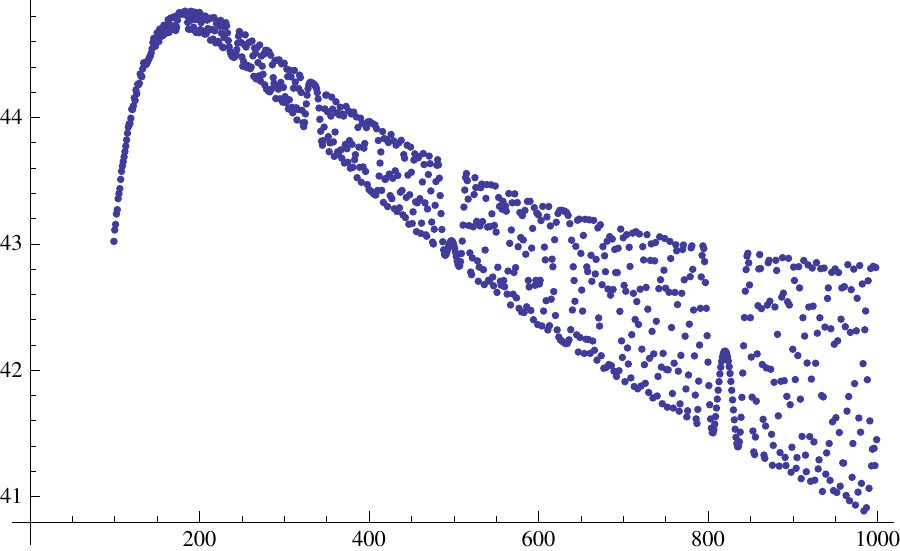} 
\includegraphics[width=8cm]{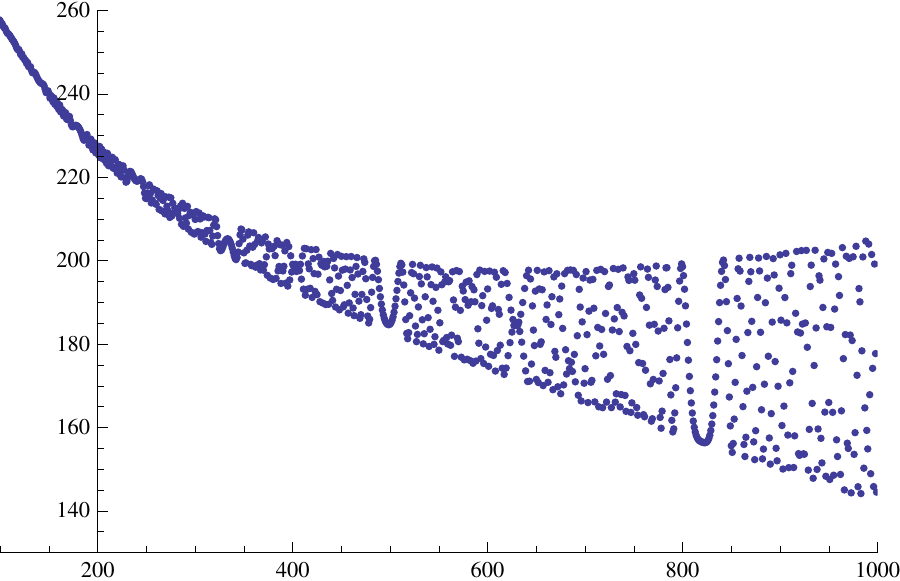} 
\end{center}
\caption{
Comparison of the QNM frequencies $\wbQNM$ given by the asymptotic expression Eq.(\ref{eq:QNM s=1}), which we will denote by $\wbQNM^a$,
and the numerical data in~\cite{QNMBerti} (after complex-conjugating it in order to account for the different definition
of the frequency), which we will denote by $\wbQNM^n$. For $s=1$ and $\ell=1$. Figs.(a)--(d) are log-plots.
Horizontal axis: overtone index $n$.
(a) In blue: $\left|\text{Re}(\wbQNM^n)\right|$; in green:  $\left|\text{Re}(\wbQNM^a)\right|$.
(b) In (dark) blue: $\left|\text{Im}(\wbQNM^n)+\frac{n}{2}\right|$; in green: the leading order for $\left|\text{Im}(\wbQNM^a)+\frac{n}{2}\right|$, 
i.e., $\frac{\lambda^2}{2n}$; in light-blue: $\left|\text{Im}(\wbQNM^a)+\frac{n}{2}\right|$.
(c): `Relative error' 
$\left| \left| \text{Re}(\wbQNM^n)/ \text{Re}(\wbQNM^a)\right|-1\right|$.
(d): Same as (c) but for the imaginary part.
(e) Plot of $n^{2}\cdot \text{Re}\left(\wbQNM^n-\wbQNM^a\right)$.
(f) Plot of $n^{5/2}\cdot \text{Im}\left(\wbQNM^n-\wbQNM^a\right)$.
Figs.(e) and (f) show, respectively, that the difference between the asymptotic Eq.(\ref{eq:QNM s=1}) and the numerical data in~\cite{QNMBerti} 
is only at order  $n^{-2}$ for the real part and $n^{-5/2}$ for the imaginary part.
 }
\label{fig:QNM numeric}
\end{figure}

For $s=1$,~\cite{Musiri:2003bv,Musiri:2007zz} show that the $O(1)$ and $O(n^{-1/2})$ terms in the QNM frequencies are zero as $n\to\infty$.
In~\cite{Cardoso:2003vt} they find numerical indications that for $s=1$ the highly-damped quasinormal modes go like
\begin{equation} \label{eq:QNM numerics s=1}
\wbQNM
\sim \frac{in}{2}+\frac{a_3 \lambda ^3+a_2 \lambda ^2+a_1 \lambda }{n^{3/2}},\quad n\to \infty
\end{equation}
with undetermined polynomial coefficients $a_1$, $a_2$ and $a_3$.
For $s=1$, from Eq.(\ref{eq:Wronsk large-nu MvdB}) we would find the QNM condition to be
\begin{equation} \label{eq:QNM cond s=1}
1+O\left(\nb^{-3/2}\right)+\left[\frac{2\aa}{\sqrt{\nb}}-\frac{2
i\aa^2
}{\nb}+O\left(\nb^{-3/2}\right)\right]\ca(\nb)e^{2\pi i\nb}=0
\end{equation}
Let us try first with the asymptotic expression $2\pi \nb\sim n\pi+a/n^{3/2}$, for some undetermined coefficient $a$. We then obtain,
from Eq.(\ref{eq:c_a MvdB}), that $\ca(\nu)=O(n)=O(\nb)$, but in Eq.(\ref{eq:QNM cond s=1}) we obtain a leading order $O(\sqrt{n})$,
which cannot be cancelled.
Trying then with $2\pi \nb\sim n\pi+a+\dots$ or $2\pi \nb\sim n\pi+a\sqrt{n}+\dots$ yields $\ca(\nu)=O(n^{-1/2})$ but the $1$ in 
Eq.(\ref{eq:QNM cond s=1}) cannot be cancelled.
Trying next with $2\pi \nb\sim n\pi+a/\sqrt{n}+\dots$ yields $\ca(\nu)=O(1)$, but then the $1$ in 
Eq.(\ref{eq:QNM cond s=1}) cannot be cancelled either.
Finally, let us try with $2\pi \nb\sim n\pi+a/n+b/n^{3/2}$, and then we obtain, from Eq.(\ref{eq:c_a MvdB}), that
 $\ca(\nu)=O(n^{1/2})=O(\nb^{1/2})$ and from Eq.(\ref{eq:QNM cond s=1}) it follows that:
$ a=4\aa^2$ and $b=
-4
\sqrt{2}
(1+i)\aa^3
$.
Therefore, we have that the highly-damped electromagnetic QNM frequencies are given by
 \begin{equation} \label{eq:QNM s=1}
 \wbQNM
= -\frac{in}{2}-\frac{i\lambda^2}{2n}+\frac{\pi^{1/2}(1-i)\lambda^3}{2^{3/2}n^{3/2}}
+O\left(\frac{1}{n^{2}}\right),\qquad
s=1
 \end{equation}
This form agrees with the form of the numerics of Eq.(\ref{eq:QNM numerics s=1}) (considering that only the leading order
in the imaginary part is shown) and
Fig.\ref{fig:QNM numeric} shows that 
it also
agrees with the numerical data in~\cite{QNMBerti}.

\section{High-frequency BC response to a perturbation} \label{Perturbation}

In this section we consider a perturbation to a Schwarzschild black hole and we investigate 
 the high-frequency BC contribution of the $\ell$-mode  to the response to such perturbation.
If the perturbation is given by
 some initial conditions $u_{\ell}^{ic}(r'_*)\equiv  u_{\ell}(r'_*, t'=0)$ and 
 $\dot u_{\ell}^{ic}(r'_*)\equiv  \partial_t u_{\ell}(r'_*, t'=0)$,
 then
the full $\ell$-mode response is given by
\begin{align} \label{eq:perturbation}
u_{\ell}(r_*, t)=\int_{-\infty}^{\infty}dr'_*\left[ G^{ret}_{\ell}(r,r'; t)\dot u_{\ell}^{ic}(r'_*)+u_{\ell}^{ic}(r'_*)\partial_tG^{ret}_{\ell}(r,r'; t)\right]
\end{align}
while the BC contribution to the response is given by
\begin{equation}\label{eq:perturbation BC}
u_{\ell}^{BC}(r_*, t)\equiv \int_{-\infty}^{\infty}dr'_*\left[ G_{\ell}^{BC}(r,r'; t)\dot u_{\ell}^{ic}(r'_*)+u_{\ell}^{ic}(r'_*)\partial_t G_{\ell}^{BC}(r,r'; t)\right]
\end{equation}
We will consider the case $\dot u_{\ell}^{ic}(r'_*)=\partial_{r'_*}u_{\ell}^{ic}(r'_*)$, corresponding to an initial  wave-packet moving towards the left. In addition, for convenience, we take a normalization given by $ \int_{-\infty}^{\infty}dr'_* \left|u_{\ell}^{ic}(r'_*)\right|=1$.
Let us define $\F(r,\omega)\equiv  \f(r,\omega) e^{-i\omega r_*}$, and by analogy with $\fb$ define  $\Fb(r,-i\nu)= - \sin ( 2 \pi \bar \nu) F(r,-i\nu)$.
Introducing Eq.(\ref{eq:DeltaG in terms of Deltag}) into Eq.(\ref{eq:perturbation BC}) with these initial conditions, we obtain:
\begin{align} \label{eq:perturbation BC excitation}
&
u_{\ell}^{BC}(r_*, \Delta t)=
\int_0^{\infty}d\nu\ I(r,\nu),\quad
I(r,\nu)\equiv
 -e^{-\nu (\Delta t-r_*)}
\frac{\nu q(\nu)\Fb(r,-i\nu)}{\pi\left|\hat W_{\pm}\right|^2}\hat C_{\ell}(\nu)
\\&
C_{\ell}(\nu)\equiv \int_{-\infty}^{+\infty}dr_*\ \F(r,-i\nu)
e^{\nu r'_*}
\left[-\nu+\partial_{r_*}\right]
u_{\ell}^{ic}(r_*),
\qquad \hat C_{\ell}(\nu)\equiv -\sin(2\pi\nb)C_{\ell}(\nu)  .
\nonumber
\end{align}
We have defined the function $C_{\ell}(\nu)$ so that, when evaluated at the quasinormal mode frequencies,
 it corresponds to the  `excitation coefficients' $\mathcal{C}_{\ell n}$ defined below for the QNM series.

We will now investigate the behaviour for large-$\nb$.
From Eq.(\ref{eq:f large-nu MvdB}), 
\begin{align} \label{eq:large-nu quantities in I}
\F(r,-i\nu)\sim  c(\nu)+e^{-2\nu r_*} \sim
\begin{cases}
&\dfrac{  (-1)^{s/2}}{\sin(2\pi\nb)}+e^{-2\nu r_*}
,\quad\quad\quad s=0,2 \\ &
\dfrac{\lambda \sqrt\pi}{2\sqrt{\nb}\sin(2\pi\nb)}+e^{-2\nu r_*}
,\quad s=1
\end{cases}
\end{align}
for $\nb\to \infty$.
 From Eqs.(\ref{eq:q for large-nu})  and (\ref{eq:Wronsk large-nu MvdB}), we have that as $\nb\to\infty$
\begin{align} \label{eq:nu q/W^2 large-nu}
s=0,2: & \quad
\frac{-2\nu q(\nu)}{|\hat W_{\pm}|^2}\sim \frac{(-1)^{s/2}2\cos(2\pi \nb)}
{\nu\left[1+3\cos^2(2\pi\nb)\right]}
,
\\ s=1: & \quad
\frac{-2\nu q(\nu)}{|\hat W_{\pm}|^2}\sim
\frac{-\lambda\sqrt\pi r_h}{\nb^{3/2}\sin(2\pi\nb)\left|1-\frac{\lambda^2\pi e^{2\pi i\nb}}{2\nb\sin(2\pi\nb)}\right|^2} .
\nonumber
\end{align}
 In order to study the large-$\nu$ asymptotics for $C_{\ell}(\nu)$, we first re-express it as 
 \begin{equation} \label{eq:large-nu exc.coeff. gral.}
 C_{\ell}(\nu)=-2\nu c(\nu) \int_{-\infty}^{\infty}
 dr_*\ e^{\nu r_*}u_{\ell}^{ic}(r_*)
 \end{equation}
 after integration by parts,
 where we have used Eq.(\ref{eq:large-nu quantities in I})
 and we have assumed that the initial conditions are such that the boundary terms are zero, i.e., that
$\F(r_*)e^{\nu r_*}u_{\ell}^{ic}(r_*)\to 0$ as $r_*\to \pm \infty$.

It is clear that if the initial conditions have compact support, with $u_{\ell}^{ic}(r_*)=0$ for all $r_*> R_*$, then
$\left| C_{\ell}(\nu)\right|\leq 2\left|c(\nu)\right|\nu e^{\nu R_*}$.
Together with the leading-order Eq.(\ref{eq:nu q/W^2 large-nu}),
this determines  that the integrand $I(r,\nu)$ in Eq.(\ref{eq:perturbation BC excitation})
goes, at most, like $e^{\nu \left(-\Delta t+\left|r_*\right|+R_*\right) }$ as $\nb\to \infty$ (ignoring powers of $\nb$).
Therefore, if the initial data is of compact support, the $\nu$-integral in the BC contribution to the perturbation response
will converge after a certain time: for $\Delta t>\left|r_*\right|+R_*$.

\begin{figure}[h!]
\begin{center}
\includegraphics[width=8cm]{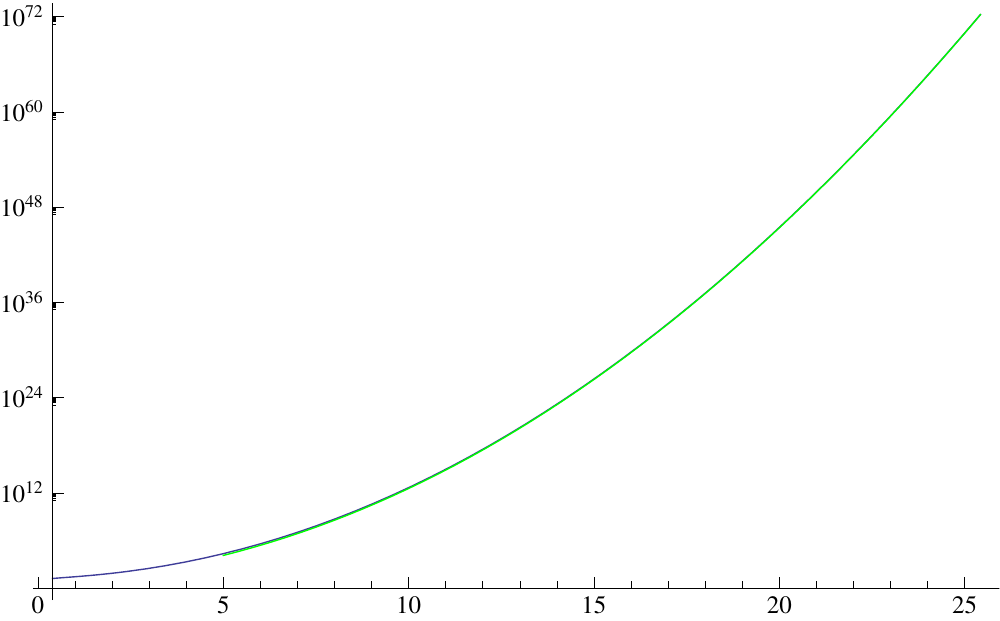} 
\end{center}
\caption{Log-plot of $\left|\hat C_{\ell}(\nu)\right|$ in Eq.(\ref{eq:perturbation BC excitation}) as a function of $\nu M$
 for $s=0$ and $\ell=1$.
 The blue curve is obtained with the method in~\cite{Casals:Ottewill:2011midBC}.
The overlapping curve in green is obtained using the large-$\nu$ asymptotics of Eq.(\ref{eq:f large-nu MvdB}) for $ \Fb$ in Eq.(\ref{eq:perturbation BC excitation}).
The coefficient $\hat C_{\ell}(\nu)$, $\forall \nu$, for 
both curves 
has been obtained by integrating using the built-in function {\it NIntegrate}
from $r=r_h$ up to $\infty$ in the computational software program {\it Mathematica}.
}
\label{fig:excitation}
\end{figure}

\begin{figure}[h!]
\begin{center}
\includegraphics[width=8cm]{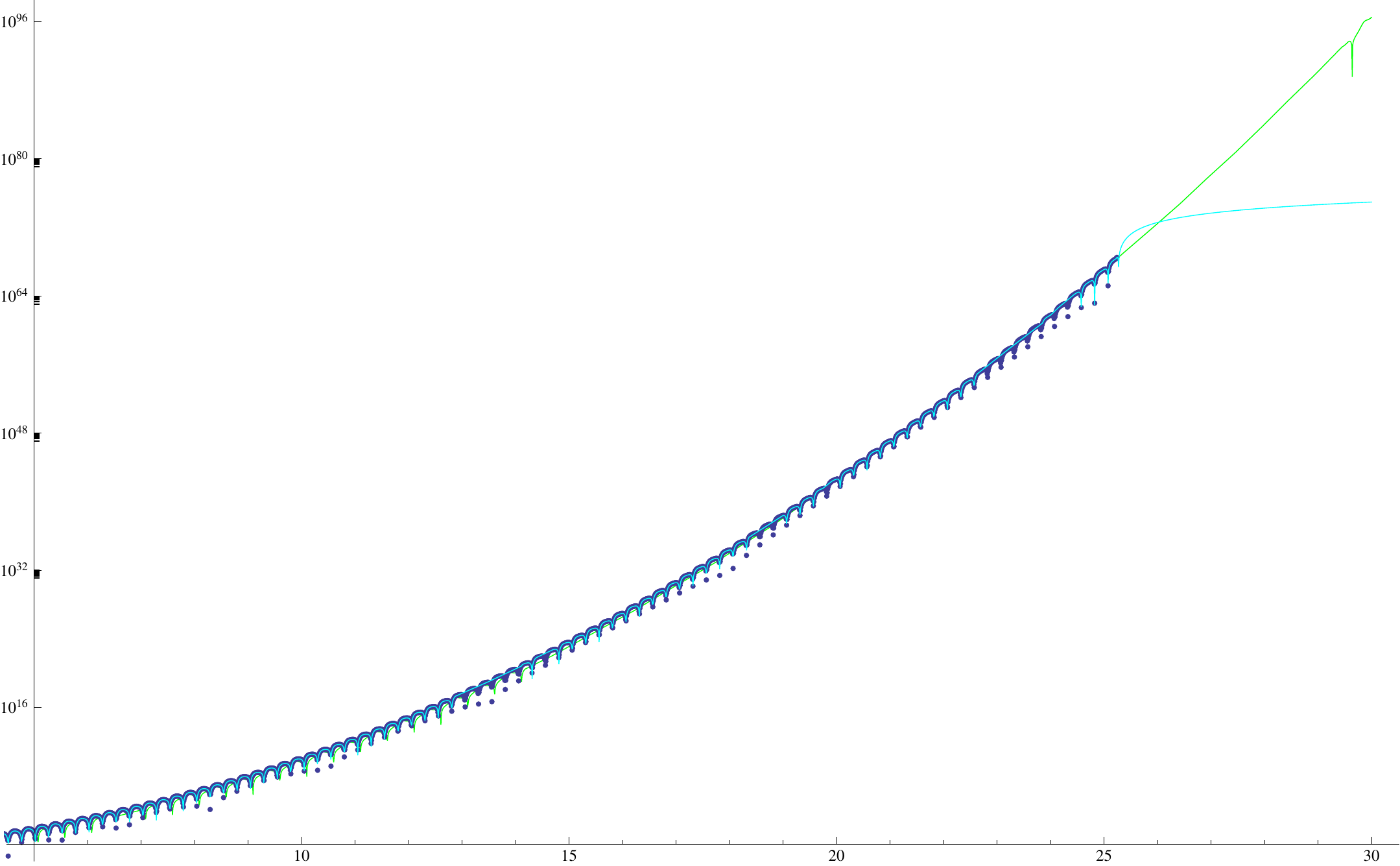} 
\end{center}
\caption{Log-plot as a function of $\nu M$ of
$|I(r=10M,\nu)|$ in Eq.(\ref{eq:perturbation BC excitation}) 
but without including the factor $e^{-\nu T}$. For $s=0$ and $\ell=1$. 
The darker blue curve is obtained with the method in~\cite{Casals:Ottewill:2011midBC} and
the light-blue curve is the interpolation of this data.
The (partly overlapping) green curve is obtained with the large-$\nu$ asymptotics: Eq.(\ref{eq:q for large-nu}) for $q(\nu)$,
Eq.(\ref{eq:f large-nu MvdB}) for $ \Fb$, Eq.(\ref{eq:Wronsk large-nu MvdB}) for the Wronskian and $C_{\ell}(\nu)$
as in the green curve in Fig.\ref{fig:excitation}.
}
\label{fig:integrand NoExpT}
\end{figure}

Let us now consider the case of non-compact initial conditions, specifically the case
 of a Gaussian distribution centered at $r_*=x_0$ and moving towards the black hole, which has frequently been used
in the literature:
\begin{equation} \label{eq:ic}
u_{\ell}^{ic}(r_*)=\frac{1}{\sqrt{2\pi}\sigma}\text{exp}\left(\frac{-\left(r_*-x_0\right)^2}{2\sigma^2}\right),
\quad
\dot u_{\ell}^{ic}(r_*)=-\frac{\left(r_*-x_0\right)}{\sigma^2}u_{\ell}^{ic}(r_*)
\end{equation}
for some $\sigma\in\mathbb{R}$.
Note that Eq.(\ref{eq:ic}) is the case of Fig.2 of Leaver~\cite{Leaver:1986} with $\sigma=2^{-1/2}(2M)^2$
(apart from a different overall constant factor).
The high-frequency asymptotics of $C_{\ell}(\nu)$ for the initial conditions (\ref{eq:ic}) are then given by:
\begin{align}\label{eq:large-nu exc.coeff.}
s=0,2: &\qquad
\hat C_{\ell}(\nu)\sim
(-1)^{s/2
} \frac{2 \nb}{r_h}
e^{\sigma^2\nu^2/2} e^{\nu x_0},\qquad \nb\to \infty,
\\
s=1: &\qquad
\hat C_{\ell}(\nu)\sim
\frac{\lambda\sqrt{\nb\pi}}{r_h
} e^{\sigma^2\nu^2/2} e^{\nu x_0}.
\nonumber
\end{align}
In Fig.\ref{fig:excitation} we plot $|\hat C_{\ell}(\nu)|$ as a function of $M\nu$.
The leading-orders in Eqs.(\ref{eq:nu q/W^2 large-nu}) and (\ref{eq:large-nu exc.coeff.}) together yield that the integrand $I(r,\nu)$
is of the order of $e^{\sigma^2\nu^2/2-\nu \left(\Delta t-\left|r_*\right|-x_0\right)}$ as $\nb\to \infty$ (ignoring powers of $\nb$).
This asymptotic behaviour implies that
the $\nu$-integral in Eq.(\ref{eq:perturbation BC excitation}) will not converge for any given values of $\Delta t $ and $r_*$.
 See Fig.\ref{fig:integrand NoExpT} for a plot of $|I(r,\nu)e^{\nu T}|$.
Leaver's Sec.III.A~\cite{Leaver:1986} considers the particular initial perturbation Eq.(\ref{eq:ic}) that we have used here. Eqs.54--56~\cite{Leaver:1986}, however, only investigate
the `late-time response', i.e., they are obtained via Eqs.38--44~\cite{Leaver:1986}, which are obtained for $\rb\gg 1$ and $\nb\ll 1$.
This is probably the reason why the large-$\nb$ divergence observed here went unnoticed in~\cite{Leaver:1986}. 
This  large-$\nb$ divergence for any $\Delta t $ and $r_*$ for Gaussian initial data
is in contrast with the convergence for $\Delta t>\left|r_*\right|+R_*$ in the case of initial data with compact support that we have seen above.

We  expect a similar behaviour for the highly-damped QNM's in the overtone $n$-sum for the QNM contribution to the perturbation response:
\begin{equation}\label{eq:perturbation QNM excitation}
u_{\ell}^{QNM}(r_*,t)=
 2\sum_{n=0}^{\infty}u_{\ell,n}^{QNM}(r_*,t),\quad
u_{\ell,n}^{QNM}(r_*,t)\equiv 
\text{Re}\left(
\frac{\mathcal{B}_{\ell n}}{\left(A^{out}_{\ell,\wQNM}\right)^2}\mathcal{C}_{\ell n}
 \F(r_*,\wQNM)
e^{-i \wQNM (\Delta t-r_*)}
\right)
\end{equation}
where $\mathcal{C}_{\ell n}\equiv C_{\ell}(\nuQNM)$,
the  QNM `excitation factors' are defined by $\mathcal{B}_{\ell  n}\equiv A^{out}_{\ell,\wQNM}/(\wQNM \alpha_{\ell n})$,
and $\alpha_{\ell n}$ is defined via $\Ain\sim (\omega-\wQNM)\alpha_{\ell n}$ as $\omega\to \wQNM $.
The $n$-sum in Eq.(\ref{eq:perturbation QNM excitation}) is over all QNMs in the fourth quadrant of the complex-$\omega$ plane. 
We note that the radial function $\g$ does not appear in Eq.(\ref{eq:perturbation QNM excitation}) because we may replace it by 
$\f/\Aout$ at a QNM frequency: the two quantities are equal when $\omega=\wQNM$, as follows from the fact that
$A^{in}_{\ell,\wQNM}=0$ and from the boundary conditions  (\ref{eq:f,near hor}) and (\ref{eq:f inf}).

Comparing Eq.(\ref{eq:perturbation QNM excitation}) with the BC contribution Eq.(\ref{eq:perturbation BC excitation}), we can  say that: 
$i\omega_{\ell n}$, $2\text{Re}\sum_{n=0}^{\infty}$, $\frac{\mathcal{B}_{\ell n}}{\left(A^{out}_{\ell,\wQNM}\right)^2}$, 
$\mathcal{C}_{\ell n}$
 in the QNM contribution `play the r\^ole' of, respectively:
$\nu$, $\int_0^{\infty}d\nu$, 
$\frac{-\nu q(\nu)}{\pi| W_{\pm}|^2}$, $C_{\ell}(\nu)$
in the BC contribution.
This is particularly true in the high-damping limit, $n\to \infty$.

Let us
find the large-$n$ asymptotics of $\mathcal{B}_{\ell n}$.
By comparing Eqs.(\ref{eq:f inf}) and (\ref{eq:f large-nu MvdB}) for $r\to \infty$ we have that $\Aout \sim c(\nu)$ as $\nb \to \infty$,
which is valid since $\Aout$ is the coefficient of the dominant solution for $r_*>0$.
From Eq.(\ref{eq:c_a MvdB}) it then follows that, to leading order as $\nb\to\infty$,
\begin{align} \label{eq:Aout large-nu}
s=0,2: & \quad
\Aout\sim\frac{(-1)^{s/2}}{\sin(2\pi\nb)},\qquad
\\s=1: & \quad
\Aout\sim\frac{\sqrt\pi\lambda}{2\sqrt{\nb}\sin(2\pi\nb)},\quad
\nonumber
\end{align}
in the 4th quadrant of the complex-$\omega$ plane.
In order to compare with~\cite{Neitzke:2003mz}, let us note that our definition of tortoise coordinate $r_*$ equals
that in Eq.B.1~\cite{Neitzke:2003mz} plus the constant `$i\pi$'.
Therefore, our coefficients $\Ain$ and $\Aout$  correspond, respectively, to $1/T_N$ and $e^{-2\pi i\nb}R_N/T_N$, where 
$R_N$ and $T_N$ are the reflection and transmission coefficients defined by Neitzke via Eq.2.10~\cite{Neitzke:2003mz}.
Eq.(\ref{eq:Aout large-nu}) for $s=0$ and $2$ then 
agrees with Eqs.2.17 and 2.18~\cite{Neitzke:2003mz},
after also taking into account a different sign in the definition of $\omega$ and converting expression (\ref{eq:Aout large-nu})
from the 4th quadrant into the 3rd quadrant by using the symmetries (\ref{eq:symms f,g}).
Note that the different constant of integration `$i\pi$' in $r_*=r_*(r)$ is probably the reason why $\Aout$ in Eq.(\ref{eq:Aout large-nu}) for $s=0$ and $2$ 
differs by a factor $e^{-2\pi i\nb}$ from  Eq.A.2~\cite{Berti:Cardoso:2006}.
In the Appendix~\ref{sec:appendix} we calculate the relationship between the coefficients $\Aout$ and $\Ain$ following a different contour
in the complex-$\omega$ plane, thus providing a 
check of Eq.(\ref{eq:Aout large-nu}).
We can now calculate that as $n\to\infty$
\begin{align}\label{eq:exc factor n->inf}
s=0,2: & \quad
\alpha_{\ell n}\sim 3\pi r_h, \quad
\mathcal{B}_{\ell n}\sim \frac{(-1)^{n+s/2}i}{\sqrt{3}\pi n},\quad 
\\
s=1: & \quad
\alpha_{\ell n}\sim \frac{2i r_h n}{\lambda^2}, \qquad
\mathcal{B}_{\ell n}\sim \frac{(-1)^{n}\lambda}{\sqrt{2\pi} n^{3/2}},\nonumber
\end{align}
where we have used Eq.(\ref{eq:Wronsk large-nu MvdB}) to obtain $\alpha_{\ell n}$ and Eqs.(\ref{eq:QNM s=0,2}) and (\ref{eq:QNM s=1}) for the 
highly-damped QNM frequencies.
The behaviour $\mathcal{B}_{\ell n}=O(n^{-1})$ for $s=0$
seems to roughly agree with Fig.2~\cite{Andersson:1997}  (although those results are not really meant to be valid for large-$n$).
To the best of our knowledge, the expressions in (\ref{eq:Aout large-nu}) and (\ref{eq:exc factor n->inf}) for the $s=1$ case are given here for the first time
in the literature.

 The asymptotics as $n\to\infty$ for $ \F(r_*,\wQNM)$ and  $\mathcal{C}_{\ell n}$ are those of 
$ \F(r_*,-i\nu)$ and $C_{\ell}(\nu)$ in Eqs.(\ref{eq:large-nu quantities in I}) and (\ref{eq:large-nu exc.coeff. gral.}), 
respectively, with the replacement $\nu\to i\omega_{\ell n}$.
Combining the asymptotics of Eqs.(\ref{eq:large-nu quantities in I}), (\ref{eq:Aout large-nu}) and 
(\ref{eq:exc factor n->inf})
we find that, for $n\to \infty$,
\begin{align} \label{eq:QNM n-term large-n gral}
& s=0,2:  \quad
u_{\ell,n}^{QNM}(r_*,t)
\sim
\frac{(-1)^{n+s/2}4}{3\sqrt 3\pi n}\ \text{Re} \left[ i
\mathcal{C}_{\ell n}
\left(c(\nuQNM)e^{-\nuQNM (\Delta t - r_*)}+e^{-\nuQNM (\Delta t +r_*)}\right)\right]
\\ & s=1:  \quad
u_{\ell,n}^{QNM}(r_*,t)
\sim
\frac{(-1)^{n}\lambda^3\sqrt{2\pi}}{n^{5/2}}\ \text{Re} \left[
\mathcal{C}_{\ell n}
\left(c(\nuQNM)e^{-\nuQNM (\Delta t - r_*)}+e^{-\nuQNM (\Delta t +r_*)}\right)\right]
\nonumber
\end{align}
where a right-moving and a left-moving wave in the radial direction can be seen, both exponentially-damped with time.
In particular, if the initial data is of compact support vanishing for  $r_* > R_*$, then, for large-$n$,
$\left| \mathcal{C}_{\ell n}\right| <\left|c(\nuQNM)\right| n e^{n R_*/2}$ and
it then follows from Eq.(\ref{eq:QNM n-term large-n gral}) that
the QNM $n$-sum in Eq.(\ref{eq:perturbation QNM excitation})
will be convergent for $\Delta t>\left|r_*\right|+R_*$, just like the corresponding BC contribution.

Let us now look at the case of Gaussian initial data, Eq.(\ref{eq:ic});
from Eq.(\ref{eq:large-nu exc.coeff.}) we find as $n\to \infty$
\begin{align}\label{eq:large-nu exc.coeff. QNM}
s=0,2:&\quad
\mathcal{C}_{\ell n}\sim
\frac{(-1)^{s/2+1+n}\sqrt{3} n}{ 2 r_h}
e^{\bar\sigma^2\left[N^2+\frac{\ln 3}{\pi}iN-\frac{\ln^2 3}{2\pi^2}\right]/8}e^{N\bar x_0/2+i\bar x_0\ln 3/(4\pi)},\qquad N\equiv n+1/2
\\  s=1:&\quad
\mathcal{C}_{\ell n}\sim
\frac{(-1)^{n+1} n^{3/2}}{\sqrt{2\pi}\lambda r_h}
e^{\bar \sigma^2 n^2/8}e^{n\bar x_0/2}
\nonumber
\end{align}
The divergence of $\mathcal{C}_{\ell n}$ for large-$n$ is not cancelled out by any other quantity in Eq.(\ref{eq:QNM n-term large-n gral}),
 so that the QNM $n$-sum for the non-compact initial data (\ref{eq:ic}) is not convergent, just like the $\nu$-integral in the corresponding BC contribution.
Specifically, from Eqs.(\ref{eq:QNM n-term large-n gral}) and (\ref{eq:large-nu exc.coeff. QNM}),
\begin{align} \label{eq:QNM n-term large-n}
& s=0,2:  \quad
u_{\ell,n}^{QNM}(r_*,t)
\sim
-\frac{2
}{3
\pi
r_h}e^{\bar\sigma^2\left(N^2-\frac{\ln^2 3}{4\pi^2}\right)/8+N x_0/2}
\times \\ &
\left[ \dfrac{  (-1)^{s/2+n}\sqrt 3}{2}e^{-N(\Delta \bar t-\bar r_*)/2}\sin\left(\frac{\ln 3}{4\pi}\left(\frac{\bar\sigma^2N}{2}-\Delta \bar t+\bar r_*+\bar x_0\right)\right)+
e^{-N(\Delta \bar t+\bar r_*)/2}\sin\left(\frac{\ln 3}{4\pi}\left(\frac{\bar\sigma^2N}{2}-\Delta \bar t-\bar r_*+\bar x_0\right)\right)\right]
\nonumber
\\ & s=1:  \quad
u_{\ell,n}^{QNM}(r_*,t)
\sim
-\frac{
\lambda^2 
}{r_h n}
e^{\bar\sigma^2n^2/8+n \bar x_0/2}
\left[\dfrac{  (-1)^{n}\sqrt{n}}{\sqrt{2\pi} \lambda}e^{-n  (\Delta \bar t - \bar r_* )/2}+e^{-n  (\Delta \bar t +\bar r_*)/2}\right]
\nonumber
\end{align}
A na\"ive attempt at the calculation of the corresponding large-frequency divergence in the BC contribution indicates that it does not cancel out
the large-$n$ divergence in the QNM contribution in the case of Gaussian initial data, although it is hard to be definitive given the
numerous simultaneous asymptotic limits and integrals involved.
On the other hand, the full perturbation response Eq.(\ref{eq:perturbation}) is known to be well defined,
therefore, we expect the high-frequency arc contribution together with the BC and QNM contributions to be regular, with the divergences in the different contributions cancelling each other out.
We also expect a similar cancellation between the high-frequency divergences from 
the BC, QNM and high-frequency arc contributions in the case of compact initial data
 for $\Delta t<\left|r_*\right|+x_0$.

\section{Conclusions} \label{sec:conclusions}

The branch cut contribution to the Green function in the Schwarzschild spacetime has scarcely been investigated in the literature, except in the small-frequency regime.
The challenging technical difficulties for the calculation away from that asymptotic regime is probably a reason for it.
However, in order to have an understanding of the full response of the black hole to an initial perturbation or, in general, to know the 
Green function globally, we also require the non-small-frequency contribution.
In~\cite{CDOWa}, the self-force was calculated in the particular case of a static particle in (a static patch of) the Nariai spacetime.
In such spacetime the `retarded' Green function possesses no BC (since the radial potential falls off exponentially near the `cosmological horizons');
however, in the Schwarzschild spacetime the Green function does possess a BC, the knowledge of which is 
required if we wish to perform a similar calculation of the self-force to the one
performed in~\cite{CDOWa} in Narai.
In this paper we have derived the large-frequency asymptotics for the BC contribution to the Green function modes for spins $0$, $1$ and $2$ in 
the Schwarzschild spacetime --  see Eq.(\ref{eq:DG large-nu MvdB}) for the leading order.
These asymptotics lead to a divergence of the BC Green function at the `very early' times $\Delta t<|r_*|+|r'_*|$ and
they lead to a  convergence of the $\nu$-integral in the upper limit when $\Delta t>|r_*|+|r'_*|$.
In  Sec.\ref{Perturbation} we have investigated the contributions of the high-frequency BC
and the high-frequency QNM to the perturbation response given an initial perturbation.
We have shown that if the initial data is of compact support within $r_*<R_*$, 
then both contributions separately converge when $\Delta t>|r_*|+R_*$; however,
when the initial data is a non-compact Gaussian distribution, both contributions diverge separately for any fixed 
time $\Delta t $ and radius $r_*$.
We expect that the divergences from all contributions to the Green function (i.e., from the QNM series, the BC and the high-frequency arc)  to cancel
each other out so that the full perturbation response is finite -- we leave this study for future work.

We have also obtained in Eq.(\ref{eq:QNM s=1})
 the highly-damped QNM frequencies for spin-$1$ (and reproduced existing results for the corresponding spin-$0$ and spin-$2$ frequencies)
  in Schwarzschild for the first time in the literature (other than the previously-known leading-order for the imaginary part).
In order to obtain the leading order of the spin-1 asymptotics we had to go up to two orders higher for large-$\nb$ than is necessary for $s=0$ and $2$.
The real part of these spin-1 frequencies approaches the NIA, unlike for $s=0$ and  $2$, and it does so faster (like $n^{-3/2}$) than for
$s=1/2$ and $5/2$  (which go like $n^{-1/2}$).


\begin{acknowledgments}
We thank Emanuele Berti and V\'itor Cardoso for making available the numerical data in~\cite{QNMBerti}.
M.C. acknowledges funding support from the Irish Research Council for Science, Engineering and Technology, co-funded by Marie Curie Actions under FP7. A.O. acknowledges support from Science Foundation Ireland under grant no 10/RFP/PHY2847.
\end{acknowledgments}


\appendix

\section{Check on the relationship between $\Aout$ and $\Ain$ }
\label{sec:appendix}

In this appendix we calculate the relationship between the coefficients $\Aout$ and $\Ain$ by following closely the method in~\cite{Neitzke:2003mz}: we will analytically
continue the radial function $\f$ starting at point A in Fig.\ref{fig:antiStokes}, then down to point B, where $\arg(r)=3\pi/4$,  then around an argument 
of `$-3\pi/2$' to reach the anti-Stokes line at $\arg(r)=-3\pi/4$, then down that anti-Stokes line to radial infinity (i.e., to the `reflection
point' of point A), and finally all around radial infinity anticlockwise back to point A (this is Fig.1~\cite{Neitzke:2003mz}). We will then impose the exact
monodromy Eq.(\ref{eq:f monodromy}). 
This is a check on the relationship between $\Aout$ and $\Ain$ found in Eqs.(\ref{eq:Wronsk large-nu MvdB})
and (\ref{eq:Aout large-nu}) following a different contour.
We will do it here only for $s=2$.

We start with the boundary condition $\f(r,-i\nu)\sim \Aout g_a(r,-i\nu)+\Ain g_a(r,+i\nu)$ at the point A in Fig.\ref{fig:antiStokes}, from Eq.(\ref{eq:f inf}).
At  $\arg(r)=3\pi/4$ we have
\begin{align} \label{eq:psi_1,2 argr=3pi/4 s=2}
\frac{\sqrt{\pi\nb^3}}{4}\psi_1^{(0)}(t)&\sim -e^{t^2/2}-ie^{-t^2/2}\sim 
 -g_a(r,i\nu)e^{i\pi\nb}-ig_a(r,-i\nu)e^{-i\pi\nb}
\\
\frac{4}{\sqrt{\pi\nb}}\psi_2^{(0)}(t)&\sim -3ie^{t^2/2}+e^{-t^2/2}\sim
-3ig_a(r,i\nu)e^{i\pi\nb}+g_a(r,-i\nu)e^{-i\pi\nb}
\nonumber
\end{align}
The first step follows from Eqs.(\ref{eq:psi_1,2^0,1(it)}) and (\ref{eq:psi_1,2 for 1<<t<<nu^1/6});
in the second step we have used Eq.(\ref{eq:g t/nu^1/6<<1}).
We solve for $g_a(r,\pm i\nu)$ and replace in the above  boundary condition to obtain
\begin{equation} \label{eq:f argr=-3pi/4 s=2}
\f(r,-i\nu) \sim \psi_1^{(0)}(t) \frac{\sqrt{\pi\nb^3}}{8}\left[3 i\Aout e^{i\pi\nb}+\Ain e^{-i\pi\nb}\right]+
\psi_2^{(0)}(t) \frac{2}{\sqrt{\pi\nb}}\left[-\Aout e^{i\pi\nb}+i\Ain e^{-i\pi\nb}\right]
\end{equation}
We have obtained this expression for $\arg(r)=3\pi/4$, but it is valid $\forall \arg(r)$ by analytic continuation.
We now explicitly evaluate it at $\arg(r)=-3\pi/4$.
From Eqs.(\ref{eq:Bessel func cont}) together with Eq.(\ref{eq:psi_1,2 argr=3pi/4 s=2}) we obtain
\begin{align}  \label{eq:psi_1,2 argr=-3pi/4 s=2}
\psi_1^{(0)}(te^{-3\pi i/2})&=-i\psi_1^{(0)}(t), &\qquad 
\frac{\sqrt{\pi\nb^3}}{4}\psi_1^{(0)}(t)&\sim -e^{t^2/2}+ie^{-t^2/2}\sim
-g_a(r,i\nu)e^{-i\pi\nb}+ig_a(r,-i\nu)e^{i\pi\nb}
\\
\psi_2^{(0)}(te^{-3\pi i/2})&=-i\psi_2^{(0)}(t)-\frac{3\nb^2\pi}{8}\psi_1^{(0)}(t),&\qquad 
\frac{4}{\sqrt{\pi\nb}}\psi_2^{(0)}(t)&\sim 5ie^{t^2/2}+3e^{-t^2/2}\sim
5ig_a(r,i\nu)e^{-i\pi\nb}+3g_a(r,-i\nu)e^{i\pi\nb}
\nonumber
\end{align}
at  $\arg(r)=-3\pi/4$, where in the last step we have again used the top equation in (\ref{eq:g t/nu^1/6<<1}). 
Introducing these expressions into Eq.(\ref{eq:f argr=-3pi/4 s=2}) we obtain
\begin{equation} \label{eq:f argr=-3pi/4 s=2 r inf}
\f (r,-i\nu)\sim g_a(r,-i\nu)e^{-i\pi\nb}\left[-3\Aout e^{i\pi\nb}+2i\Ain e^{-i\pi\nb}\right]+
g_a(r,i\nu)e^{i\pi\nb}\left[i\Aout e^{i\pi\nb}-3\Ain e^{-i\pi\nb}\right]
\end{equation}
at the `reflection point' of point A in Fig.~\ref{fig:antiStokes}.
We can analytically continue this expression all around infinity anticlockwise and back to point A.
We then equate this expression to the original boundary condition given above after applying the monodromy  Eq.(\ref{eq:f monodromy}),
i.e., we equate Eq.(\ref{eq:f argr=-3pi/4 s=2 r inf}) to $\f\left((r-r_h)e^{2\pi i},-i\nu\right)\sim e^{-2\pi i\nb}\Aout g_a(r,-i\nu)+e^{-2\pi i\nb}\Ain g_a(r,+i\nu)$.
Equating the coefficient of $g_a(r,-i\nu)$ gives precisely the relationship between $\Ain$ and $\Aout$ that follows from Eqs.(\ref{eq:Wronsk large-nu MvdB})
and (\ref{eq:Aout large-nu}). 
Note that  the coefficient of $g_a(r,i\nu)$  is not to be trusted since when closing the contour at infinity it is $\text{Re}(r_*)>0$ and so $g_a(r,i\nu)$ is the subdominant solution there.


\bibliographystyle{apsrev}


\end{document}